\documentclass[11pt, a4paper]{article}
\usepackage[utf8]{inputenc}
\usepackage{authblk}
\usepackage{enumitem}
\usepackage[babel=true]{csquotes}
\usepackage{caption}
\usepackage{subcaption}
\usepackage{color}
\usepackage{dsfont}
\usepackage{booktabs}
\usepackage{capt-of}
\usepackage{multicol}

\usepackage{appendix}
\usepackage{float}
\usepackage{url}
\usepackage[colorlinks=true,bookmarks=false,linkcolor=blue,citecolor=blue,urlcolor=blue]{hyperref}
\usepackage{array,tabu,multirow,makecell}
\usepackage{threeparttable}
\usepackage{array}
\usepackage{longtable}
\usepackage{rotating}
\usepackage{tabularx}
\usepackage{lastpage}
\usepackage{lscape}
\usepackage{stackengine}
\usepackage{amsmath} 
\usepackage{amsfonts}
\usepackage[T1]{fontenc}
\usepackage{amsthm}
\usepackage{geometry}
\usepackage{graphicx}
\usepackage{bbm}
\usepackage{moreverb}
\usepackage{color, pdfcolmk}
\usepackage{fancyhdr}
\usepackage{csquotes}  
\usepackage{etoolbox}
\makeatletter
\def\@footnotecolor{blue}
\define@key{Hyp}{footnotecolor}{%
 \HyColor@HyperrefColor{#1}\@footnotecolor%
}
\patchcmd{\@footnotemark}{\hyper@linkstart{link}}{\hyper@linkstart{footnote}}{}{}
\makeatother
\hypersetup{footnotecolor=blue}
\theoremstyle{plain}% default
\usepackage[round]{natbib}
\usepackage[english]{babel}
\usepackage{pgfplots}
\pgfplotsset{compat=1.18}
\usetikzlibrary{arrows.meta,decorations.pathreplacing,positioning}

\title{Predictive Accuracy versus Interpretability in Energy Markets:
A Copula-Enhanced TVP-SVAR Analysis}

\author[1]{Fredy POKOU} %\thanks{\texttt{fredypokou@gmx.fr}}}

\author[2]{Jules SADEFO KAMDEM} %\thanks{\texttt{jules.sadefo-kamdem@umontpellier.fr}}}

\author[3]{Emmanuel GNANDI} %\thanks{\texttt{kpanteemmanuel@gmail.com}}
\affil[1]{Inria, CNRS, Univ. of Lille, Centrale Lille, UMR 9189 - CRIStAL, F-59000 Lille, France}

\affil[2]{MRE UR 209 and Faculty of Economics. Montpellier University, France}

\affil[3]{INSA de Toulouse, Département de Génie Mathématique, Toulouse, France}

\begin{document}

\maketitle

\begin{abstract}
This paper investigates whether structural econometric models can rival machine learning in forecasting energy–macro dynamics while retaining causal interpretability. Using monthly data from 1999 to 2025, we develop a unified framework that integrates Time-Varying Parameter Structural VARs (TVP-SVAR) with advanced dependence structures, including DCC-GARCH, t-copulas, and mixed Clayton–Frank–Gumbel copulas. These models are empirically evaluated against leading machine learning techniques Gaussian Process Regression (GPR), Artificial Neural Networks, Random Forests, and Support Vector Regression across seven macro-financial and energy variables, with Brent crude oil as the central asset.
The findings reveal three major insights. First, TVP-SVAR consistently outperforms standard VAR models, confirming structural instability in energy transmission channels. Second, copula-based extensions capture non-linear and tail dependence more effectively than symmetric DCC models, particularly during periods of macroeconomic stress. Third, despite their methodological differences, copula-enhanced econometric models and GPR achieve statistically equivalent predictive accuracy (t-test p = 0.8444). However, only the econometric approach provides interpretable impulse responses, regime shifts, and tail-risk diagnostics.
We conclude that machine learning can replicate predictive performance but cannot substitute the explanatory power of structural econometrics. This synthesis offers a pathway where AI accuracy and economic interpretability jointly inform energy policy and risk management.
\end{abstract}


\section{Introduction}
\label{sec1}
\paragraph{Context: }
Over the past two decades, the integration between global energy markets and macro-financial systems has intensified to an unprecedented degree, fundamentally reshaping how shocks propagate across sectors and borders. Major crises such as the 2008 global financial collapse, the 2014–2015 oil price breakdown, the COVID-19 pandemic, and recent geopolitical disruptions have revealed that oil price dynamics is not simply driven by linear interactions with monetary policy, industrial demand, or global uncertainty. Instead, they exhibit regime shifts, volatility clustering, and asymmetric spillovers, especially during episodes of financial distress and geopolitical stress.
These events have exposed the limitations of conventional econometric tools grounded in parameter constancy, Gaussian innovations, and symmetric dependence structures.\\
For policymakers, investors, and risk managers, the challenge is no longer limited to predicting energy prices but to understanding the structural mechanisms through which macroeconomic shocks propagate to commodity markets under evolving regimes and extreme tail conditions. This dual requirement forecast accuracy and interpretability under non-linear dependence—defines one of the central empirical frontiers in modern energy economics.
\paragraph{Problem Statement: }
Vector Autoregressive (VAR) models remain a foundational tool for analysing macro–energy interdependence, yet their stationarity and Gaussian assumptions inhibit their capacity to capture structural breaks, asymmetric contagion, and tail co-movement. Even Time-Varying Parameter SVARs (TVP-SVAR), which accommodate parameter drift, continue to rely on elliptical error structures and primarily capture mean propagation, neglecting dynamic volatility and dependence. In contrast, machine learning models particularly Gaussian Process Regression (GPR) can flexibly approximate highly non-linear relationships but offer no structural interpretation, making them unsuitable for economic policy or risk attribution.
This leads to a fundamental question at the crossroads of econometrics and artificial intelligence:\\
\textbf{Can structurally grounded econometric models rival state-of-the-art machine learning approaches in predictive accuracy, while preserving interpretability over macro-financial energy dynamics?} \\
Addressing this question requires an integrated empirical framework combining:
\begin{itemize}
\item[](i) time-varying structural transmission (TVP-SVAR),
\item[](ii) dynamic volatility and dependence (DCC-GARCH, Copulas), and
\item[](iii) nonparametric predictive benchmarks (GPR, ANN, RF, SVM).
\end{itemize}

\paragraph{Main Contributions: }
This paper advances the empirical modelling of energy–macro linkages in four major dimensions:
\begin{enumerate}
\item Hybrid Structural–Nonlinear Framework\\
   We construct a unified modelling architecture that nests VAR and TVP-SVAR within volatility-aware extensions (DCC-GARCH, t-copulas, and Clayton–Frank–Gumbel mixed copulas), enabling joint analysis of causal transmission, regime shifts, and tail dependence.

\item Dynamic Copula Diagnostics and Tail Contagion\\
   Using global and tail-focused goodness-of-fit tests, we provide one of the first formal assessments of dependence structures on VAR versus TVP-SVAR residuals. Results demonstrate that time variation improves central dependence but fails to capture upper-tail contagion, justifying the use of mixed copulas.

\item Predictive Parity with Machine Learning\\
   Through a comparative evaluation of RMSE distributions and t-tests, we show that copula-enhanced econometric models can match the predictive accuracy of leading ML models (GPR, ANN, RF, SVM), challenging the notion that black-box methods dominate structural modelling.

\item Econometric Interpretability as Structural Surrogate for ML\\
   We demonstrate that when GPR replicates the TVP-SVAR’s forecasting performance, it does so by implicitly learning the same regime dynamics. Thus, econometric models serve as structural interpreters of machine learning outputs, reconciling predictive strength with economic meaning.
\end{enumerate}

\paragraph{Literature review: }
Empirical research on energy–macro dynamics has historically been dominated by Vector Autoregressive (VAR) models \citep{sims1980macroeconomics}, which remain a cornerstone for analysing lagged interactions among macroeconomic and commodity variables. However, conventional VARs are limited by their assumption of time-invariant parameters, which is incompatible with markets characterised by frequent structural breaks, monetary regime shifts, and geopolitical shocks \citep{hamilton2009causes}. 
\newpage
\noindent Structural VARs (SVAR), incorporating economically motivated restrictions \citep{blanchard1988dynamic,kilian2014role}, improved shock identification but nevertheless retained the rigidity of constant structural coefficients.\\

\noindent Recognising the prevalence of evolving policy regimes and market behaviour, Time-Varying Parameter SVAR (TVP-SVAR) emerged as a breakthrough \citep{cogley2005drifts,primiceri2005time}, allowing transmission mechanisms to evolve through state-dependent parameters. These models have been widely applied in energy markets to detect dynamic connectedness \citep{mishra2022dynamic}, however, mainly mean-based frameworks remain, assuming Gaussian disturbances and ignoring nonlinear or asymmetric dependence on volatility. As a result, they often fail to capture tail-driven contagion, particularly during oil price collapses (2008, 2014–2015) or energy shocks amplified by global uncertainty \citep{zhang2025unravelling}.\\

\noindent To incorporate heteroskedasticity and volatility transmission, multivariate GARCH models, particularly Dynamic Conditional Correlation (DCC) by \cite{engle2002dynamic}, have become a standard tool in financial econometrics. The asymmetric DCC extension (ADCC) by \citet*{cappiello2006asymmetric} introduced the ability to differentiate between positive and negative shocks, an essential feature for modeling crisis contagion in oil and equity markets \citep{hafner2006volatility}. However, these models rely on elliptical distributions (Gaussian or Student-t), imposing symmetric dependence and rendering them inadequate for upper- and lower-tail asymmetries, frequently observed in energy markets \citep{baumeister2013time,bu2025extreme}.\\

\noindent In response to these limitations, copula-based dependence modeling has gained prominence. Archimedean copulas Clayton for downside risk, Gumbel for speculative booms, and Frank for symmetric co-movement provide greater flexibility in capturing tail dependence and nonlinear linkages \citep{patton2012review,genest2009goodness}. Recent contributions in commodity markets have used time-varying or mixture copulas to identify endogenous regime switching \citep{pokou2024empirical,bu2025extreme}, but these models remain parametric and static in structure, often incapable of adapting to real-time regime changes without external state indicators. \\

\noindent Parallel to these econometric advances, machine learning models, including Artificial Neural Networks (ANN), Support Vector Regression (SVR), Random Forests (RF), and Gaussian Process Regression (GPR), have demonstrated powerful predictive performance in energy price forecasting and volatility modeling \citep{wu2014gaussian,han2016gaussian,medeiros2022forecasting}. Nevertheless, their inherent black-box nature precludes causal inference and policy interpretation, making it a challenge for regulatory and strategic decision making \citep{zheng2017econometrics}.\\

\noindent Against this backdrop, our work positions itself at the intersection of structural interpretability and nonparametric predictive power. We advance the literature by establishing that structural econometric models, when augmented with copula-based dependence, can achieve predictive performance comparable to state-of-the-art ML, while uniquely preserving the ability to analyze impulse responses, volatility spillovers, and regime-specific contagion. Rather than opposing econometrics and ML, our findings demonstrate their complementarity:
\textbf{machine learning validates empirical relevance, while econometrics reveal structural meaning.}

\paragraph{Paper Organization: }
The remainder of the paper is structured as follows. Section \ref{sec2} presents the methodological framework, covering VAR, SVAR, TVP-SVAR, DCC-GARCH, copula models, and their hybrid extension with Gaussian Process Regression. Section \ref{sec3} reports the empirical findings, including structural stability tests, impulse responses, dependence diagnostics, and comparative predictive evaluation against machine learning benchmarks. Section \ref{sec4} discusses the implications of predictive parity and structural interpretability between econometric and ML approaches. Section \ref{sec5} concludes with policy insights and directions for future research.

\section{Methodology}
\label{sec2}
This section outlines the econometric and machine learning framework adopted to model and forecast the interconnected dynamics between energy markets and macro-financial variables. The methodology integrates four key pillars: (i) linear benchmark models (VAR and SVAR), (ii) time-varying structural models (TVP-SVAR), (iii) volatility-aware dependence structures (DCC and Copula-GARCH), and (iv) hybrid extensions via Gaussian Process Regression (GPR). This layered architecture enables a comprehensive decomposition of mean dynamics, volatility spillovers, and nonlinear dependence a methodological synthesis rarely implemented simultaneously in energy economics.

\subsection{Vector Autoregression (VAR)}
We begin with a Vector Autoregression (VAR) of order $p$, the canonical framework introduced by \cite{sims1980macroeconomics} to model joint dynamics among multiple endogenous variables without imposing strong theoretical priors. Formally, the system is written as:
\begin{equation}
Y_t = c + A_1 Y_{t-1} + \cdots + A_p Y_{t-p} + \varepsilon_t,
\qquad \varepsilon_t \sim \mathcal{N}(0,\Sigma),
\label{e1}
\end{equation}
where $Y_t \in \mathbb{R}^k$ is a vector of endogenous macro-financial and energy series, $c$ is a vector of deterministic components, $A_i$ are autoregressive coefficient matrices, and $\varepsilon_t$ denotes reduced-form disturbances with covariance matrix $\Sigma$.
The VAR provides a flexible empirical structure capable of capturing feedback effects and lagged transmission across variables. However, it operates under two stringent assumptions:
\begin{itemize}
\item[](i) parameter constancy over time, and
\item[](ii) Gaussian innovations, implying symmetric linear dependence.
\end{itemize}
Such restrictions are known to be violated in periods of structural reorganization, changes in the policy regime, or turbulence in the energy market \citep{lutkepohl1993introduction}.
\subsection{Structural VAR (SVAR) and Shock Identification}
To uncover economically meaningful shocks, the VAR is extended into a Structural VAR (SVAR) by imposing identification restrictions on contemporaneous interactions. The structural form is given by:
\begin{equation}
A Y_t = B_1 Y_{t-1} + u_t,
\qquad E(u_t u_t') = \Lambda,
\label{e2}
\end{equation}
where $A$ encodes contemporaneous relations among variables, and $u_t$ denotes orthogonal structural innovations. Identification is typically achieved through short-term recursive restrictions, often implemented via Cholesky decomposition, which enforce a lower-triangular structure in $A^{-1}$.\\
This transformation allows recovered structural shocks to be interpreted within distinct macro-economic and energy transmission channels. However, conventional SVARs retain the assumption of time-invariant structural matrices, making them ill suited for empirical environments characterized by evolving policy regimes, shifting geopolitical dynamics, and episodic energy crises \citep{kilian2014role}. Consequently, their causal interpretation through impulse response functions (IRFs) can become misleading in the presence of structural instability.

\subsection{Time-Varying Parameter SVAR (TVP-SVAR)}
To accommodate structural evolution in transmission mechanisms, we adopt the Time-Varying Parameter SVAR (TVP-SVAR) framework of \cite{primiceri2005time}, allowing both coefficients and shock propagation matrices to evolve stochastically over time:
\begin{equation}
Y_t = A_t Y_{t-1} + \varepsilon_t,
\qquad
\varepsilon_t = L_t^{-1} \Sigma_t \eta_t,
\quad \eta_t \sim \mathcal{N}(0,I),
\label{e3}
\end{equation}
where the autoregressive matrices $A_t$, the structural impact matrix $L_t$, and the volatility terms $\Sigma_t$ follow independent random walk processes.\\ 
The TVP-SVAR framework captures gradual regime changes, episodic instability, and evolving impulse transmission, thereby addressing the limitations of constant-parameter SVARs.
Despite this flexibility in the mean equation, the model remains silent on the second-moment dynamics of volatility and dependence phenomena that are empirically prominent in financial and energy markets due to volatility clustering and crisis-induced contagion.\\

To fully capture the dynamic co-movements and tail-driven dependence among innovations, it becomes necessary to augment the TVP-SVAR framework with a second-stage model for conditional variance and correlation, which we address through multivariate GARCH structures such as DCC and ADCC.

\subsection{Conditional Volatility and Dynamic Correlation: DCC, ADCC, and Copula-Based Extensions}
\label{sec2.4}
Although the VAR and TVP-SVAR frameworks capture mean dynamics and structural transmission mechanisms, they do not model conditional heteroskedasticity or cross-market volatility spillovers. To address this limitation, we adopt a multivariate GARCH specification that captures the evolving dependence between markets over time. Specifically, we employ the Dynamic Conditional Correlation (DCC) model of \cite{engle2002dynamic} and its asymmetric extension (ADCC) of \citet*{cappiello2006asymmetric} , both of which provide a tractable yet flexible framework for modeling time-varying correlations.\\

\noindent Let $\varepsilon_t$ denote a $K \times 1$ vector of residuals from the mean equation (VAR or TVP-SVAR). The conditional return process is expressed as:
\begin{equation}
\varepsilon_t = H_t^{1/2} z_t, \quad z_t \sim F_z,
\label{e4}
\end{equation}
where $H_t$ is the conditional covariance matrix and $F_z$ represents the multivariate distribution of standardized innovations that are typically Gaussian or Student-(t).
\subsubsection{Univariate GARCH Marginals}

Each marginal variance $h_{i,t}$ is modeled using a GARCH(1,1) process:
\begin{equation}
h_{i,t} = \omega_i + \alpha_i \varepsilon_{i,t-1}^2 + \beta_i h_{i,t-1},
\label{e5}
\end{equation}
where $\omega_i > 0,\, \alpha_i,\, \beta_i \geq 0$ and $\alpha_i + \beta_i < 1$. \\
This formulation captures volatility clustering through lagged shocks and persistence. Although GARCH (1,1) remains the benchmark specification, other variants such as EGARCH or GJR-GARCH may be applied to accommodate leverage effects or nonlinear asymmetries.
Stacking the marginal volatilities yields the diagonal matrix $D_t = \text{diag}(\sqrt{h_{1,t}}, \dots, \sqrt{h_{K,t}})$, which defines the full conditional covariance matrix:
\begin{equation}
H_t = D_t R_t D_t,
\label{e6}
\end{equation}
where $R_t$ is the time-varying correlation matrix.
\subsubsection{Dynamic Conditional Correlation (DCC-GARCH)}
The DCC model decomposes the conditional covariance into dynamic variances and correlations. The correlation structure evolves according to:
\begin{equation}
Q_t = (1 - a - b)\bar{Q} + a (z_{t-1} z_{t-1}^\top) + b Q_{t-1},
\label{e7}
\end{equation}
where $\bar{Q} = \mathbb{E}[z_t z_t^\top]$ is the unconditional correlation matrix and $a, b > 0$ with $a + b < 1$. The dynamic correlation matrix is then obtained through normalization:
\begin{equation}
R_t = \text{diag}(Q_t)^{-1/2} \, Q_t \, \text{diag}(Q_t)^{-1/2}.
\label{e8}
\end{equation}
This framework flexibly captures time-varying co-movements among markets, though it assumes symmetric dependence.

\subsubsection{Asymmetric DCC (ADCC-GARCH)}
Empirical evidence suggests that correlations often intensify after negative shocks, a phenomenon known as the leverage effect. To capture this asymmetry, the ADCC specification augments the DCC equation with an additional term:

\begin{equation}
Q_t = (1 - a - b - g)\bar{Q} + a(z_{t-1}z_{t-1}^\top) + bQ_{t-1} + g(n_{t-1}n_{t-1}^\top),
\label{e9}
\end{equation}
where $n_t = \min(z_t, 0)$ and $g> 0$ measure the degree of correlation asymmetry. The ADCC-GARCH model therefore allows for crisis-driven contagion, where correlations strengthen disproportionately during downturns.

\subsubsection{Elliptical Distributions and the Role of Copulas}
\label{sec2.4.4}
Standard DCC and ADCC models rely on elliptical distributions, usually Gaussian or multivariate Student-(t). These distributions impose symmetric dependence structures:
\begin{itemize}
\item The Gaussian case implies a zero-tail dependence.
\item The Student-(t) allows for symmetric tail dependence, but cannot differentiate between upper- and lower-tail contagion.
\end{itemize}
To overcome these constraints, we invoke \citeauthor{sklar1959fonctions}’s theorem \citep{sklar1959fonctions}, which states that any continuous multivariate distribution $F(z_1, \dots, z_K)$ can be decomposed into its marginal cumulative distributions $F_i(z_i)$ and a copula function $C$:
$$
F(z_1, \dots, z_K) = C(F_1(z_1), \dots, F_K(z_K)).
$$
Each marginal transformation $u_i = F_i(z_i)$ maps observed data into uniform variables $u_i \in [0,1]$ via the Probability Integral Transform, such that:
$$
u_i = F_i(z_i) \quad \Rightarrow \quad u_i \sim \mathcal{U}(0,1).
$$
Substituting these transformations yields the copula representation:
$$
F(z_1, \dots, z_K) = C(u_1, \dots, u_K),
$$
where the copula $C$ captures the entire dependence structure, independently of the marginal dynamics. This separation is fundamental: marginal distributions can be modeled through univariate GARCH processes, while the copula governs the joint dependence structure.\\

\noindent For elliptical dependence, the Student-(t) copula provides a natural extension to the DCC framework by introducing tail dependence. It is defined as:
\begin{equation}
C_{\nu, R}(u_1, \dots, u_K) =
t_{\nu, R}\left(t_\nu^{-1}(u_1), \dots, t_\nu^{-1}(u_K)\right),
\label{e10}
\end{equation}
where $t_{\nu, R}(\cdot)$ denotes the multivariate Student-(t) CDF with correlation matrix $R$ and degrees of freedom $\nu$, and $t_\nu^{-1}(\cdot)$ is its univariate inverse.\\
This specification effectively links the multivariate Student-t distribution with the DCC-GARCH model, allowing for time-varying, symmetric tail dependence. Consequently, the t-copula DCC-GARCH serves as a bridge between traditional econometric volatility models and modern dependence structures. \\

\noindent Despite its flexibility, the t-copula remains symmetric and cannot distinguish directional contagion, an important feature of energy and financial markets. To capture asymmetric dependence, we extend our analysis in the next section to Archimedean copulas, including Clayton (lower-tail), Gumbel (upper-tail), and Frank (symmetric central dependence).
Mixtures of these copulas further enhance flexibility in modeling regime-specific tail comovements.

\subsection{Copula-Based Dependence Modelling}

While DCC and t-copula frameworks capture symmetric and elliptical dependence, they fall short in modeling asymmetric tail comovements, particularly during extreme market conditions such as oil crashes or speculative rallies. To address these limitations, we adopt copula-based GARCH models, where marginal volatilities are modeled through univariate GARCH(1,1) processes, as in Section \ref{sec2.4}, while the joint dependence structure is governed by Archimedean copulas, which allow for directional asymmetry across the joint distribution.

\subsubsection{Archimedean Copulas}
Archimedean copulas are defined by a generator function $\varphi: [0,1] \rightarrow [0,\infty)$, completely monotone, such that:
$$
C(u_1, \dots, u_K) = \varphi^{-1}\left( \varphi(u_1) + \cdots + \varphi(u_K) \right).
$$
They provide closed-form specifications for non-exchangeable, asymmetric dependence, critical for modelling financial and energy contagion. We retain three canonical families:
\begin{itemize}
\item Clayton Copula (Lower Tail Dependence)
\begin{equation}
C_{\text{\tiny{Clayton}}}(u_1, \dots, u_K \mid \theta) =
\left( u_1^{-\theta} + \cdots + u_K^{-\theta} - (K - 1) \right)^{-1/\theta},
\quad \theta > 0.
\label{e11}
\end{equation}
It captures strong co-movement during downside crashes (e.g. oil price collapses).

\item Gumbel Copula (Upper Tail Dependence)
\begin{equation}
C_{\text{\tiny{Gumbel}}}(u_1, \dots, u_K \mid \theta) =
\exp \Big( - \left[
(-\ln u_1)^{\theta} + \cdots + (-\ln u_K)^{\theta}
\right]^{1/\theta} \Big),
\quad \theta \ge 1.
\label{e12}
\end{equation}
It captures extreme upward co-movement, consistent with speculative energy rallies.

\item Frank Copula (Symmetric Central Dependence)
\begin{equation}
C_{\text{\tiny{Frank} }}(u_1, \ldots, u_K; \theta) =
-\frac{1}{\theta} ,
\ln \left[
1 +
\frac{
\prod_{i=1}^K \left( e^{-\theta u_i} - 1 \right)
}{
(e^{-\theta} - 1)^{K-1}
}
\right],
  \quad \theta \ne 0.
\label{e13}
\end{equation}
It models moderate dependence around the center—useful in stable regimes, but with no tail dependence.
\end{itemize}
However, no single Archimedean family can jointly capture both lower- and upper-tail contagion. We therefore construct a mixture copula:
\begin{equation}
\begin{split}
C(u_1, \dots, u_K,\boldsymbol{\theta}) &= w_1\, C_{\text{\tiny{Clayton}}}(u_1, \dots, u_K \mid \theta_1) + w_2\, C_{\text{\tiny{Frank} }}(u_1, \ldots, u_K; \theta_2) \\
& \qquad \qquad + w_3\, C_{\text{\tiny{Gumbel}}}(u_1, \dots, u_K \mid \theta_3)\\
& \quad w_j \ge 0,\quad \sum_{j=1}^{3} w_j = 1.
\end{split}
\label{e14}
\end{equation}
\paragraph{Estimation and Bootstrap Inference: }
Parameters $\boldsymbol{\theta} = (\theta_1, \theta_2, \theta_3)$ and weights $(w_1, w_2, w_3)$ are estimated using maximum likelihood estimation (MLE):

$$
\mathcal{L}(\boldsymbol{\theta}, w) =
\sum_{t=1}^{T} \ln
\left[
\sum_{j=1}^{3} w_j \, c_j(u_{1,t}, \dots, u_{K,t} \mid \theta_j)
\right],
$$
where $c_j(\cdot)$ denotes the copula density of family $j$.\\
To ensure robust inference under tail dependence, we compute parametric bootstrap confidence intervals, following \cite{genest2009goodness}.
\newpage
\noindent In the broader context of macro-financial and energy modeling, mixed copulas provide a unified and flexible mechanism to characterize evolving dependence structures across both constant-parameter VAR and time-varying TVP-SVAR frameworks. By combining Clayton, Frank, and Gumbel components within a single mixture specification, this approach is capable of representing crisis-induced synchronization during macro-financial stress, while simultaneously capturing speculative amplification frequently observed during oil price booms.\\ 

\noindent Furthermore, the capacity of mixed copulas to accommodate transitions between low-volatility regimes and high-volatility contagion states allows them to reproduce nonlinear and state-dependent co-movements that traditional correlation-based models cannot replicate. As such, when coupled with GARCH-type marginal dynamics, copula mixtures offer a natural complement to both VAR and TVP-SVAR mean structures, jointly modeling structural transmission, conditional heteroskedasticity, and asymmetric tail dependence, thus achieving a level of integration in dependence modeling that neither parametric SVAR nor standard DCC frameworks can attain in isolation.
\subsection{Hybrid Econometric–Machine Learning Models} 
The hybrid forecasting framework builds on the premise that macro–energy time series embody both linear autoregressive structure and nonlinear stochastic components which no single model can capture entirely \citep{zhang2003time}. Formally, a multivariate series $Y_t$ may be decomposed as:
\begin{equation}
Y_t = f_{\text{\tiny{L}}}(Y_{t-1}, Y_{t-2}, \ldots) + f_{\text{\tiny{NL}}}(\varepsilon_{t-1}, \varepsilon_{t-2}, \ldots),
\label{e15}
\end{equation}

where $f_{\text{\tiny{L}}}(\cdot)$ represents the linear transmission mechanism and $f_{\text{\tiny{NL}}}$ captures hidden nonlinear dependencies.\\ 
In our empirical framework, the deterministic linear mapping $f_{\text{\tiny{L}}}$ is specified through VAR or TVP-SVAR, generating structural forecasts:

$$
\widehat{Y}_{t+1}^{L} = \widehat{Y}_{t+1}^{\text{\tiny{VAR or TVP-SVAR}}}.
$$
The residuals from the structural model,
$$
\varepsilon_t = Y_t - \widehat{Y}_t^{L},
$$
serve as inputs to nonlinear machine learning models, which estimate the stochastic component $f_{\text{\tiny{NL}}}$. The full hybrid predictive model is therefore defined by:
\begin{equation}
\widehat{Y}_{t+1} = \widehat{Y}_{t+1}^{L} + \widehat{\varepsilon}_{t+1}^{\text{\tiny{ML}}}
\label{e16}
\end{equation}
This approach generalizes \cite{zhang2003time}’s hybrid ARIMA–ANN scheme to a multivariate structural econometric setting, where machine learning augments, rather than replaces, causal dynamics.\\

\noindent Among competing nonlinear models such as Artificial Neural Networks (ANN), Random Forests (RF), and Support Vector Regression (SVR) widely used in energy and financial prediction \citep{hippert2002neural,chen2020multivariate,medeiros2022forecasting,pokou2024hybridization} we focus on Gaussian Process Regression (GPR) due to its fully probabilistic formulation and analytical tractability.\\

\noindent A Gaussian Process (GP) defines a prior over functions:
$$
g(x) \sim \mathcal{GP}(m(x), k(x, x')),
$$
with mean function $m(x) = 0$ and covariance kernel $k(\cdot, \cdot)$, encoding smoothness and nonlinear interactions. Given training data $\mathcal{D} = {(x_i, y_i)}_{i=1}^N$, representing lagged residuals and error targets, the joint prior to observations and the values of the test function is:

\begin{equation}
\begin{bmatrix}
\mathbf{y} \\
g(x_*)
\end{bmatrix}
\sim \mathcal{N}
\left(
0,
\begin{bmatrix}
K(X,X) + \sigma_n^2 I & K(X, x_*) \\
K(x_*, X) & K(x_*, x_*)
\end{bmatrix}
\right),
\label{e17}
\end{equation}
where $K(X,X)$ is the kernel matrix with entries $k(x_i, x_j)$.
\newpage
The predictive posterior for $g(x_*)$ is:
\begin{equation*}
\small
g(x_*) ,|, X, \mathbf{y} \sim \mathcal{N}\left(
K(x_*,X)[K(X,X) + \sigma_n^2 I]^{-1}\mathbf{y},,
K(x_*,x_*) - K(x_*,X)[K(X,X) + \sigma_n^2 I]^{-1}K(X,x_*)
\right)
\end{equation*}

\paragraph{Kernel Specification: The Role of the RBF Kernel: }
Nonlinearity in GPR is governed by the kernel function, which defines similarity between observation points. The Radial Basis Function (RBF) kernel—also known as the Gaussian kernel—is defined as:
$$
k_{\text{RBF}}(x_i, x_j) =
\exp\left( -\frac{|x_i - x_j|^2}{2\ell^2} \right),
$$
where $\ell$ controls the smoothness scale.\\ 
The RBF kernel is particularly suited for macro–energy forecasting, as it captures continuous regime transitions, mean-reverting cycles, and smooth nonlinear effects, all of which characterise commodity markets.
Other kernels (Matérn, Polynomial, Rational Quadratic) may be considered, but the RBF kernel offers a universal approximation capacity under modest hyperparameter tuning \citep{rasmussen2006gaussian}, making it ideal for benchmarking against structured econometric models.\\

\noindent In the hybrid context, the GPR model operates not on raw data but on structural residuals:
$$
\widehat{\varepsilon}_{t+1}^{\text{GPR}} = g(\varepsilon_{t}, \varepsilon_{t-1}, \dots).
$$
By construction, the final prediction:
$$
\widehat{Y}_{t+1} =
\widehat{Y}_{t+1}^{\text{VAR/TVP-SVAR}} +
g(\varepsilon_{t}, \varepsilon_{t-1}, \dots)
$$
provides a synthesis of economic structure (mean dynamics) and statistical flexibility (nonlinear correction), enabling fair comparison with volatility-aware copula models.

\section{Empirical results}
\label{sec3}
This section presents the empirical framework of the study, including the dataset description, preliminary unit root and cointegration analyses, the specification of the benchmark VAR model, and its diagnostic assessment. These steps establish the foundation for subsequent structural and time-varying investigations using SVAR and TVP-SVAR models.
\subsection{Data Description}
The empirical analysis relies on monthly observations spanning January 1999 to July 2025, compiled from established macro-financial sources: Federal Reserve Economic Data (FRED), the University of Michigan Survey of Consumers, and the Global Economic Policy Uncertainty (GEPU) database. The data set comprises seven series selected to capture financial conditions, real activity, uncertainty, international transmission, final energy costs, and the global crude benchmark. DGS10 (10-Year Treasury Constant Maturity Rate), UMCSENT (Consumer Sentiment Index), GEPUCURRENT (Global Economic Policy Uncertainty Index), INDPRO\_US (Industrial Production Index), USD/EUR (dollar–euro exchange rate), APE (average U.S. electricity price), and BRENT crude oil (the focal variable).
%
%--------------------%
%      Tableau 1     %
%--------------------%
\begin{table}[H]
\centering
\resizebox{18.3cm}{!}{
\begin{tabular}{|l|l|l|l|} 
\hline
Symbol         &  Variable name                                      & Description                                                                       & Source \\
\hline
DGS10          & 10-Year Treasury Constant Maturity Rate             & U.S. long-term interest rate, proxy for financing conditions and monetary stance  & FRED   \\
UMCSENT        & University of Michigan Consumer Sentiment Index     & Indicator of household expectations and confidence, proxy for consumption demand  & FRED   \\
GEPUCURRENT    & Global Economic Policy Uncertainty Index            & Global index of economic policy-related uncertainty, proxy for uncertainty shocks & FRED   \\
INDPRO$\_$US      & Industrial Production Index (United States)         & Proxy for U.S. real economic activity and industrial energy demand                & FRED   \\
USD/EUR        & U.S. Dollar to Euro Exchange Rate                   & Measures the relative value of the USD vs EUR, proxy for global oil affordability & FRED   \\
APE            & Average Price: Electricity per Kilowatt-Hour (U.S.) & Final energy price reflecting household and industrial electricity costs          & FRED   \\
BRENT          & Brent Crude Oil Price                               & Global benchmark for crude oil prices, dependent variable of interest             & FRED   \\
\hline
\end{tabular}
}
\caption{Variables, Definitions, and Data Sources}
\label{tab1}
\end{table}

For forecasting evaluation, the sample is strictly partitioned into an in-sample (estimation) window and an out-of-sample (evaluation) window. The in-sample period runs from January 1999 through February 2023 and is used exclusively to estimate the baseline linear models (VAR/SVAR), the time-varying specification (TVP-SVAR), and the volatility/dependence layers (DCC/ADCC-GARCH and copulas), including any hyperparameter selection. The out-of-sample period extends from March 2023 through July 2025 and is reserved for genuine ex post prediction and performance assessment (e.g., rolling one-step-ahead RMSE), with no reestimation contamination from future information. This split yields a long historical window for stable parameter learning and a sufficiently turbulent evaluation window encompassing recent macro-energy shocks, thereby providing a stringent test of predictive robustness.
The economic definitions and primary sources for each series are summarized in Table \ref{tab1}.

\subsection{Stationarity and Cointegration Tests}
Preliminary descriptive statistics for raw series (Table \ref{tab2}) exhibit pronounced deviations from Gaussianity, substantial skewness, and excess kurtosis, consistent with well-known stylized facts for macrofinancial data \citep{cont2001empirical}. Visual diagnostics also suggest stochastic trends and potential structural breaks, motivating a formal assessment of integration and long-term comovement.
We first apply Augmented Dickey–Fuller (ADF) tests \citep{dickey1979distribution,dickey1981likelihood} to the level series. For each variable, the test equation includes an intercept (and, where visually warranted, a linear trend), with the lag order of the augmentation selected by an information criterion to mitigate residual autocorrelation. Across all series, the ADF statistics fail to reject the unit-root null at conventional levels, indicating non-stationarity in levels. We then evaluate cointegration using the Johansen trace procedure \citep{johansen1991estimation,johansen1995likelihood}, implemented in a VAR with intercept and lag length chosen by information criteria within the in-sample window. The trace statistics (Table \ref{tab3}) do not reject the nullity of no cointegrating vectors, implying the absence of stable long-term equilibria among the variables during the estimation period.

%--------------------%
%      Tableau 2     %
%--------------------%
\begin{table}[H]
\centering
\resizebox{15cm}{!}{
\begin{tabular}{|l|l|l|l|l|l|l|l|l|} 
\hline
Series        & Mean       &  Std       & Skewness  &  Kurtosis   & Min          &  Max        &   ADF       &  Jarque-Bera \\
\hline
DGS10         & 3.317285   & 1.386878   & 0.229098  & -0.803341   &  0.623636    & 6.661       &  -1.81845   &  10.334876   \\
UMCSENT       & 84.851724  & 13.501094  &-0.356188  & -0.570812   &  50.0        & 112.0       &  -2.297992  &  10.069098   \\
GEPUCURRENT   & 138.640188 & 70.496671  & 1.180152  &  1.173406   &  47.86227    & 420.861165  &  -0.509207  &  83.953981   \\
INDPRO$\_$US     & 96.507948  & 5.079894   &-0.403135  & -1.031767   &  84.6746     & 104.1038    &  -2.154909  &  20.718257   \\
USD/EUR       & 1.19214    & 0.158025   &-0.061841  & -0.414056   &  0.852538    & 1.575864    &  -1.935017  &  2.256438    \\
APE           & 0.121079   & 0.020995   &-0.325424  & -0.878049   &  0.084       & 0.168       &  -0.114879  &  14.434424   \\
BRENT         & 63.552504  & 30.406868  & 0.317887  & -0.927976   &  10.271579   & 132.718182  &  -2.680076  &  15.289602   \\
\hline
\end{tabular}
}
\caption{Descriptive statistics of raw data}
\label{tab2}
\end{table}

%--------------------%
%      Tableau 3     %
%--------------------%
\begin{table}[H]
\centering
\resizebox{6cm}{!}{
\begin{tabular}{|l|l|l|l|} 
\hline
$r_0$   &  $r_1$  & test statistic  & critical value \\
\hline
0       &   7     &  103.2          &   125.6 \\
\hline
\end{tabular}
}
\caption{Johansen cointegration test using trace test statistic with 5\% significance level}
\label{tab3}
\end{table}
In accordance with these findings and standard multivariate practice, we transform each series into logarithmic returns
$$
r_t \equiv \log\left(\frac{p_t}{p_{t-1}}\right),
$$
thus removing unit roots and placing the variables on a comparable scale. Post-transformation diagnostics (Table \ref{tab4}) show ADF statistics that strongly reject the null unit root for all series, confirming the weak stationarity suitable for VAR-type modeling. Although residual nonnormality persists, heavy tails and asymmetry remain evident, such features are typical for macrofinancial returns and are compatible with robust estimation and inference in the subsequent models \citep{cont2001empirical}.

%--------------------%
%      Tableau 4     %
%--------------------%
\begin{table}[H]
\centering
\resizebox{15cm}{!}{
\begin{tabular}{|l|l|l|l|l|l|l|l|l|} 
\hline
Series        & Mean       &  Std      & Skewness   &  Kurtosis   & Min        &  Max       &   ADF        &  Jarque-Bera \\
\hline
DGS10         & -0.000876  & 0.083387  & -1.172895  &  7.749062   & -0.54753   &  0.255591  &  -10.68701   &  792.070894   \\
UMCSENT       & -0.00178   & 0.054196  & -0.535004  &  1.217508   & -0.215875  &  0.127624  &  -14.999348  &  31.745831    \\
GEPUCURRENT   &  0.004203  & 0.186282  &  0.61043   &  1.124548   & -0.482292  &  0.750697  &  -6.498415   &  33.290886    \\
INDPRO$\_$US  &  0.000567  & 0.011931  & -5.741803  &  72.271427  & -0.142045  &  0.063771  &  -12.95621   &  64706.640473 \\
USD/EUR       & -0.000272  & 0.022268  & -0.003325  &  0.455581   & -0.077944  &  0.061864  &  -10.854986  &  2.508474     \\
APE           &  0.002349  & 0.02039   &  0.612083  &  2.141035   & -0.052368  &  0.074901  &  -3.89402    &  73.498276    \\
BRENT         &  0.006738  & 0.106786  & -1.059152  &  5.757671   & -0.554909  &  0.469097  &  11.952445   &  454.79233    \\
\hline
\end{tabular}
}
\caption{Descriptive statistics of log-returns}
\label{tab4}
\end{table}

\subsection{Benchmark VAR Model Specification and Diagnostics}
The specification of the benchmark Vector Autoregressive model (VAR) is guided by a rigorous selection of the lag order using well-established information criteria. Specifically, we computed the Akaike Information Criterion (AIC) \citep{akaike2003new}, the Hannan-Quinn Criterion (HQ) \citep{hannan1979determination}, the Schwarz Bayesian Information Criterion (SC) \citep{schwarz1978estimating}, and the Final Prediction Error (FPE) in alternative lag structures. As reported in Table \ref{tab5}, all criteria consistently reach their minimum at one lag, indicating that a VAR(1) provides the most parsimonious yet informative representation of the joint dynamics:
$$
\text{Lag}^* = 1.
$$
Accordingly, the benchmark model is specified as:
$$
Y_t = A_0 + A_1 Y_{t-1} + \varepsilon_t,\quad \varepsilon_t \sim \mathcal{N}(0, \Sigma),
$$
where $\small{Y_t=\left[DGS10_t,\ UMCSENT_t,\ GEPU_t,\ INDPRO\_US_t,\ USD/EUR_t,\ APE_t,\ BRENT_t \right]^{\top}}$ denotes the vector of endogenous macro-financial and energy variables observed at time $t$. This formulation captures short-term feedback mechanisms and dynamic interdependencies between variables through the autoregressive structure.
%--------------------%
%      Tableau 5     %
%--------------------%
\begin{table}[H]
\centering
\resizebox{8.5cm}{!}{
\begin{tabular}{|l|llll|} 
\hline
lag  &   AIC(n)      &   HQ(n)      &    SC(n)     &    FPE(n)       \\
\hline
\textbf{1}    &  \textbf{-43.92898}   &  \textbf{-43.67176}  &  \textbf{-43.28791}  &  \textbf{8.354466e-20} \\
2    &  -43.87225    &  -43.35781   &  -42.59011   &   8.846778e-20  \\
3    &  -43.83892    &  -43.06726   &  -41.91571   &   9.159749e-20  \\
4    &  -43.79453    &  -42.76564   &  -41.23024   &   9.602524e-20  \\
5    &  -43.64827    &  -42.36216   &  -40.44291   &   1.116674e-19  \\
6    &  -43.51229    &  -41.96896   &  -39.66586   &   1.288295e-19  \\
7    &  -43.37966    &  -41.57911   &  -38.89216   &   1.485545e-19  \\
8    &  -43.53633    &  -41.47855   &  -38.40776   &   1.287009e-19  \\
9    &  -43.40600    &  -41.09100   &  -37.63636   &   1.491506e-19  \\
10   &  -43.32164    &  -40.74942   &  -36.91092   &   1.658335e-19  \\
11   &  -43.27619    &  -40.44675   &  -36.22440   &   1.782621e-19  \\
12   &  -43.87234    &  -40.78568   &  -36.17948   &   1.014710e-19  \\
\hline
\end{tabular}
}
\caption{VAR lag order selection}
\label{tab5}
\end{table}

\noindent Although the VAR(1) framework is well suited for modeling linear intertemporal interactions, it rests on strong assumptions, most notably the constancy of parameters over time and Gaussian residual innovations. These assumptions may be restrictive in environments characterized by structural regime shifts, volatility clustering, or nonlinear dependence, all of which are significant in energy and financial markets.\\

\noindent Consequently, this benchmark serves not only as a reference point, but also as a diagnostic platform from which more flexible structures, SVAR for structural identification and TVP-SVAR for time variation, will be critically evaluated in subsequent sections.
By establishing the VAR(1) as the foundation of our empirical framework, we ensure coherence in model comparison, allowing us to assess the incremental value of structural identification, time variation, and non-linear dependence introduced in advanced econometric and hybrid models.
\newpage
\subsection{Structural Stability and Justification for the TVP-SVAR Framework}
\label{sec3.4}
A fundamental requirement for reliable structural inference within the VAR/SVAR paradigm is the assumption of time-invariant coefficients. Traditional linear frameworks implicitly presume that the relationships among macroeconomic, financial, and energy variables remain constant throughout the estimation horizon. However, the past two decades have been characterized by recurrent changes in the monetary regime, dislocations of the oil market, and global uncertainty shocks, conditions under which such an assumption is unlikely to hold \citep{hamilton2009causes,kilian2014role}.
%
%--------------------%
%     Figure 1       %
%--------------------%
\begin{figure}[H]
%\begin{figure}
\centering
\begin{subfigure}[b]{0.4\textwidth}
    \centering
    \includegraphics[width=1.15\hsize]{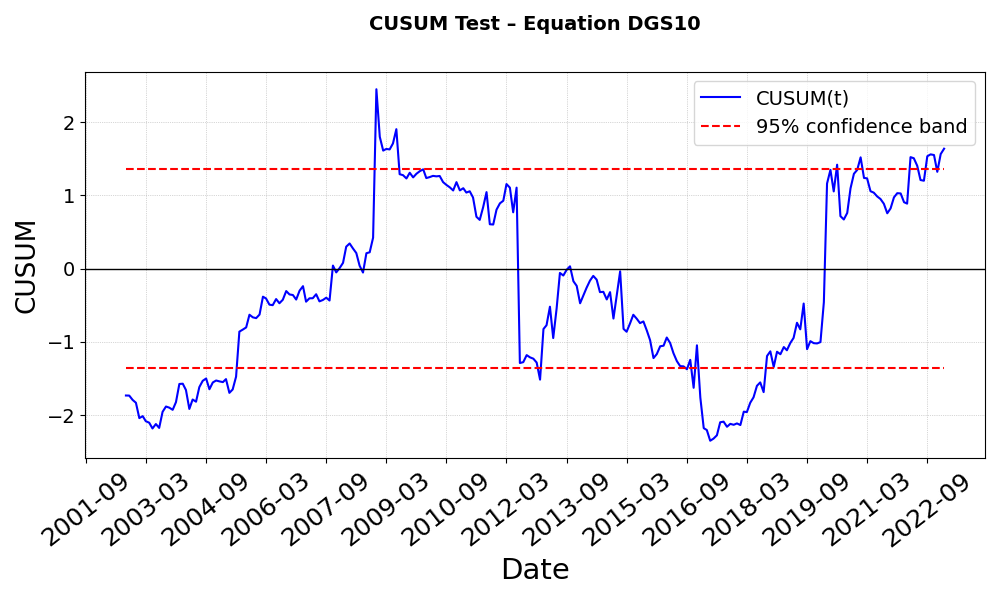}
    \caption{DGS10}
    \label{fig1_m1}
\end{subfigure}
\hfil
\begin{subfigure}[b]{0.4\textwidth}
    \centering
    \includegraphics[width=1.15\hsize]{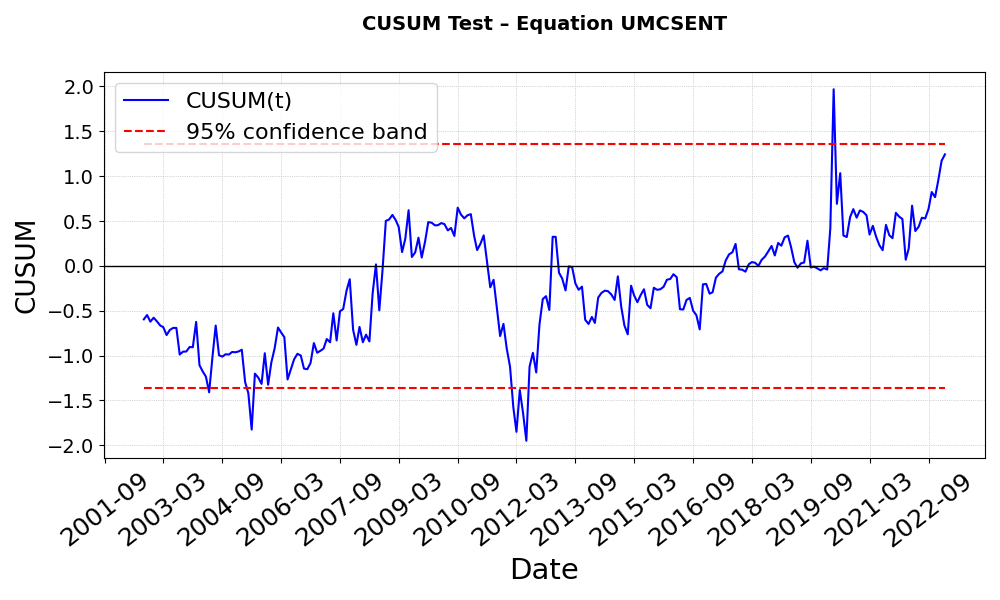}
    \caption{UMSCENT}
    \label{fig1_m2}
\end{subfigure}
\hfil
\begin{subfigure}[b]{0.4\textwidth}
    \centering
    \includegraphics[width=1.15\hsize]{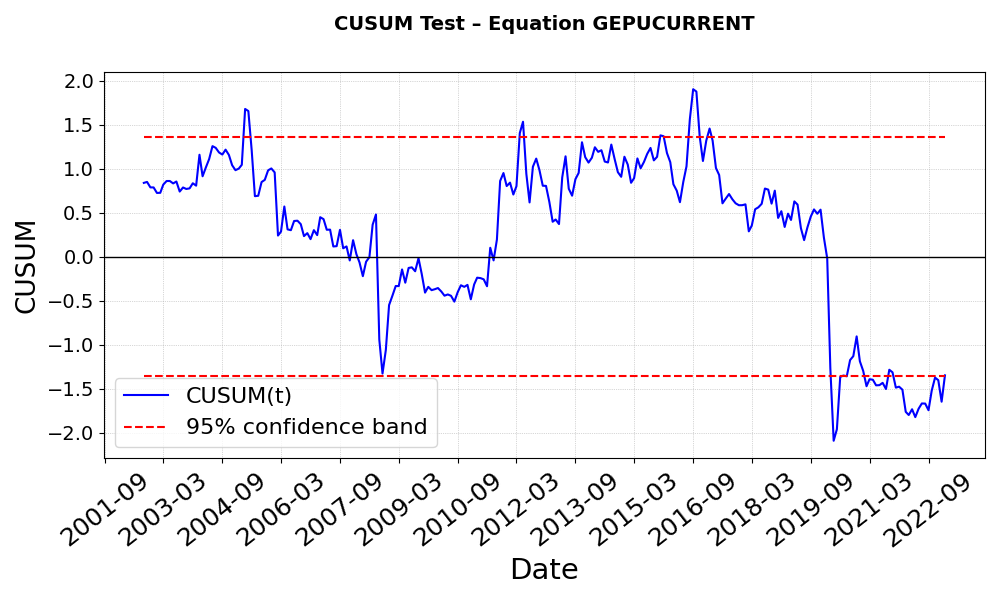}
    \caption{GEPUCURRENT}
    \label{fig1_m3}
\end{subfigure}
\hfil
\begin{subfigure}[b]{0.4\textwidth}
    \centering
    \includegraphics[width=1.15\hsize]{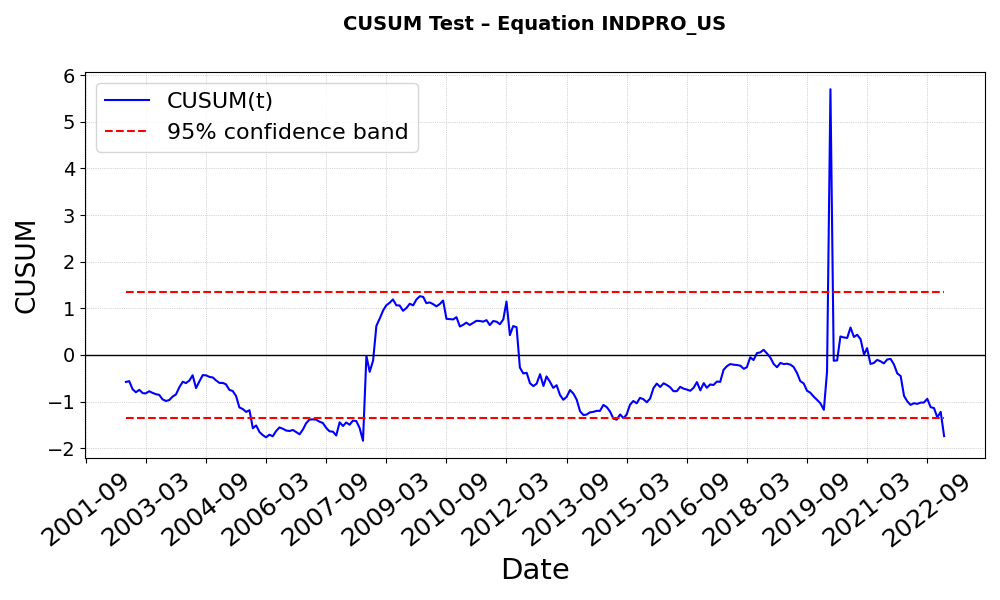}
    \caption{INDPRO$\_$US}
    \label{fig1_m4}
\end{subfigure}
\hfil
\begin{subfigure}[b]{0.4\textwidth}
    \centering
    \includegraphics[width=1.15\hsize]{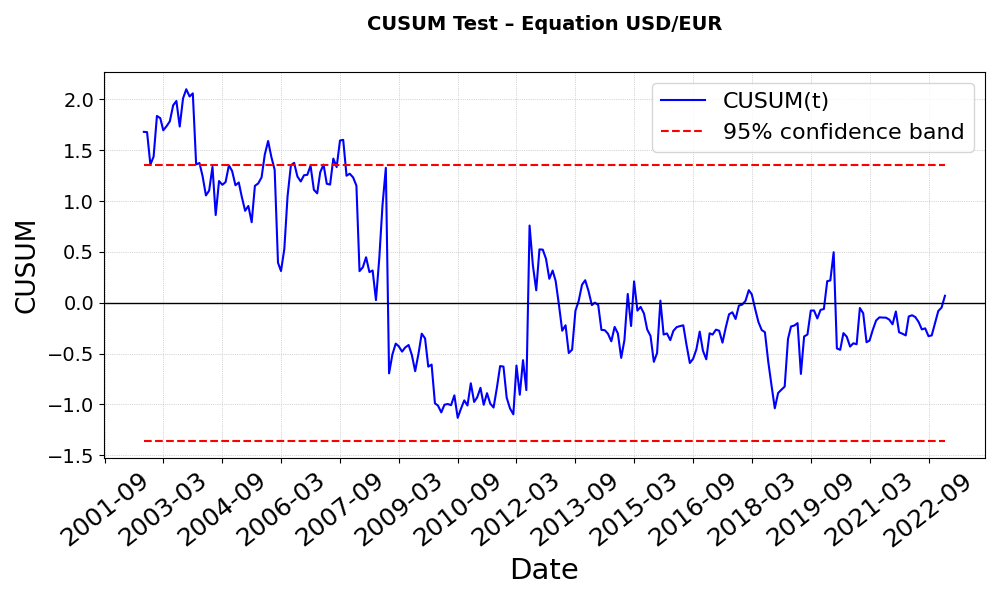}
    \caption{USD/EUR}
    \label{fig1_m5}
\end{subfigure}
\hfil
\begin{subfigure}[b]{0.4\textwidth}
    \centering
    \includegraphics[width=1.15\hsize]{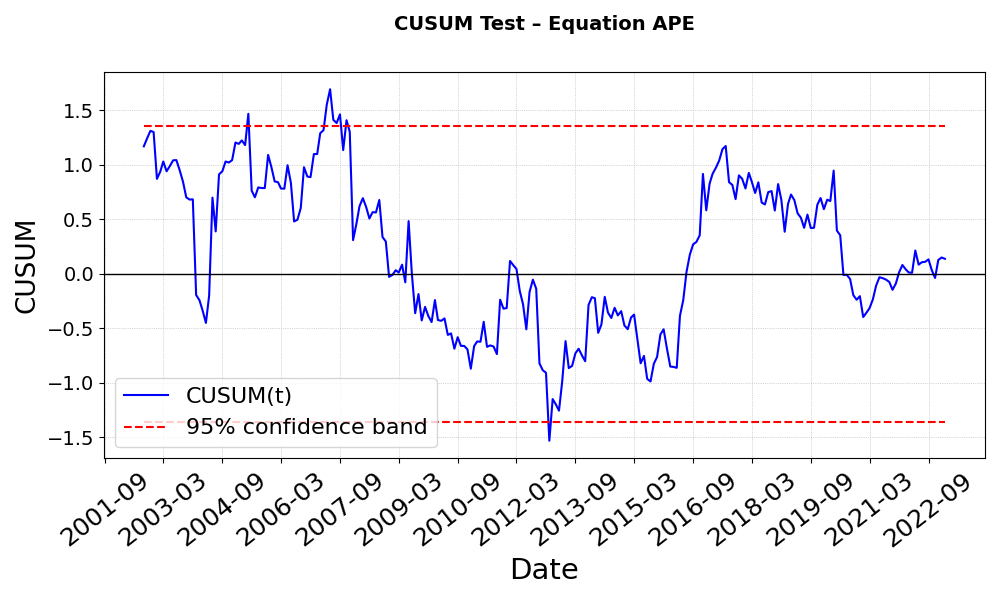}
    \caption{APE}
    \label{fig1_m6}
\end{subfigure}
\hfil
\begin{subfigure}[b]{0.4\textwidth}
    \centering
    \includegraphics[width=1.15\hsize]{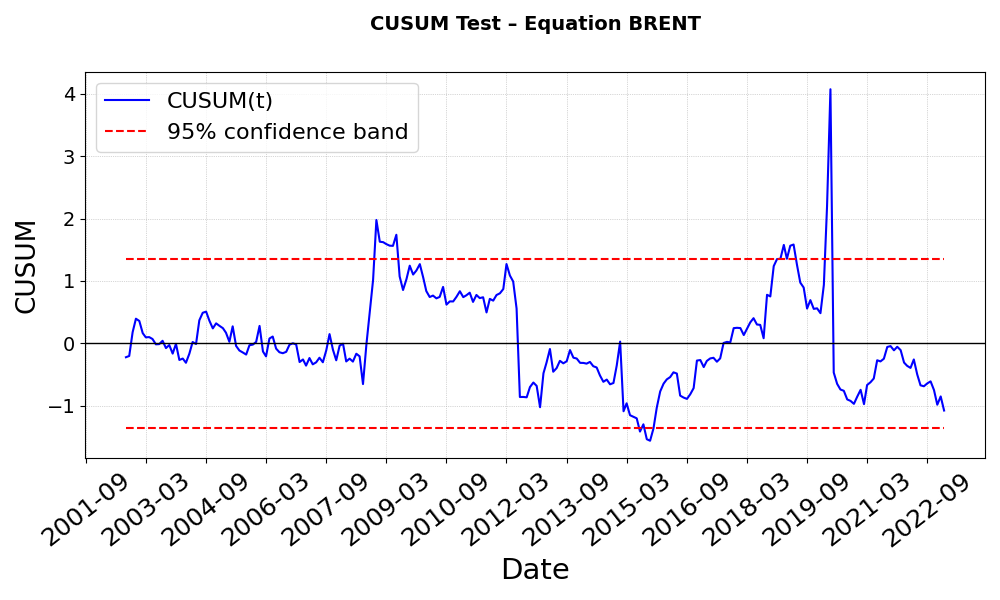}
    \caption{BRENT}
    \label{fig1_m7}
\end{subfigure}
%------------
\caption{Structural stability diagnosis}
\label{fig::fig_stab}
%------------
\end{figure}
\newpage
\noindent To empirically assess this concern, we perform CUSUM stability tests based on recursive residuals for each equation in the estimated VAR(1) system.
Figure \ref{fig::fig_stab} displays the CUSUM statistics and associated confidence bands 95\% for the seven endogenous variables. Persistent deviations beyond critical thresholds constitute formal evidence of structural instability, indicating that the underlying parameters evolve rather than remain fixed.

\paragraph{Evidence of Evolving Macroeconomic Transmission Mechanisms: }
The CUSUM diagnostics reveal marked instability in several key variables. The equation for long-term interest rates (DGS10) shows significant parameter drift, particularly coinciding with major shifts in US monetary policy, such as the implementation of quantitative easing (QE), the transition to a zero interest rate policy (ZIRP), and the subsequent policy normalization phase after 2015. This instability suggests that the monetary transmission mechanism has undergone profound regime transitions.
Similarly, the equation associated with global policy uncertainty (GEPUCURRENT) exhibits pronounced breaks during internationally disruptive episodes, including the 2008 financial crisis, the 2014-2015 collapse of the oil price, and the COVID-19 pandemic. These fluctuations reflect the increasingly state-dependent nature of uncertainty shocks in influencing macro–energy dynamics.
The Brent crude oil equation displays some of the strongest departures from stability, particularly around the 2014 price collapse and the 2020 energy demand shock. These episodes underscore that oil price dynamics is governed by non-linear responses to both demand- and supply-driven shocks, which static linear models cannot adequately capture.

\paragraph{Methodological Implications for Model Design: }
The evidence provided by the CUSUM analysis directly challenges the constancy assumption inherent in traditional VAR and SVAR models \citep{lutkepohl2013introduction}. When considered alongside earlier signs of non-normal residuals and conditional heteroskedasticity, these findings collectively imply that the macro–energy nexus operates through evolving regimes rather than stable linear propagation.\\

\noindent To account for this, we transition to a Time-Varying Parameter SVAR (TVP-SVAR) framework, based on \cite{primiceri2005time} and \cite{cogley2005drifts}. Unlike static SVAR models, the TVP formulation allows both autoregressive coefficients and variance–covariance structures to evolve stochastically:
$$
Y_t = A_t Y_{t-1} + \varepsilon_t,\qquad
\varepsilon_t = L_t^{-1} \Sigma_t e_t,\qquad e_t \sim \mathcal{N}(0, I),
$$
thereby accommodating:
\begin{itemize}
\item Structural regime shifts induced by macroeconomic crises and policy realignments,
\item Time-varying impulse responses and evolving shock transmission channels,
\item Interaction with higher-order volatility dynamics, later modelled via DCC-GARCH and copulas.
\end{itemize}
Given this clear empirical evidence of instability, the impulse responses derived from a static SVAR must be interpreted only as benchmark dynamics. In the following section, we first present conventional IRFs with parameter constancy before extending the analysis to the TVP-SVAR, where the responses are time-indexed and regime-sensitive. This progression enables a direct comparison between constant and evolving causal structures, which is essential to understand the transmission mechanisms that govern the macro-energy system.

\subsection{Impulse Response Function Analysis (SVAR Model)}
\label{sec3.5}
The impulse response functions (IRFs) shown in Figures \ref{figA1}-\ref{figA4} provide a systematic evaluation of how structural shocks propagate within the macroenergy framework under a constant-parameter SVAR specification. Computed using orthogonalized shocks of one standard deviation and a recursive identification scheme with Brent crude oil ordered last, these IRFs capture the sequential transmission of disturbances across financial conditions, macroeconomic fundamentals, and global energy markets over a 24-month horizon. This benchmark analysis is essential to assess the directional impact of shocks before incorporating time variation in the subsequent TVP-SVAR framework.
\newpage
\noindent A key finding emerges from the response to monetary policy shocks, identified through innovations in the 10-year Treasury yield (DGS10). As seen in Figures \ref{figA1_m1}, \ref{figA2_m7} and \ref{figA4_m1}, a positive interest rate shock generates a short-lived increase in Brent prices. This initial uptick reflects a signaling channel in which higher long-term yields are interpreted as evidence of economic strength and anticipated energy demand. However, this effect rapidly decays and occasionally turns negative, aligning with the contractionary implications of tighter financial conditions and higher discount rates. This dual mechanism is consistent with the literature that emphasizes the ambiguous role of monetary policy in commodity price \citep{kilian2011does}.\\

\noindent In contrast, shocks in consumer sentiment, represented by UMCSENT, exert only marginal and transient effects on Brent, as documented in Figures \ref{figA1_m7}-\ref{figA1_m12}, \ref{figA2_m8}  and \ref{figA4_m2}. The muted response indicates that household expectations, while informative for domestic consumption cycles, have limited direct influence on global commodity markets. Energy prices remain predominantly tied to real economic activity and risk conditions rather than soft indicators of confidence.\\

\noindent Shocks to global economic policy uncertainty (GEPUCURRENT) produce the most pronounced and persistent reactions. Figures \ref{figA1_m2}, \ref{figA2_m1},\ref{figA2_m3} and \ref{figA4_m3} reveal sharp and persistent declines in Brent prices after increases in uncertainty, consistent with increased risk aversion, reduced speculative demand, and precautionary inventory reductions. This deflationary effect corroborates recent evidence that uncertainty shocks operate as dominant contractional forces in commodity markets \citep{caldara2016macroeconomic}.\\

\noindent Industrial production shocks (INDPRO\_US), depicted in Figures \ref{figA2_m7}-\ref{figA2_m12} and \ref{figA4_m5}, produce moderate but procyclical increases in Brent. These responses reflect genuine real-demand channels, where higher output stimulates energy consumption. However, the impact remains weaker than that of financial or uncertainty shocks, suggesting a decoupling between industrial output and energy intensity, likely due to structural efficiencies and sectoral diversification in production.\\

\noindent Exchange rate shocks (USD/EUR), illustrated in Figures \ref{figA3_m1}-\ref{figA3_m6}, consistently depress Brent prices. An appreciation in the dollar, making commodities based on the dollar more expensive for non-US buyers, suppresses global demand and reinforces the role of the dollar as a pricing anchor in the oil market \citep{chen2010can}. The persistence of this effect highlights the strategic importance of currency regimes in international energy valuation.\\

\noindent Shocks to electricity prices (APE), shown in Figures \ref{figA3_m7}-\ref{figA3_m12} and \ref{figA4_m5}, generate short-term positive adjustments in Brent. This substitution effect reflects inter-fuel transmission, where rising downstream energy costs increase pressure on upstream inputs, particularly under constrained energy supply conditions. Although its magnitude is limited, it underscores the systemic integration between the primary and final energy markets.\\

\noindent The own Brent shocks, reported in Figure \ref{figA4_m6}, exhibit strong persistence and slow mean reversion. The high degree of inertia underscores the structural rigidities inherent in the oil market, including inventory adjustment lags, supply chain constraints, and geopolitical risk premia. This endogenous propagation confirms Brent’s status as both a transmitter and a reflector of macrofinancial stress.\\

\noindent  Together, these IRFs provide three key insights. First, financial and uncertainty shocks dominate real activity in amplitude and persistence, highlighting the financialization of energy markets. Second, the transmission mechanisms are asymmetric, with negative shocks generating stronger effects than positive ones. Third, while the constant-coefficient SVAR framework captures essential short-run dynamics, it fails to accommodate the evolving nature of these relationships across regimes. This limitation motivates the transition to a time-varying framework (TVP-SVAR), where structural responses are allowed to evolve over time, thus offering a more realistic representation of macro–energy dynamics.
\newpage
\subsubsection{Time-Varying Impulse Response Analysis (TVP-SVAR Model)}

\begin{figure}[H]
%\begin{figure}
\centering
\begin{subfigure}[b]{0.4\textwidth}
    \centering
    \includegraphics[width=1.15\hsize]{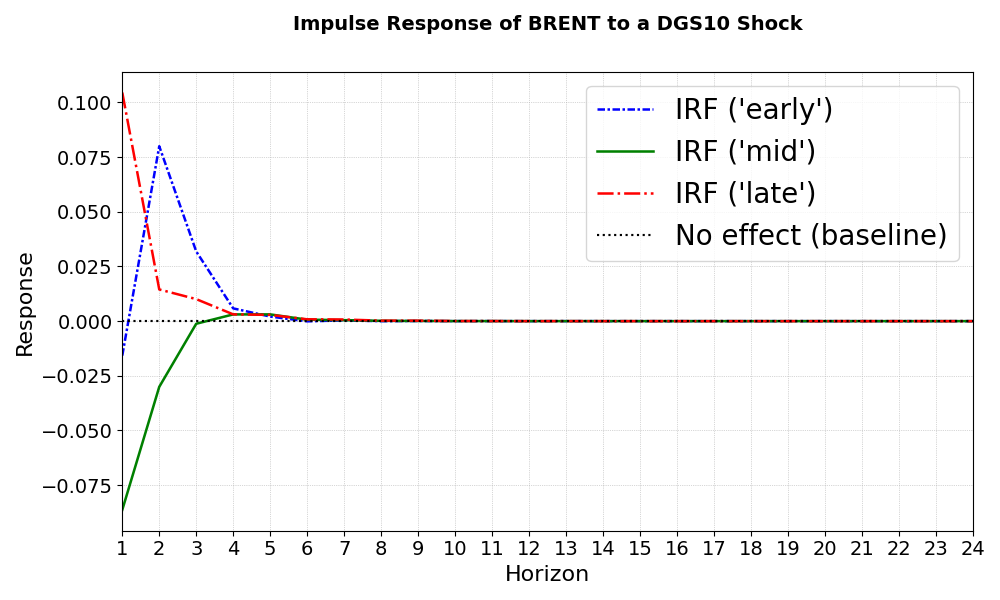}
    \caption{DGS10}
    \label{fig2_m1}
\end{subfigure}
\hfil
\begin{subfigure}[b]{0.4\textwidth}
    \centering
    \includegraphics[width=1.15\hsize]{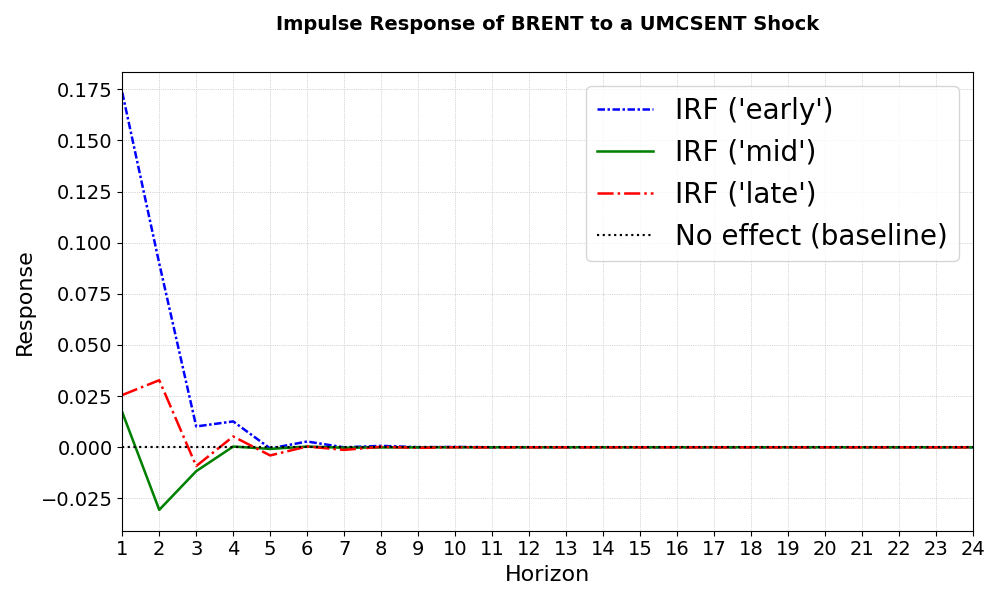}
    \caption{UMSCENT}
    \label{fig2_m2}
\end{subfigure}
\hfil
\begin{subfigure}[b]{0.4\textwidth}
    \centering
    \includegraphics[width=1.15\hsize]{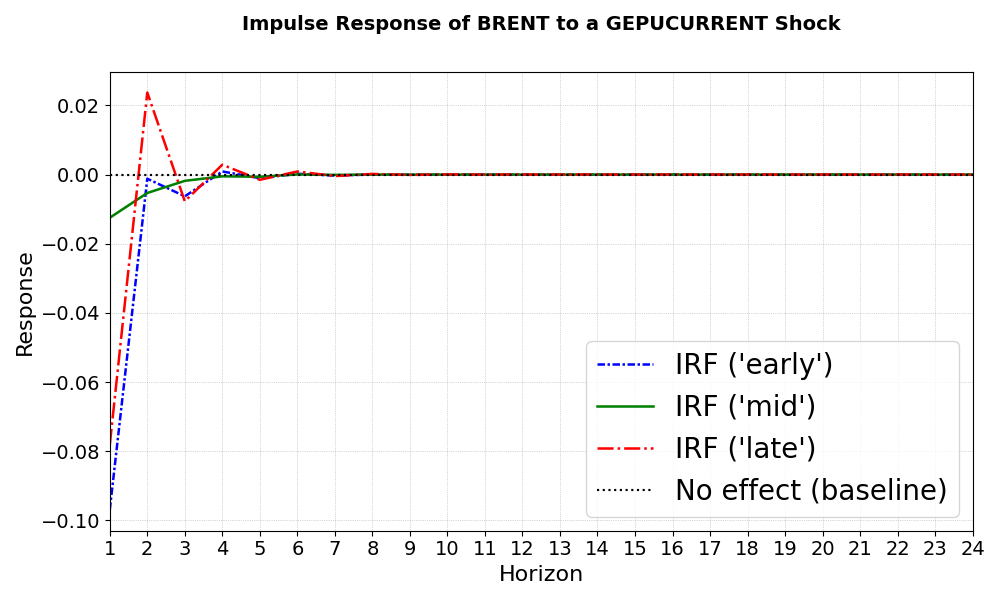}
    \caption{GEPUCURRENT}
    \label{fig2_m3}
\end{subfigure}
\hfil
\begin{subfigure}[b]{0.4\textwidth}
    \centering
    \includegraphics[width=1.15\hsize]{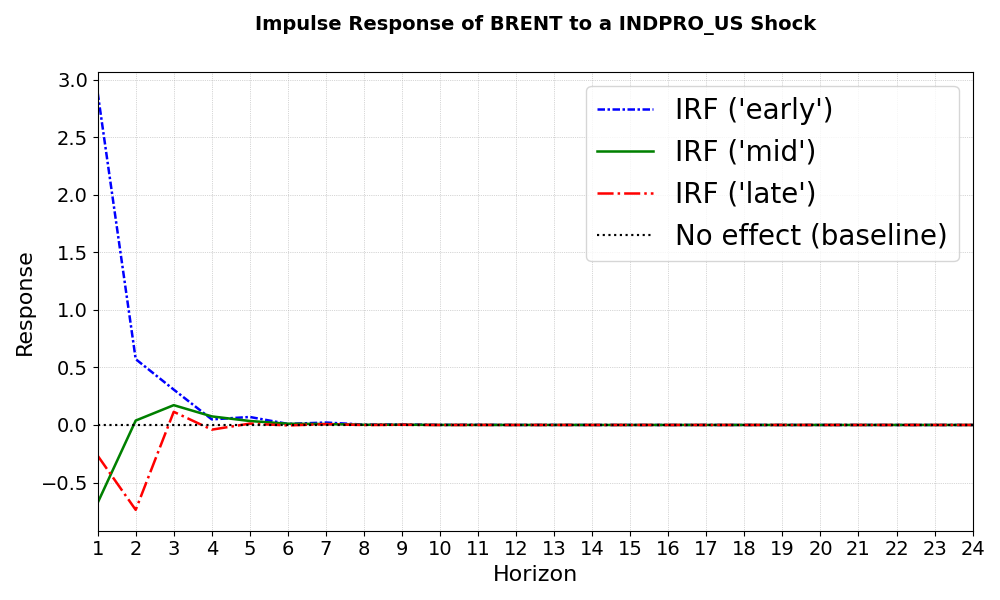}
    \caption{INDPRO$\_$US}
    \label{fig2_m4}
\end{subfigure}
\hfil
\begin{subfigure}[b]{0.4\textwidth}
    \centering
    \includegraphics[width=1.15\hsize]{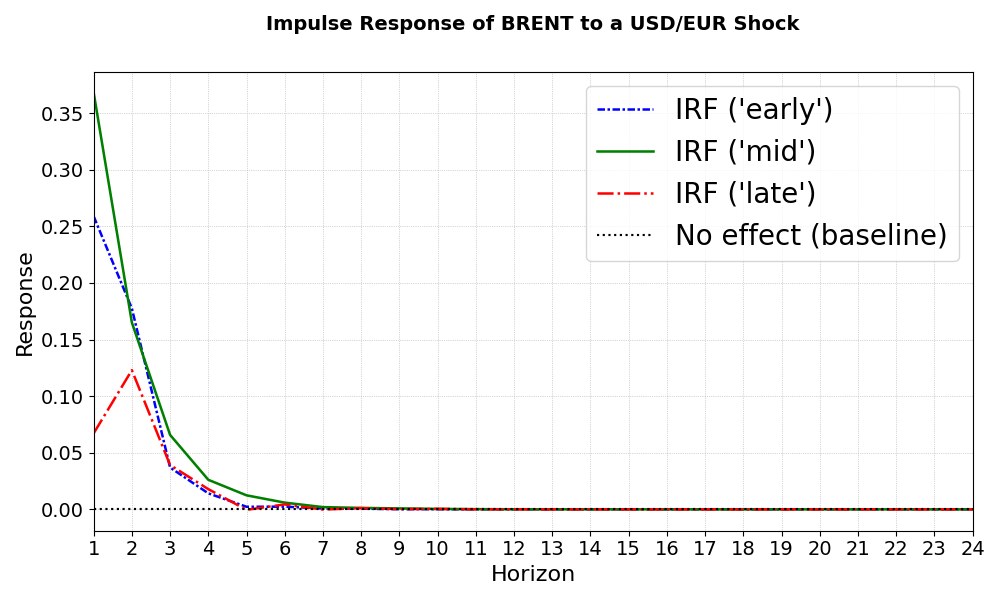}
    \caption{USD/EUR}
    \label{fig2_m5}
\end{subfigure}
\hfil
\begin{subfigure}[b]{0.4\textwidth}
    \centering
    \includegraphics[width=1.15\hsize]{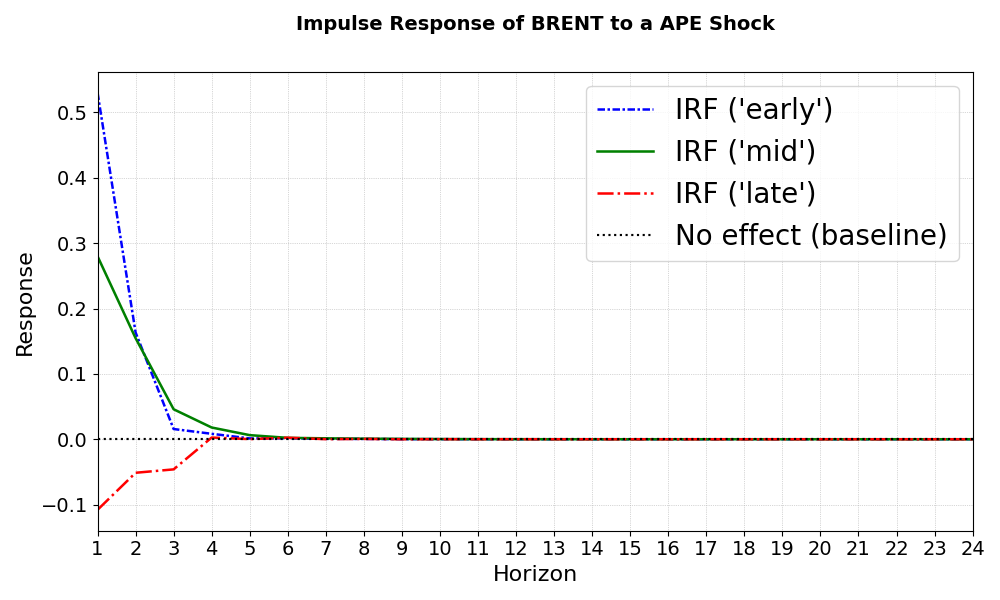}
    \caption{APE}
    \label{fig2_m6}
\end{subfigure}
%------------
\caption{Impulse Response for BRENT}
\label{fig2}
%------------
\end{figure}

Figure \ref{fig2} illustrates the evolution of the response of Brent crude oil to different structural shocks in three temporal regimes \textit{early}, \textit{mid}, and \textit{late} as estimated by the TVP-SVAR model. Unlike the constant coefficient SVAR, which assumes uniform transmission over time, TVP-SVAR reveals that macroenergy links are intrinsically regime-dependent, responding to changes in monetary policy frameworks, financial market integration, and recurrent global crises.\\

\noindent Monetary policy shocks, captured through innovations in the 10-year Treasury yield (DGS10), exhibit a distinct temporal attenuation. In the early regime, a positive interest rate shock generates an increase in Brent prices, reflecting the traditional signaling channel, where higher yields indicate stronger macroeconomic conditions and anticipated energy demand. However, this effect weakens markedly in the late regime, where oil markets appear less responsive to conventional rate signals, probably due to the post-2008 environment of quantitative easing, abundant liquidity, and diminished relevance of real interest rate expectations. This temporal shift implies a structural reconfiguration of the monetary transmission to commodity markets.
%
%--------------------%
%     Figure 2       %
%--------------------%
\newpage
\noindent Uncertainty shocks (GEPUCURRENT) demonstrate the most persistent and regime-dependent influence. In the early regime, uncertainty induces a moderate decline in Brent, consistent with precautionary withdrawal effects. However, during the \textit{mid} and \textit{late} regimes corresponding to the global financial crisis and the COVID-19 pandemic, the negative impact intensifies and becomes highly persistent. This pattern underscores an increasing dominance of global risk sentiment over physical demand fundamentals, suggesting that oil markets have become progressively more financialized and sensitive to tail risk events rather than medium-term macroeconomic conditions.\\

\noindent Household confidence shocks (UMCSENT) continue to exert limited influence across all phases, but subtle temporal asymmetries emerge. Although virtually negligible in the \textit{early} regime, a modest positive effect appears in the \textit{mid}-phase, reflecting the indirect signaling role of sentiment during recovery after the crisis. However, this influence dissipates again in the late regime, confirming that consumer confidence remains peripheral relative to financial and geopolitical drivers in shaping oil price dynamics.\\

\noindent Shocks to real economic activity, proxied by industrial production (INDPRO\_US), show a notable decline in relevance over time. In the \textit{early} regime, Brent responds positively to expansions in production, consistent with classical demand-driven energy consumption. However, this linkage weakens during the \textit{mid} regime and nearly disappears in the \textit{late} period, indicating a structural decoupling between industrial output and oil demand. This attenuation can be attributed to sectoral energy diversification, improved efficiency, and the progressive substitution of oil-intensive inputs.\\

\noindent Exchange rate shocks (USD/EUR) and electricity price shocks (APE) remain largely transitory throughout, yet their temporal profiles provide important nuance. In the \textit{early} phases, currency depreciation exerts measurable downward pressure on Brent, but this effect diminishes in the \textit{mid} and \textit{late} regimes, consistent with increased hedging of currency risk through derivatives and financial integration. APE shocks generate short-lived positive adjustments, indicative of interfuel substitution, although substitution elasticity declines over time, reflecting deeper integration across energy markets.\\

\noindent Brent’s own shocks exhibit significant inertia in all regimes but with decreasing amplitude. In the \textit{early} regime, self-exciting dynamics produce extended deviations, whereas in later phases, price corrections become more rapid, consistent with the rise of high-frequency trading, improved inventory management, and greater speculative efficiency. This decline in persistence signals a transition from supply-dominated price formation to one shaped by financial shocks and market expectations. \\

\noindent In general, the TVP-SVAR results confirm that the structural relationships governing the dynamics of the oil price are neither static nor symmetric. Financial shocks—particularly uncertainty and monetary transmission—gain prominence across regimes, while traditional real-sector determinants weaken. The evidence strongly supports the adoption of a regime-sensitive analytical framework that moves beyond constant-parameter models. These findings motivate the integration of volatility-aware extensions, such as DCC-GARCH and copula structures, to further capture contagion, tail dependence, and asymmetric co-movement in the global macro–energy system.

\subsection{Model Diagnostics and Motivation for Volatility-Aware Extensions}
\label{sec3.6}
A rigorous residual analysis is indispensable to assess the statistical adequacy of both the benchmark VAR and the time-varying parameter SVAR (TVP-SVAR) models. Tables \ref{tab6} through \ref{tab12} present a comprehensive set of diagnostic tests that cover autocorrelation, distributional normality, and conditional heteroskedasticity and consistently reveal significant violations of the classical assumptions that underlie linear Gaussian frameworks. These findings collectively establish the empirical necessity of complementing mean-equation dynamics with volatility-aware and dependence-sensitive extensions such as DCC-GARCH and copula models.
The multivariate Portmanteau test (Table \ref{tab6}) strongly rejects the null hypothesis of absent serial correlation in VAR residuals ($\chi^2 = 1011.4$, $p < 10^{-10}$), indicating that linear dynamics alone does not fully capture lagged crossdependencies among macrofinancial and energy variables. 
\newpage
\noindent Even after allowing for time-varying coefficients within the TVP-SVAR, similar evidence of serial persistence remains, suggesting that evolving means do not eliminate autocorrelated error structures. These results align with empirical observations that macroeconomic systems often experience delayed adjustment mechanisms and regime changes that are not captured by first-moment specifications alone \citep{primiceri2005time,lutkepohl2013introduction}.
%
%--------------------%
%      Tableau 6     %
%--------------------%
\begin{table}[H]
\centering
\resizebox{7.5cm}{!}{
\label{tab:portmanteau-var}
\begin{tabular}{|l| l|}
\hline
\textbf{Statistic} & \textbf{Value} \\
\hline
Data                                  & Residuals of VAR  \\
Test                                  & Portmanteau Test (asymptotic) \\
Test statistic $\chi^{2}$             & 1011.4 \\
Degrees of freedom (df)               & 735 \\
$p$-value                             & $4.584 \times 10^{-11}$ \\
\bottomrule
\end{tabular}
}
\caption{Portmanteau Test for Residual Autocorrelation (VAR)}
\label{tab6}
\end{table}

\noindent The normality diagnostics reported in Tables \ref{tab7} and \ref{tab10} further underscore the inadequacy of Gaussian assumptions. The multivariate Jarque–Bera statistics yield extreme values (VAR: $\chi^2 = 11.282$; TVP-SVAR: $\chi^2 = 22.082$; both $p < 10^{-16}$), and the decomposition into skewness and kurtosis confirms substantial asymmetry and heavy-tailed behavior. This distinct non-normality is consistent with financial and energy markets exposed to episodic crises, tail events, and nonlinear contagion \citep{cont2001empirical}. Under such conditions, Gaussian residuals are empirically implausible, and volatility must be modeled via heavy-tailed or flexible conditional distributions.

%--------------------%
%      Tableau 7     %
%--------------------%
\begin{table}[H]
\centering
\resizebox{18.5cm}{!}{
\begin{subtable}[c]{0.45\textwidth}
\centering
\begin{tabular}{|l|l|}
\toprule
\textbf{Statistic} & \textbf{Value} \\
\midrule
Data & Residuals of VAR  \\
Test & Jarque--Bera (multivariate) \\
$\chi^2$ & 11\,282 \\
df & 14 \\
$p$-value & $< 2.2\times10^{-16}$ \\
\bottomrule
\end{tabular}
\caption{JB-Test (Multivariate)}
\end{subtable}
\quad
\begin{subtable}[c]{0.45\textwidth}
\centering
\begin{tabular}{|l|l|}
\toprule
\textbf{Statistic} & \textbf{Value} \\
\midrule
Data & Residuals of VAR  \\
Test & Skewness (multivariate) \\
$\chi^2$ & 455.97 \\
df & 7 \\
$p$-value & $< 2.2\times10^{-16}$ \\
\bottomrule
\end{tabular}
\caption{Skewness Only (Multivariate)}
\end{subtable}
\begin{subtable}[c]{0.45\textwidth}
\centering
\begin{tabular}{|l|l|}
\toprule
\textbf{Statistic} & \textbf{Value} \\
\midrule
Data & Residuals of VAR \\
Test & Kurtosis  (multivariate) \\
$\chi^2$ & 10\,827 \\
df & 7 \\
$p$-value & $< 2.2\times10^{-16}$ \\
\bottomrule
\end{tabular}
\caption{Kurtosis Only (Multivariate)}
\end{subtable}
}
\begin{tablenotes}\footnotesize
\item \textit{Note.} \textit{The multivariate normality test (Jarque–Bera) breaks down non-normality into components of skewness and kurtosis. P-values $\ll$ 0.001 indicate a rejection of the normality of VAR residuals}.
\end{tablenotes}
\caption{Normality Test for VAR Residuals}
\label{tab7}
\end{table}

%--------------------%
%     Table 8        %
%--------------------%
\begin{table}[H]
\centering
\begin{threeparttable}
\resizebox{15cm}{!}{
\begin{tabular}{|lccccccc|}
\toprule
 & \multicolumn{7}{|c|}{\textbf{Values}} \\
\cmidrule(lr){2-8}
\textbf{Statistic} & \textbf{DGS10} & \textbf{UMCSENT} & \textbf{GEPUCURRENT} & \textbf{INDPRO\_US} & \textbf{USD/EUR} & \textbf{APE} & \textbf{BRENT} \\
\midrule
Data     & Residual & Residual & Residual & Residual & Residual & Residual & Residual \\
$\chi^{2}$ & 13.283 & 8.1761 & 7.4064 & 8.0773 & 20.500 & 19.558 & 10.433 \\
df       & 12 & 12 & 12 & 12 & 12 & 12 & 12 \\
$p$-value & 0.3488 & 0.7712 & 0.8296 & 0.7791 & 0.05821 & 0.07591 & 0.5780 \\
\bottomrule
\end{tabular}
}
\end{threeparttable}
\caption{Univariate ARCH-LM Test on VAR Residuals}
\label{tab8}
\end{table}

\noindent The evidence of time-varying volatility is further corroborated by the univariate ARCH-LM tests in Tables 8 and 11, which reveal significant autoregressive conditional heteroskedasticity for key variables such as USD/EUR, APE, and BRENT. More critically, the multivariate ARCH tests of \cite{tsay2005analysis}  (Tables \ref{tab9} and \ref{tab12}) provide overwhelming rejection of homoscedasticity across all systems. The rejection of the null hypothesis of constant variance, coupled with strong cross-volatility dependence, signals the presence of systemic volatility spillovers a feature that cannot be captured by mean-only VAR or SVAR specifications.

%--------------------%
%     Table 9        %
%--------------------%
\begin{table}[H]
\centering
\resizebox{18cm}{!}{
\begin{tabular}{|l|llll|}
\hline
\textbf{Test} & \textbf{Null Hypothesis ($H_0$)} & \textbf{Test Statistic} & \textbf{p-value} & \textbf{Conclusion} \\
\hline
Q(m) (LM test) & No multivariate ARCH effects & 81.71 & $3.3\times10^{-16}$ & Reject $H_0$ : conditional heteroskedasticity present \\
Rank-based test & No ARCH effects (robust version) & 21.49 & 0.00065 & Reject $H_0$ : robust evidence of time-varying variance \\
$Q_k(m)$ of squared series & No cross-volatility dependence & 625.92 & 0.00000 & Reject $H_0$ : strong volatility interdependence \\
Robust test (5\%) & Constant variance (non-Gaussian robust) & 317.25 & 0.00127 & Reject $H_0$ : heteroskedasticity confirmed \\
\hline
\end{tabular}
}
\caption{Multivariate ARCH test results for VAR residuals \citep{tsay2005analysis}}
\label{tab9}
\end{table}

%--------------------%
%     Tableau 10     %
%--------------------%
\begin{table}[H]
\centering
\resizebox{18.5cm}{!}{
\begin{subtable}[c]{0.45\textwidth}
\centering
\begin{tabular}{|l|l|}
\toprule
\textbf{Statistic} & \textbf{Value} \\
\midrule
Data & Residuals of TVP-SVAR  \\
Test & Jarque--Bera (multivariate) \\
$\chi^2$ & 22\,082 \\
df & 14 \\
$p$-value & $< 2.2\times10^{-16}$ \\
\bottomrule
\end{tabular}
\caption{JB-Test (Multivariate)}
\end{subtable}
\quad
\begin{subtable}[c]{0.45\textwidth}
\centering
\begin{tabular}{|l|l|}
\toprule
\textbf{Statistic} & \textbf{Value} \\
\midrule
Data & Residuals of TVP-SVAR  \\
Test & Skewness (multivariate) \\
$\chi^2$ & 788.919 \\
df & 7 \\
$p$-value & $< 2.2\times10^{-16}$ \\
\bottomrule
\end{tabular}
\caption{Skewness Only (Multivariate)}
\end{subtable}
\begin{subtable}[c]{0.45\textwidth}
\centering
\begin{tabular}{|l|l|}
\toprule
\textbf{Statistic} & \textbf{Value} \\
\midrule
Data & Residuals of TVP-SVAR \\
Test & Jarque--Bera (multivariate) \\
$\chi^2$ & 9\,395 \\
df & 7 \\
$p$-value & $< 2.2\times10^{-16}$ \\
\bottomrule
\end{tabular}
\caption{Kurtosis Only (Multivariate)}
\end{subtable}
}
\begin{tablenotes}\footnotesize
\item \textit{Note.} \textit{The multivariate normality test (Jarque–Bera) breaks down non-normality into components of skewness and kurtosis.\\ P-values $\ll$ 0.001 indicate a rejection of the normality of TVP-SVAR residuals}.
\end{tablenotes}
\caption{Normality Test for TVP-SVAR residuals}
\label{tab10}
\end{table}

\noindent Importantly, while TVP-SVAR successfully accounts for structural instability in transmission coefficients, it does not address second-moment dynamics. The persistence of non-Gaussian, heteroskedastic, and interdependent residuals indicates that an adequate representation of macro–energy systems requires explicit modeling of conditional covariance through multivariate GARCH or copula-based frameworks \citep{engle2002dynamic, bauwens2006multivariate}.\\

\noindent Taken together, the diagnostic evidence conveys a clear methodological mandate:
\begin{itemize}
\item[](i) mean-based models (VAR or TVP-SVAR) alone are statistically insufficient,
\item[](ii) higher-order dependencies via volatility clustering and nonlinear co-movement—must be explicitly modeled, and
\item[](iii) flexible frameworks such as DCC-GARCH, t-DCC, and copulas are required to capture contagion, systemic risk, and tail dependence.
\end{itemize}
This empirical motivation forms the conceptual transition to the next section, where volatility-aware extensions are integrated with the VAR and TVP-SVAR structures to establish a comprehensive modeling framework for macro–energy dynamics.

%--------------------%
%     Table 11       %
%--------------------%
\begin{table}[H]
\centering
\begin{threeparttable}
\resizebox{15cm}{!}{
\begin{tabular}{|lccccccc|}
\toprule
 & \multicolumn{7}{|c|}{\textbf{Values}} \\
\cmidrule(lr){2-8}
\textbf{Statistic} & \textbf{DGS10} & \textbf{UMCSENT} & \textbf{GEPUCURRENT} & \textbf{INDPRO\_US} & \textbf{USD/EUR} & \textbf{APE} & \textbf{BRENT} \\
\midrule
Data     & Residual & Residual & Residual & Residual & Residual & Residual & Residual \\
$\chi^{2}$ & 18.748 & 11.127 & 11.531 & 15.344 & 20.372 & 17.319 & 14.194 \\
df       & 12 & 12 & 12 & 12 & 12 & 12 & 12 \\
$p$-value & 0.0948 & 0.5181 & 0.4841 & 0.2232 & 0.06037 & 0.138 & 0.2885 \\
\bottomrule
\end{tabular}
}
\end{threeparttable}
\caption{Univariate ARCH-LM Test on TVP-SVAR residuals}
\label{tab11}
\end{table}

%--------------------%
%     Table 12       %
%--------------------%
\begin{table}[H]
\centering
\resizebox{18.3cm}{!}{
\begin{tabular}{|l|llll|}
\hline
\textbf{Test} & \textbf{Null Hypothesis ($H_0$)} & \textbf{Test Statistic} & \textbf{p-value} & \textbf{Conclusion} \\
\hline
Q(m) (LM test) & No multivariate ARCH effects & 125.6907 & 0.00000 & Reject $H_0$ : conditional heteroskedasticity present \\
Rank-based test & No ARCH effects (robust version) & 188.8244 & 0.00000 & Reject $H_0$ : robust evidence of time-varying variance \\
$Q_k(m)$ of squared series & No cross-volatility dependence & 1461.109 & 0.00000 & Reject $H_0$ : strong volatility interdependence \\
Robust test (5\%) & Constant variance (non-Gaussian robust) & 1485.041 & 0.00000 & Reject $H_0$ : heteroskedasticity confirmed \\
\hline
\end{tabular}
}
\caption{Multivariate ARCH test results for TVP-SVAR residuals \citep{tsay2005analysis}}
\label{tab12}
\end{table}
\subsection{Modelling Dynamic Dependence and Contagion: from DCC-GARCH to Copulas}
The diagnostic evidence in Section \ref{sec3.6} establishes pronounced serial dependence, conditional heteroskedasticity, and heavy-tailed innovations in both VAR and TVP-SVAR residuals. Modeling mean dynamics alone is therefore insufficient. We proceed in two complementary steps. First, we estimate symmetric and asymmetric Dynamic Conditional Correlation GARCH models (DCC and ADCC) on the standardized residuals to recover time-varying second moments and correlation persistence \citep{engle2002dynamic, cappiello2006asymmetric,hafner2006volatility}. 
\newpage
\noindent Second, we move beyond elliptical dependence and study non-linear co-movement and tail linkage using copula-based specifications, namely a dynamic Student-t copula and a bootstrap-estimated mixture of Archimedean copulas (Clayton–Frank–Gumbel) \citep{patton2012review,genest2009goodness}.

\noindent \subsubsection{Symmetric and Asymmetric DCC-GARCH}
All conditional variances are modeled by univariate margins of GARCH (1,1) with Student-t innovations; the estimated $\alpha_{1}$ quantify the \textit{news} in volatility, while $\beta_{1}$ measure its persistence. In our data $\beta_{1,i}$ typically exceeds 0.80, confirming pronounced volatility clustering in macro-financial and energy returns.
The DCC layer governs the dynamics of conditional correlations. Its two core parameters, $\mathrm{dccA}_1$ and $\mathrm{dccB}_1$, correspond to the coefficients $a$ and $b$ in Eq. \eqref{e7}. The first parameter $\mathrm{dccA}_1$, captures the immediate response of the correlations to new standardized shocks (“\textit{news in correlations}”), while $\mathrm{dccB}_1$ governs their persistence over time. Stationarity requires $\mathrm{dccA}_1 + \mathrm{dccB}_1 < 1$. When the innovation vector follows a multivariate Student-t distribution, the degree-of-freedom parameter ($\mathrm{mshape} \equiv \nu$) reflects the thickness of the joint tails.\\

\noindent Table \ref{tab13} reports the estimates for the symmetric DCC specification. Using VAR residuals, we obtain $\mathrm{dccA}_1 = 0.0159$, $\mathrm{dccB}_1 = 0.7016$, and $\mathrm{mshape} = 9.12$. When the model is re-estimated with TVP-SVAR residuals, these parameters increase to $\mathrm{dccA}_1 = 0.0900$ and $\mathrm{dccB}_1 = 0.8459$, with $\mathrm{mshape} = 11.15$. This shift indicates that allowing for time variation in the conditional mean makes correlations substantially more reactive to shocks and more persistent over time, while the joint distribution remains heavy-tailed, though slightly thinner. Economically, this finding suggests that structural instability in the mean reveals enduring co-movements consistent with contagion mechanisms typically observed during episodes of financial or energy market stress.

%--------------------%
%     Table 13       %
%--------------------%
\begin{table}[H]
\centering
\resizebox{8.cm}{!}{
\begin{subtable}[c]{0.7\textwidth}
\centering
\begin{tabular}{lccc}
\toprule
 & \multicolumn{3}{c}{\textbf{Parameters}} \\
\cmidrule(lr){2-4}
\textbf{Series} & $\alpha_1$ & $\beta_1$ & $\text{shape}$ \\
\midrule
DGS10        & 0.0389  & 0.9601  & 4.6420  \\
UMCSENT      & 0.1119  & 0.8862  & 5.7115  \\
GEPUCURRENT  & 0.0000  & 0.9990  & 5.5334  \\
INDPRO$\_$US   & 0.3913  & 0.2424  & 5.3048  \\
USD/EUR      & 0.0668  & 0.9010  & 12.5275 \\
APE          & 0.0170  & 0.9619  & 2.8526  \\
BRENT        & 0.3479  & 0.0866  & 17.2996 \\
\midrule
\textbf{Joint DCC Parameters:} & &  $\text{dccA}_1 = 0.0159$ & \\
 & &  $\text{dccB}_1 = 0.7016$ & \\
 & &  $\text{mshape} = 9.1215$ & \\
\bottomrule
\end{tabular}
\caption{VAR}
\end{subtable}
}
\quad
\resizebox{8.cm}{!}{
\begin{subtable}[c]{0.7\textwidth}
\centering
\begin{tabular}{lccc}
\toprule
 & \multicolumn{3}{c}{\textbf{Parameters}} \\
\cmidrule(lr){2-4}
\textbf{Series} & $\alpha_1$ & $\beta_1$ & $\text{shape}$ \\
\midrule
DGS10        & 0.1200  & 0.8392  & 4.7142   \\
UMCSENT      & 0.4432  & 0.4252  & 99.9999  \\
GEPUCURRENT  & 0.3051  & 0.6710  & 100.0000 \\
INDPRO$\_$US & 0.3744  & 0.6246  & 3.4808   \\
USD/EUR      & 0.6096  & 0.3894  & 13.6896  \\
APE          & 0.0000  & 0.9990  & 8.8900   \\
BRENT        & 0.5112  & 0.4571  & 99.9996  \\
\midrule
\textbf{Joint DCC Parameters:} & &  $\text{dccA}_1 = 0.0900$ & \\
 & &  $\text{dccB}_1 = 0.8459$ & \\
 & &  $\text{mshape} = 11.1490$ & \\
\bottomrule
\end{tabular}
\caption{TVP-SVAR}
\end{subtable}
}
\caption{DCC-GARCH Parameter Estimates}
\label{tab13}
\end{table}

%--------------------%
%     Table 14       %
%--------------------%
\begin{table}[H]
\centering
\resizebox{8.cm}{!}{
\begin{subtable}[c]{0.7\textwidth}
\centering
\begin{tabular}{lccc}
\toprule
 & \multicolumn{3}{c}{\textbf{Parameters}} \\
\cmidrule(lr){2-4}
\textbf{Series} & $\alpha_1$ & $\beta_1$ & $\text{shape}$ \\
\midrule
DGS10        & 0.0389  & 0.9601  & 4.6420   \\
UMCSENT      & 0.1119  & 0.8862  & 5.7115  \\
GEPUCURRENT  & 0.0000  & 0.9990  & 5.5334  \\
INDPRO\_US   & 0.3913  & 0.2424  & 5.3048  \\
USD/EUR      & 0.0668  & 0.9010  & 12.5275 \\
APE          & 0.0170  & 0.9619  & 2.8526  \\
BRENT        & 0.3479  & 0.0866  & 17.2996 \\
\midrule
\textbf{Joint DCC Parameters:} & &  $\text{dccA}_1 = 0.0155$ & \\
 & &  $\text{dccB}_1 = 0.7007$ & \\
 & &  $\text{dccG}_1 = 0.0010$ & \\
 & &  $\text{mshape} = 9.1193$ & \\
\bottomrule
\end{tabular}
\caption{VAR}
\end{subtable}
}
\quad
\resizebox{8.cm}{!}{
\begin{subtable}[c]{0.7\textwidth}
\centering
\begin{tabular}{lccc}
\toprule
 & \multicolumn{3}{c}{\textbf{Parameters}} \\
\cmidrule(lr){2-4}
\textbf{Series} & $\alpha_1$ & $\beta_1$ & $\text{shape}$ \\
\midrule
DGS10        & 0.1200  & 0.8392  & 4.7142  \\
UMCSENT      & 0.4432  & 0.4252  & 99.9999 \\
GEPUCURRENT  & 0.3051  & 0.6710  & 100.0000 \\
INDPRO\_US   & 0.3744  & 0.6246  & 3.4808  \\
USD/EUR      & 0.6096  & 0.3894  & 13.6896 \\
APE          & 0.0000  & 0.9990  & 8.8900  \\
BRENT        & 0.5112  & 0.4571  & 99.9996 \\
\midrule
\textbf{Joint DCC Parameters:} & &  $\text{dccA}_1 = 0.0825$ & \\
 & &  $\text{dccB}_1 = 0.8329$ & \\
 & &  $\text{dccG}_1 = 0.0285$ & \\
 & &  $\text{mshape} = 11.2582$ & \\
\bottomrule
\end{tabular}
\caption{TVP-SVAR}
\end{subtable}
}
\caption{Asymmetric DCC-GARCH Parameter Estimates}
\label{tab14}
\end{table}
\newpage
\noindent The ADCC extension augments the DCC recursion with an asymmetry term. Its coefficient, $\mathrm{dccG}_1$, corresponding to $g$ in Eq. \eqref{e9}, captures the differential response of correlations to negative versus positive shocks (component-wise leverage). A positive $\mathrm{dccG}_1$ implies downward synchronization, which means that correlations increase more after adverse shocks than they decrease after favorable ones.\\

\noindent Table \ref{tab14} presents the results for the asymmetric DCC specification. Under VAR, $\mathrm{dccG}_1 = 0.0010$ is economically negligible, suggesting that the correlation dynamics is largely symmetric. 
However, in TVP-SVAR, $\mathrm{dccG}_1 = 0.0285$ becomes meaningful, while $\mathrm{dccA}_1 = 0.0825$ and $\mathrm{dccB}_1 = 0.8329$ confirm strong responsiveness and persistence. The associated $\mathrm{mshape} = 11.26$ again points to heavy, but not extreme, tails. Introducing time variation in the mean therefore uncovers asymmetric dependence patterns characteristic of contagion under stress, where correlations spike after adverse shocks, undermining diversification exactly when it is most needed, a well-documented phenomenon in financial-energy linkages.

%--------------------%
%     Table 15       %
%--------------------%
\begin{table}[H]
\centering
\resizebox{8.cm}{!}{
\begin{subtable}[c]{0.75\textwidth}
\centering
\begin{tabular}{|l|ccc|}
\toprule
 & \multicolumn{3}{c|}{\textbf{Parameters}} \\
\cmidrule(lr){2-4}
\textbf{Series} & $\alpha_1$ & $\beta_1$ & $\text{shape}$ \\
\midrule
DGS10        & 0.0983 & 0.8550 & 5.0557 \\
UMCSENT      & 0.1149 & 0.8814 & 4.6643 \\
GEPUCURRENT  & 0.0000 & 0.9990 & 5.3217 \\
INDPRO\_US   & 0.4899 & 0.1865 & 4.0015 \\
USD/EUR      & 0.0676 & 0.8884 & 11.0791 \\
APE          & 0.0135 & 0.9854 & 2.9553 \\
BRENT        & 0.4040 & 0.0000 & 9.9794 \\
\midrule
\textbf{Joint DCC Copula Parameters:} & &  $\text{dccA}_1 = 0.0088$ & \\
 & &  $\text{dccB}_1 = 0.6699$ & \\
 & &  $\text{mshape} = 28.6404$ & \\
\bottomrule
\end{tabular}
\caption{VAR}
\end{subtable}
}
\quad
\resizebox{8.cm}{!}{
\begin{subtable}[c]{0.75\textwidth}
\centering
\begin{tabular}{|l|ccc|}
\toprule
 & \multicolumn{3}{c|}{\textbf{Parameters}} \\
\cmidrule(lr){2-4}
\textbf{Series} & $\alpha_1$ & $\beta_1$ & $\text{shape}$ \\
\midrule
DGS10        & 0.1526 & 0.7747 & 5.2209 \\
UMCSENT      & 0.0285 & 0.9471 & 5.1827 \\
GEPUCURRENT  & 0.0957 & 0.6517 & 22.6798 \\
INDPRO\_US   & 0.3545 & 0.6288 & 3.6390 \\
USD/EUR      & 0.2523 & 0.7103 & 6.0256 \\
APE          & 0.0000 & 0.9990 & 9.3484 \\
BRENT        & 0.0202 & 0.9642 & 4.8221 \\
\midrule
\textbf{Joint DCC Copula Parameters:} & &  $\text{dccA}_1 = 0.0075$ & \\
 & &  $\text{dccB}_1 = 0.8454$ & \\
 & &  $\text{mshape} = 22.2284$ & \\
\bottomrule
\end{tabular}
\caption{TVP-SVAR}
\end{subtable}
}
\caption{Dynamic DCC Student Copula Parameter Estimates}
\label{tab15}
\end{table}
\noindent To move beyond elliptical and tail-symmetric dependence structures, the marginal GARCH(1,1) processes are coupled with a dynamic Student-t copula whose correlation matrix follows the DCC recursion. The parameters retain their standard interpretation: $\mathrm{dccA}_1$ captures the sensitivity of conditional correlations to new standardized shocks (“\textit{news}”), $\mathrm{dccB}_1$ reflects their persistence, and $\mathrm{mshape}$ denotes the degrees of freedom of the copula, governing the strength of the dependence of the common tail across all pairs of variables.\\

\noindent Table \ref{tab15} reports the estimation results for the dynamic t-copula DCC specification. Using VAR residuals, we obtain $\mathrm{dccA}_1 = 0.0088$, $\mathrm{dccB}_1 = 0.6699$, and $\mathrm{mshape} = 28.64$, suggesting moderately persistent correlations and tails approaching Gaussian behavior. When using TVP-SVAR residuals, $\mathrm{dccA}_1$ remains similar at 0.0075, but $\mathrm{dccB}_1$ rises markedly to 0.8454, with $\mathrm{mshape} = 22.23$. This pattern indicates a highly persistent dependence and stronger joint tail thickness than in the VAR framework. Economically, introducing time variation in the mean reveals a regime in which extreme co-movements become more enduring, consistent with the predominance of episodic crisis dynamics and contagion effects in energy-financial interactions.\\

\noindent Across Tables \ref{tab13}–\ref{tab15}, the joint evolution of the DCC parameters ($\mathrm{dccA}_1,\mathrm{dccB}_1,\mathrm{dccG}_1,\mathrm{mshape}$) corresponding to the coefficients ($a,b,g,\nu$) in Eqs. \eqref{e7} and \eqref{e9} provides a coherent view of the dependence structure across models. Under the TVP-SVAR specification, both $\mathrm{dccA}_1$ and $\mathrm{dccB}_1$ increase noticeably, indicating that conditional correlations respond more strongly to new shocks and exhibit greater persistence. 
This enhanced reactivity and memory suggest that the effects of contagion are not only sharper but also more durable once the structural variation in the mean is taken into account. The emergence of a positive $\mathrm{dccG}_1$ further signals an asymmetry in correlation dynamics, with negative shocks inducing stronger co-movements than positive ones, a feature consistent with downside contagion observed during stress episodes.\\ 

\noindent Finally, the reduction in copula degrees of freedom $\mathrm{mshape}$ under the TVP-SVAR framework points to heavier joint tails, which implies a greater likelihood of simultaneous extreme movements across markets.
\newpage
\noindent Overall, these results suggest that incorporating time variation in the mean fundamentally reshapes the dependence structure: correlations become more reactive, more persistent, and more asymmetric, while joint tail thickness amplifies systemic interconnectedness. This configuration reflects a tightening of cross-market links and a deterioration of diversification benefits during periods of financial stress or uncertainty hallmarks of crisis-driven contagion in macro–energy interactions.

\subsubsection{From the Dynamic t-Copula to Archimedean and Mixed Copulas}
Building on the dynamic t-copula coupled with DCC-GARCH, which confirms persistent and heavy-tailed dependence but remains elliptical and hence tail symmetric, we next explore whether the data exhibit directional tail behavior specifically, stronger dependence during downturns than during upswings. To this end, we employ Archimedean copulas whose generators naturally accommodate asymmetric tail features. Individual copula families (Clayton, Frank, and Gumbel) are assessed through Cramér–von Mises goodness-of-fit statistics with parametric bootstrap inference, following the benchmark procedures of \citet*{genest2009goodness}, \cite{kojadinovic2011goodness}, and \cite{patton2012review}. When single families prove insufficient, mixture specifications are introduced to capture multiple coexisting dependence regimes.

%--------------------%
%     Table 16       %
%--------------------%
\begin{table}[H]
\centering
\resizebox{8.3cm}{!}{
\begin{minipage}{0.75\textwidth}
\centering
\begin{tabular}{|l|cccc|}
\hline
Model & Copula & Statistic & p-value & Decision (5\%) \\
\hline
             & Clayton & 0.0064 & 0.9515 & Accepted \\
\textbf{VAR} & Frank   & 0.0343 & 0.0005 & Rejected\\
             & Gumbel  & 0.0483 & 0.0005 & Rejected\\
\hline
                  & Clayton & 0.0069 & 0.7537 & Accepted \\
\textbf{TVP-SVAR} & Frank   & 0.0068 & 0.8207 & Accepted \\
                  & Gumbel  & 0.0253 & 0.0499 & Marginal (5\%)\\
\hline
\end{tabular}
\subcaption{Global Dependence}
\end{minipage}%
}
\hfill
\resizebox{8.3cm}{!}{
\begin{minipage}{0.75\textwidth}
\centering
\begin{tabular}{|l|cccc|}
\hline
Model & Copula & Statistic & p-value & Decision (5\%) \\
\hline
             & Clayton & 0.0059 & 0.6207 & Accepted \\
\textbf{VAR} & Frank   & 0.0343 & 0.0005 & Rejected\\
             & Gumbel  & 0.0413 & 0.0005 & Rejected\\
\hline
                  & Clayton & 0.0069 & 0.7318 & Accepted \\
\textbf{TVP-SVAR} & Frank   & 0.0069 & 0.7338 & Accepted \\
                  & Gumbel  & 0.0275 & 0.0509 & Marginal (5\%)\\
\hline
\end{tabular}
\subcaption{Tail Dependence}
\end{minipage}
}
\caption{Goodness-of-Fit Tests for Copulas: VAR vs TVP-SVAR Residuals}
\label{tab16}
\end{table}
%

%--------------------%
%     Table 17       %
%--------------------%
\begin{table}[H]
\centering
\resizebox{9.55cm}{!}{
\begin{tabular}{|lcccccc|}
\hline
 & \multicolumn{3}{c|}{\textbf{VAR}} & \multicolumn{3}{c|}{\textbf{TVP-SVAR}} \\
\cmidrule(lr){2-4} \cmidrule(lr){5-7}
 & Clayton & Frank & Gumbel & Clayton & Frank & Gumbel \\
\midrule
\textbf{Parameter}  & 0.0188 & 0.0484 & 1.0388 & 0.0833 & 0.2544 & 1.0310 \\
\textbf{Weight}     & 1.0000 & 0.0000 & 0.0000 & 0.7061 & 0.2939 & 0.0000 \\
\midrule
\textbf{Log-Likelihood} & \multicolumn{3}{c|}{0.9194} & \multicolumn{3}{c|}{4.6302} \\
%\bottomrule
\hline
\end{tabular}
}
\caption{Mixed Copula Parameter Estimates with Bootstrap Optimization}
\label{tab17}
\end{table}

\noindent Table \ref{tab16} reports the global goodness of fit results ($Sn$) and tail-focused ($SnC$) for the three Archimedean copulas under VAR and TVP-SVAR residuals. For the VAR model, the Clayton copula is strongly accepted in both tests, whereas the Frank and Gumbel families are decisively rejected. This outcome indicates a clear dominance of lower-tail dependence, synchronization under stress, but an inability to capture association during normal or expansionary phases. For the TVP-SVAR model, both Clayton and Frank copulas are globally accepted, while Gumbel attains only marginal significance at the 5\% level in the tail test. Allowing for time variation in the mean thus enriches the dependence structure, revealing joint co-movement during crises (Clayton) and more balanced association in tranquil periods (Frank). The borderline relevance of the Gumbel copula suggests that the upper tail dependence remains limited, consistent with evidence that positive oil or macro shocks propagate less uniformly than adverse ones \citep{baumeister2013time}. \\

\noindent Table \ref{tab17} extends the analysis to a mixed Clayton–Frank–Gumbel copula estimated by maximum likelihood with parametric bootstrap for both weights and family parameters. Under the VAR residuals, the mixture collapses to a pure Clayton component (weight 1.000), confirming that dependence is almost exclusively lower-tail. For the residuals of TVP-SVAR, the mixture assigns a dominant weight to Clayton (0.706) and a substantial share to Frank (0.294), while Gumbel remains negligible. The composite model achieves a significantly higher log-likelihood, demonstrating the statistical and economic value of allowing multiple dependence regimes. Once structural time variation is acknowledged, dependence is driven primarily by crisis-induced contagion (Clayton) but complemented by moderate, symmetric co-movement during stable periods (Frank).
\newpage

\noindent The lack of a Gumbel contribution reinforces the view that speculative upper-tail linkages are not a defining feature of the macro–energy relationship.\\

\noindent Together, Tables \ref{tab16} and \ref{tab17} provide consistent evidence that incorporating time-varying parameters is essential to recover realistic dependence beyond crisis episodes. Lower-tail synchronization remains the dominant feature, yet a mixed copula specification is statistically warranted to reconcile stress-regime contagion with business cycle co-movement within a unified dependence framework. This motivates the copula-augmented TVP-SVAR as the structural backbone for the predictive analysis that follows.

\subsection{Comparative Predictive Evaluation}
\label{sec3.8}
This section performs a systematic comparison of out-of-sample predictive performance across competing econometric and machine learning frameworks, thus addressing the central empirical question of this study: \textit{Can structurally grounded models, such as TVP-SVAR and Copula-GARCH, match or surpass the forecast accuracy of flexible nonparametric learners such as Gaussian Process Regression?}\\ 

\noindent Forecasts are generated using a 30-step rolling window, and evaluated using the Root Mean Squared Error (RMSE) across all seven macro-financial and energy variables. The results, synthesized in Figure \ref{fig3} and numerically detailed in the appendix Table \ref{tabA1}, establish a clear hierarchy in predictive performance and highlight the conditions under which structural interpretability and predictive precision converge or diverge.\\

\noindent A first and unequivocal finding is the superiority of time-varying parameter structures over their constant-coefficient counterparts. Across all variables and forecasting horizons, the TVP-SVAR significantly outperforms the standard VAR, even before any volatility or dependence augmentation is introduced. This systematic improvement visible in the 'Original' benchmark entries in Figure \ref{fig3} confirms the evidence of structural instability documented in Sections \ref{sec3.4} to \ref{sec3.6}. In particular, for variables most exposed to policy and uncertainty regimes, such as DGS10, GEPUCURRENT, and BRENT, the benefits of time variation in the mean equation are especially pronounced. These findings indicate that parameter constancy is an untenable assumption in macro-energy dynamics and that adaptative structural models are a prerequisite for credible forecasting.\\

\noindent The second set of results concerns volatility-aware extensions, namely the DCC-GARCH, ADCC-GARCH, and Copula-GARCH specifications. Incorporating dynamic conditional correlation yields modest RMSE gains relative to the baseline VAR, but these improvements remain limited when dependence structures are restricted to symmetric elliptical forms. It is only through the integration of copula-based dependence, particularly under the TVP-SVAR mean, that a substantive reduction in forecast error is observed. This is most evident in variables characterized by asymmetric shocks and tail-sensitive behavior such as exchange rates (USD/EUR), industrial production (INDPRO), and notably energy prices (BRENT and APE). These patterns confirm that forecast performance in energy–finance systems is driven not only by conditional variance, but by deeper forms of tail dependence and nonlinear contagion, especially around crisis episodes (e.g., 2008, 2014-15, 2020).\\

\noindent However, the most striking empirical result concerns the performance of Gaussian Process Regression (GPR) within the hybrid learning class. In all panels of Figure \ref{fig3}, GPR consistently achieves the lowest RMSE values, outperforming not only VAR and TVP-SVAR, but also models augmented by volatility and extended copula. Unlike feedforward neural networks (ANN) or ensemble methods such as random forests (RF), the Bayesian nonparametric GPR kernel structure allows it to approximate smooth nonlinear functions while internally regulating overfitting. Notably, its predictive behavior closely mirrors the logic of structurally time-varying models, suggesting that GPR may be implicitly learning regime shifts and transmission asymmetries usually encoded explicitly in econometric models. This convergence substantiates the argument that machine learning, when properly specified, does not contradict structural modeling but rather reproduces its latent mechanisms through the flexible function approximation.
\newpage
%--------------------%
%     Figure 3       %
%--------------------%
\begin{figure}[H]
%\begin{figure}
\centering
\begin{subfigure}[b]{0.4\textwidth}
    \centering
    \includegraphics[width=1.15\hsize]{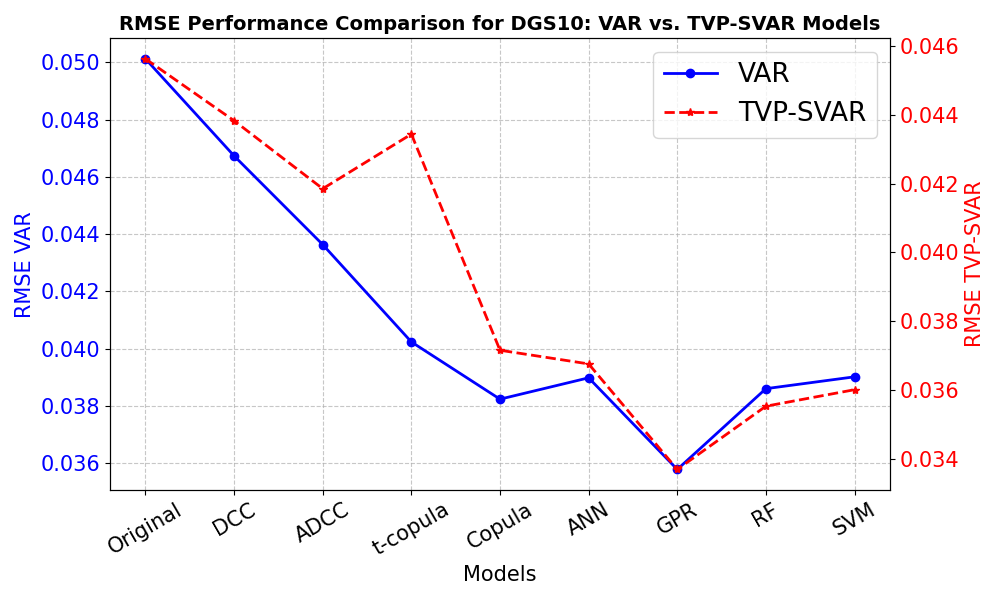}
    \caption{DGS10}
    \label{fig3_m1}
\end{subfigure}
\hfil
\begin{subfigure}[b]{0.4\textwidth}
    \centering
    \includegraphics[width=1.15\hsize]{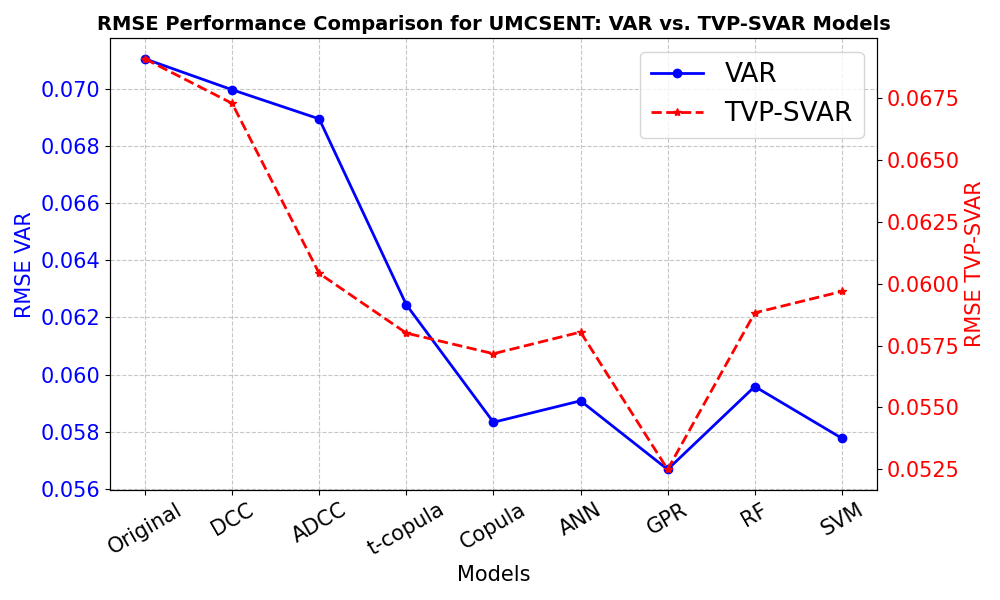}
    \caption{UMSCENT}
    \label{fig3_m2}
\end{subfigure}
\hfil
\begin{subfigure}[b]{0.4\textwidth}
    \centering
    \includegraphics[width=1.15\hsize]{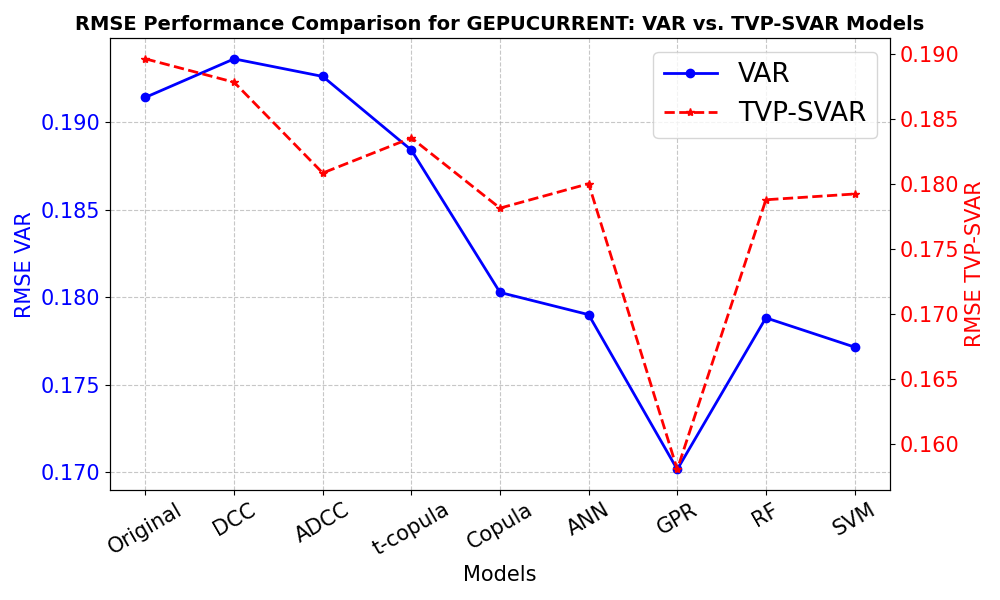}
    \caption{GEPUCURRENT}
    \label{fig3_m3}
\end{subfigure}
\hfil
\begin{subfigure}[b]{0.4\textwidth}
    \centering
    \includegraphics[width=1.15\hsize]{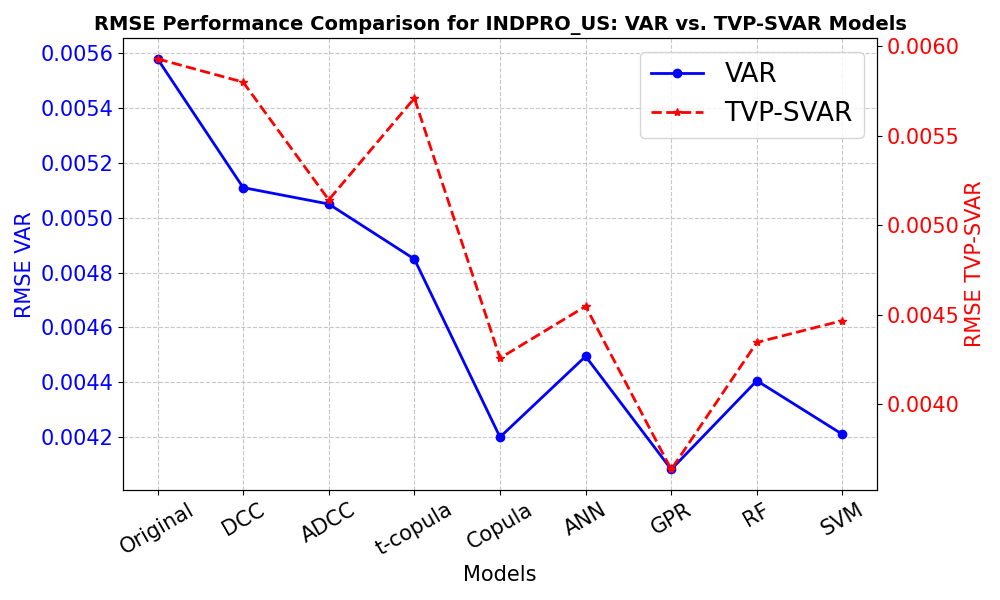}
    \caption{INDPRO$\_$US}
    \label{fig3_m4}
\end{subfigure}
\hfil
\begin{subfigure}[b]{0.4\textwidth}
    \centering
    \includegraphics[width=1.15\hsize]{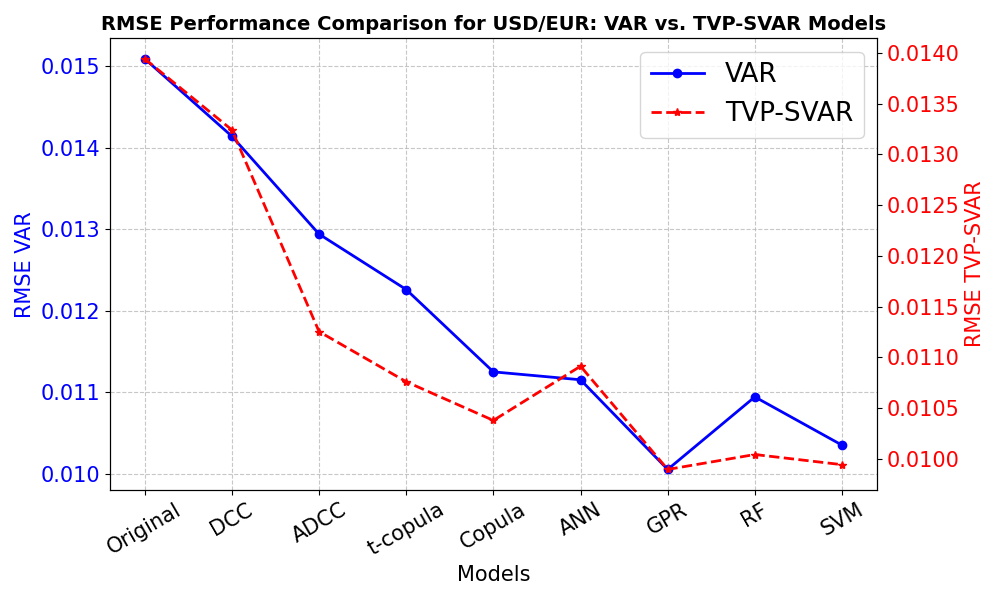}
    \caption{USD/EUR}
    \label{fig3_m5}
\end{subfigure}
\hfil
\begin{subfigure}[b]{0.4\textwidth}
    \centering
    \includegraphics[width=1.15\hsize]{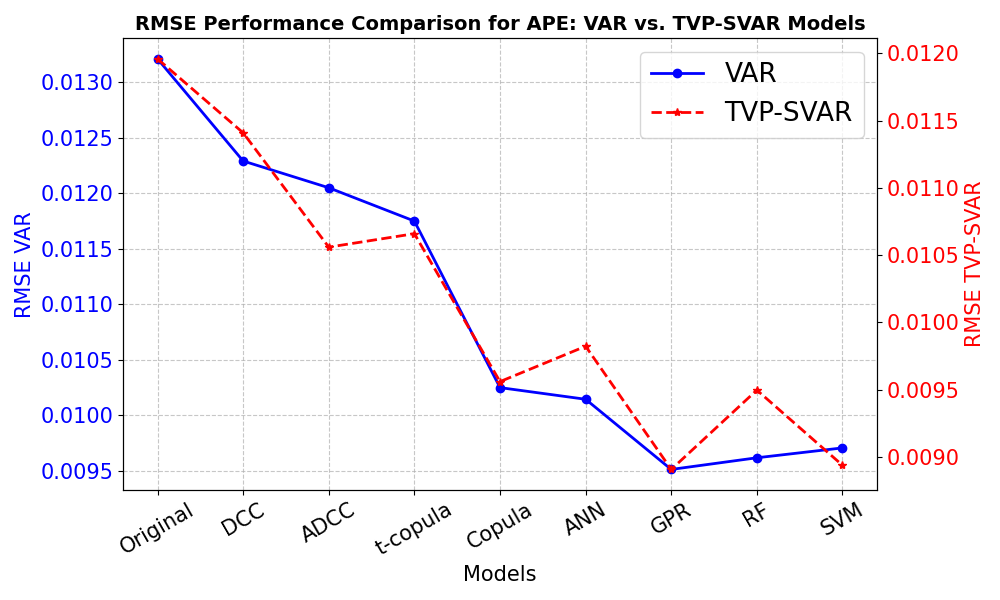}
    \caption{APE}
    \label{fig3_m6}
\end{subfigure}
\hfil
\begin{subfigure}[b]{0.4\textwidth}
    \centering
    \includegraphics[width=1.15\hsize]{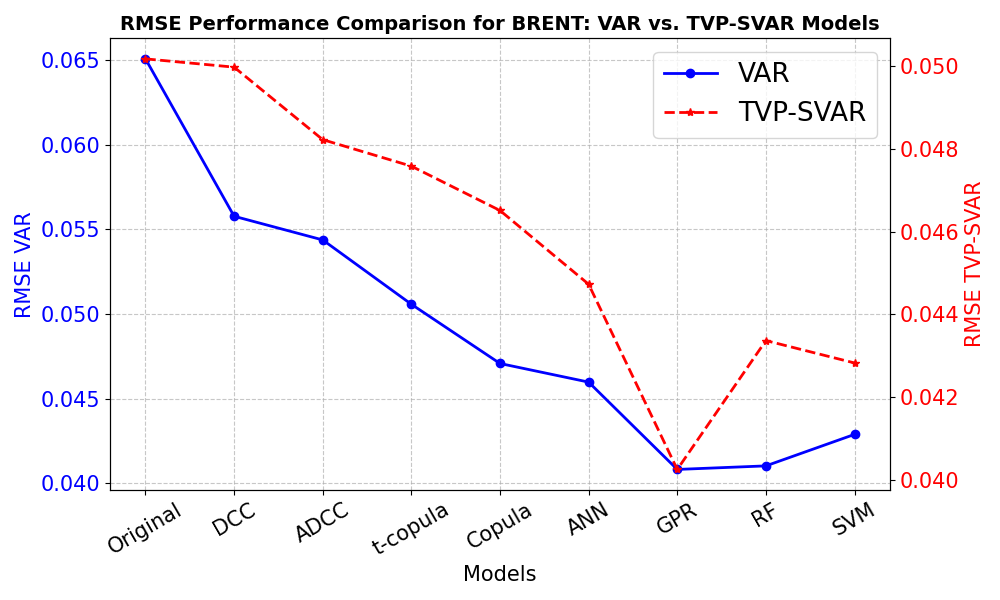}
    \caption{BRENT}
    \label{fig3_m7}
\end{subfigure}
%------------
\begin{tablenotes}\footnotesize
\item \textit{Note.} \textit{“Original” refers to baseline VAR or TVP-SVAR forecasts. “DCC” and “ADCC” correspond to VAR-DCC\\ (or TVP-SVAR-DCC) volatility extensions, while “Copula” encompasses VAR-GARCH-Clayton or \\TVP-SVAR-GARCH-Mixed Copula configurations.}.
\end{tablenotes}
%------------
\caption{RMSE-Based Predictive Performance: Copula-Enhanced Models vs. ML Hybrids}
\label{fig3}
%------------
\end{figure}

\newpage
\noindent Economically, the predictive hierarchy inferred from Figure \ref{fig3} reveals a consistent ordering:
\begin{itemize}
\item[] (i) TVP-SVAR dominates VAR by capturing evolving macro-financial regimes;
\item[] (ii) Copula-enhanced volatility models refine this further by admitting tail co-movement and contagion;
\item[] (iii) GPR achieves the highest accuracy, yet its margin over Copula-GARCH is often statistically narrow, raising the methodological question of whether flexibility truly outperforms structure, or simply approximates it.
\end{itemize}

\noindent This tension between predictive power and interpretability forms the basis of Section \ref{sec4}, where formal statistical testing \citep{diebold2002comparing, giacomini2006tests} is used to evaluate whether the gains of machine learning are significant or economically negligible. Ultimately, the results of this section illustrate a key insight of this paper: Forecast accuracy and structural interpretability are not mutually exclusive. By integrating copula dependence within a TVP-SVAR backbone, econometric models can approach the performance frontier of machine learning without sacrificing causal interpretability, thereby offering a principled bridge between statistical learning and economic theory.

\section{Discussions}
\label{sec4}
A central question in contemporary forecasting research is whether the adaptive flexibility of machine learning genuinely surpasses the explanatory rigour of econometric structures. To address this, we contrast the predictive performance of copula-augmented TVP-SVAR models with Gaussian Process Regression (GPR), not merely in terms of mean accuracy but through the full empirical distribution of forecast errors. This comparison, summarized in Figure \ref{fig4}, provides a distributional lens that complements the point estimates reported in Section \ref{sec3.8}.\\

\noindent The box plots reveal a remarkable convergence: both modeling classes produce highly comparable median RMSE values and overlapping interquartile ranges. Copula-based models display marginally heavier upper tails, reflecting occasional overreaction to volatility bursts, yet this dispersion remains within non-significant bounds. To formalize this observation, Table \ref{tab:ttest} reports a two-sample t-test on mean RMSE values. With a \textit{p-value} of 0.8444, the null hypothesis of equal predictive performance between Copula-based and GPR-based models cannot be rejected at the 5\% level. Statistically, there is no discernible superiority of machine learning over the structurally informed econometric framework.\\

\noindent
\makebox[\textwidth][c]{% Centrage global
    % -------- Figure --------
    \begin{minipage}[b]{0.48\textwidth}
        \centering
        \includegraphics[width=1.03\linewidth]{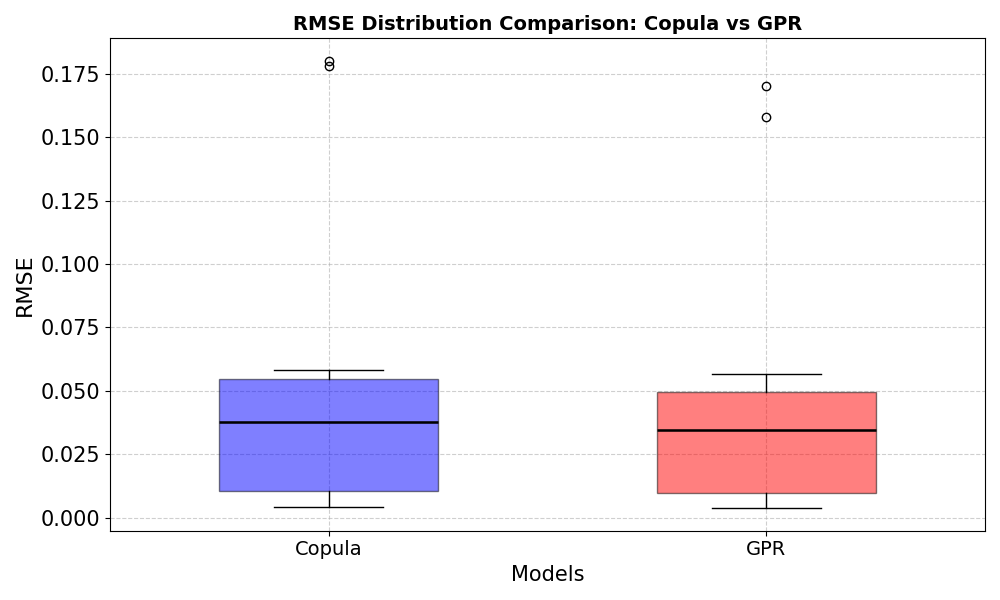}
        \captionof{figure}{Distributional Comparison of Forecast Errors: Copula-Based vs GPR Models}
        \label{fig4}
    \end{minipage}
    \hfill
    % -------- Tableau --------
\begin{minipage}[b]{0.48\textwidth}
 \centering
\resizebox{7.2cm}{!}{
\begin{tabular}{|ll|}
\hline
\textbf{Statistic}                 & \textbf{Value} \\
\hline
Null Hypothesis ($H_{0}$)         & $\mu_{\text{\tiny{Copula}}} = \mu_{\text{\tiny{GPR}}}$ \\
\textit{t}-value                  & 0.1983 \\
Degrees of Freedom               & 25.809 \\
\textit{p}-value                 & 0.8444 \\
Alternative Hypothesis           & $\mu_{\text{\tiny{Copula}}} \neq \mu_{\text{\tiny{GPR}}}$ \\
Mean (Copula Model)             & 0.04948 \\
Mean (GPR Model)                & 0.04528 \\
\textbf{Decision (5\% level)}    & \textbf{Fail to Reject $H_{0}$} \\
\hline
\end{tabular}
}
\captionof{table}{Two-Sample \textit{t}-Test: Copula-Based Models vs. GPR-Based Models}
%\caption{Welch Two-Sample \textit{t}-Test: Copula-Based Models vs. GPR-Based Models}
\label{tab:ttest}
\end{minipage}
}
\noindent This finding has important methodological consequences. Despite its parametric constraints, the TVP-SVAR–Copula model replicates the predictive accuracy of GPR, a fully nonparametric model. Such parity suggests that GPR may implicitly learn the same underlying economic regimes and dependence structures that econometric models explicitly represent through time-varying coefficients, structural shocks, and tail dependence. In this sense, while GPR learns patterns, TVP-SVAR explains mechanisms. 
\newpage
\noindent The implication is far-reaching: When a machine learning model matches a structural model in accuracy, it is likely because it has internalized, albeit opaquely, the same causal regularities.\\

\noindent The broader implication is articulated in Table \ref{tab:method_implications}, which frames the comparison along four methodological dimensions: statistical robustness, economic interpretability, predictive precision, and tail / regime sensitivity. Although both model classes score similarly in prediction, they diverge fundamentally in epistemological function. Copula augmented TVP-SVAR offers interpretable channels identifiable shocks, dynamic regimes, and tail contagion diagnostics whereas GPR operates as a black box, with limited capacity to decompose risk or attribute causality. Thus, predictive equivalence does not imply methodological equivalence.\\

\noindent Ultimately, these results advocate not for the dominance of either paradigm, but for their complementarity. Machine learning contributes adaptive generalisation, while econometric structures provide theoretical accountability. The fusion of these paradigms embodied in our hybrid TVP-SVAR–Copula approach achieves forecast parity with GPR while preserving interpretability. This confirms the empirical philosophy of this paper: the path forward in macro–energy forecasting lies not in replacing theory with algorithms, but in integrating structural insight into learning frameworks.

\begin{table}[H]
\centering
\resizebox{15.55cm}{!}{
\begin{tabular}{|l|ll|}
\toprule
\textbf{Dimension} & \textbf{TVP-SVAR / Copula-Based Models} & \textbf{GPR (Machine Learning Models)} \\ 
\midrule
Statistical Robustness   & High -- MLE \& Bootstrap Validation      & Moderate -- Data-Driven, No Constraints \\
Economic Interpretability & Strong -- IRFs \& Structural Shocks      & None -- Black-Box Mechanism             \\
Predictive Accuracy       & Comparable (t-Test Benchmark)            & Comparable (t-Test Benchmark)            \\
Tail \& Regime Sensitivity & Explicit -- Captures Extremes (Copulas) & Implicit -- Kernel-Based Learning        \\
\bottomrule
\end{tabular}
}
\caption{Methodological Comparison: TVP-SVAR/Copula Models vs. GPR Machine Learning}
\label{tab:method_implications}
\end{table}

\section{Conclusion}
\label{sec5}
This study set out to bridge a long-standing divide between two modelling traditions often seen as adversarial: structural econometrics, grounded in theory and causal interpretation, and machine learning, oriented toward flexibility and predictive performance. By focusing on the interaction between global energy prices and macro-financial dynamics, we demonstrated that these paradigms are not mutually exclusive; when each is extended to its methodological frontier, they converge both empirically and conceptually.\\

\noindent From an econometric perspective, the constant parameter VAR proved inadequate in environments characterized by structural shifts and evolving transmission mechanisms. The Time-Varying Parameter SVAR (TVP-SVAR) addressed these limitations by incorporating drifting coefficients and stochastic volatility, offering a more faithful representation of the macroeconomic conditions influencing Brent oil markets. When further enhanced with multivariate GARCH dynamics and copula-based dependence structures, the model successfully captured features that traditional correlation metrics systematically overlook, namely extreme comovements, asymmetric tail risks, and contagion during crises.\\

\noindent A central empirical result emerged from the comparative predictive analysis. When benchmarked against Gaussian Process Regression (GPR), a state-of-the-art machine learning method, the copula-enhanced TVP-SVAR framework achieved statistically indistinguishable forecast performance, as indicated by two-sample t-tests on RMSE distributions. This parity carries profound implications: it establishes that a rigorously specified, volatility-aware econometric model can rival advanced nonparametric algorithms in forecasting complex macro-financial series, without sacrificing interpretability.\\

\noindent However, predictive equivalence does not imply epistemological symmetry. Although GPR infers latent structures implicitly through kernel learning, it does not reveal the economic mechanisms that govern such dynamics. The TVP-SVAR–Copula framework, in contrast, makes these mechanisms explicit through impulse responses, structural shock decompositions, and tail-dependence diagnostics. 
\newpage
\noindent Thus, it provides not only a forecast, but a coherent account of \textit{why} and \textit{through which channels} forecasts materialize an indispensable quality in contexts involving policy design, systemic risk assessment, and market regulation.\\

\noindent The broader implication is that econometrics and machine learning should be viewed as complementary rather than competing paradigms. Machine learning validates, in a data-driven sense, the relevance of structures posited by theory; econometrics interprets and organizes the latent representations uncovered by learning algorithms. The most powerful inference, therefore, lies at their intersection. In this respect, the objective is not to replace structural models with black-box predictors, but to allow structural models to serve as interpreters of machine learning outputs.\\

\noindent Several avenues arise for future research. Formal forecast comparison procedures—such as \citet*{diebold2002comparing}  or \citet*{giacomini2006tests} tests should be applied to assess the statistical significance between regimes and horizons. The integration of Bayesian neural networks or transformer-based sequence models may offer hybrid architectures that retain structural clarity while expanding functional flexibility. Moreover, given the geopolitical centrality of energy markets, further work should investigate asymmetries under policy uncertainty, exploring not only \textit{how} prices adjust, but \textit{why} they react differently to monetary tightening, demand collapses, or supply shocks.\\

\noindent In conclusion, this study demonstrates that econometrics need not compete with machine learning; it can guide it. If GPR succeeds in prediction, it is because it learns, albeit implicitly, the same regime dynamics that the TVP-SVAR articulates explicitly. Our central proposition may therefore be stated succinctly:
\begin{itemize}
\item Machine learning predicts; structural econometrics explains.
\item Forecast parity confirms, rather than undermines, the value of economic structure.
\end{itemize}
Rather than choosing between interpretability and performance, this work illustrates the feasibility and necessity of models that offer both.

\bibliography{name}

\setcounter{figure}{0}
\renewcommand{\thefigure}{\Alph{section}\arabic{figure}}

\setcounter{table}{0}
\renewcommand{\thetable}{\Alph{section}\arabic{table}}

\newpage

\begin{appendix}
\section{Appendix 1}

\begin{table}[H]
\centering
\resizebox{18.2cm}{!}{
\begin{tabular}{|l|ccccccc|}
\hline
\textbf{Model} & \textbf{DGS10} & \textbf{UMCSENT} & \textbf{GEPUCURRENT} & \textbf{INDPRO\_US} & \textbf{USD/EUR} & \textbf{APE} & \textbf{BRENT} \\
\hline
VAR                      & 0.0501 & 0.0710 & 0.1914 & 0.0056 & 0.0151 & 0.0132 & 0.0651 \\
VAR-DCC                  & 0.0467 & 0.0700 & 0.1936 & 0.0051 & 0.0141 & 0.0123 & 0.0558 \\
VAR-ADCC                 & 0.0436 & 0.0689 & 0.1926 & 0.0051 & 0.0129 & 0.0120 & 0.0544 \\
VAR-DCC-t-copula         & 0.0402 & 0.0624 & 0.1884 & 0.0048 & 0.0123 & 0.0117 & 0.0506 \\
VAR-GARCH-Clayton        & 0.0382 & 0.0583 & 0.1803 & 0.0042 & 0.0113 & 0.0102 & 0.0471 \\
VAR-ANN                  & 0.0390 & 0.0591 & 0.1790 & 0.0045 & 0.0112 & 0.0101 & 0.0460 \\
VAR-GPR                  & 0.0358 & 0.0567 & 0.1702 & 0.0041 & 0.0101 & 0.0095 & 0.0408 \\
VAR-RF                   & 0.0386 & 0.0596 & 0.1788 & 0.0044 & 0.0109 & 0.0096 & 0.0410 \\
VAR-SVM                  & 0.0390 & 0.0578 & 0.1772 & 0.0042 & 0.0104 & 0.0097 & 0.0429 \\
\midrule
TVP-SVAR                 & 0.0456 & 0.0691 & 0.1896 & 0.0059 & 0.0139 & 0.0120 & 0.0502 \\
TVP-SVAR-DCC             & 0.0438 & 0.0673 & 0.1878 & 0.0058 & 0.0132 & 0.0114 & 0.0500 \\
TVP-SVAR-ADCC            & 0.0419 & 0.0604 & 0.1808 & 0.0051 & 0.0113 & 0.0106 & 0.0482 \\
TVP-SVAR-DCC-t-copula    & 0.0434 & 0.0580 & 0.1835 & 0.0057 & 0.0108 & 0.0107 & 0.0476 \\
TVP-SVAR-GARCH-MixCopula & 0.0372 & 0.0572 & 0.1781 & 0.0043 & 0.0104 & 0.0096 & 0.0465 \\
TVP-SVAR-ANN             & 0.0368 & 0.0580 & 0.1800 & 0.0045 & 0.0109 & 0.0098 & 0.0447 \\
TVP-SVAR-GPR             & 0.0337 & 0.0525 & 0.1580 & 0.0036 & 0.0099 & 0.0089 & 0.0403 \\
TVP-SVAR-RF              & 0.0355 & 0.0588 & 0.1788 & 0.0043 & 0.0100 & 0.0095 & 0.0434 \\
TVP-SVAR-SVM             & 0.0360 & 0.0597 & 0.1792 & 0.0045 & 0.0099 & 0.0089 & 0.0428 \\
\hline
\end{tabular}
}
\caption{Root Mean Squared Errors (RMSE) across Models and Variables}
\label{tabA1}
\begin{tablenotes}
\small
\item \textit{Notes:} \textit{The table reports root mean squared errors (RMSE) for each model–variable pair.\\ 
Lower values indicate superior predictive performance. Machine learning extensions and time-varying \\ specifications consistently outperform static VAR benchmarks, especially for energy-related variables.}
\end{tablenotes}
\end{table}

%--------------------%
%     Figure A1      %
%--------------------%
\begin{sidewaysfigure}
%\begin{figure}
\centering
\begin{subfigure}[b]{0.3\textwidth}
    \centering
    \includegraphics[width=0.7\hsize]{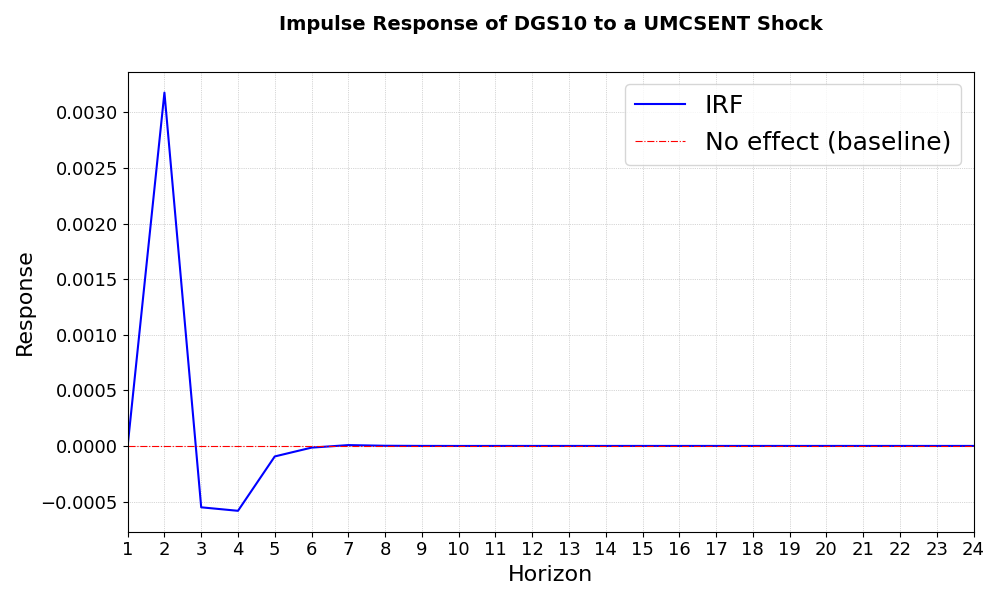}
    \caption{UMCSENT Shock on DGS10}
    \label{figA1_m1}
\end{subfigure}
\hfil
\begin{subfigure}[b]{0.3\textwidth}
    \centering
    \includegraphics[width=0.7\hsize]{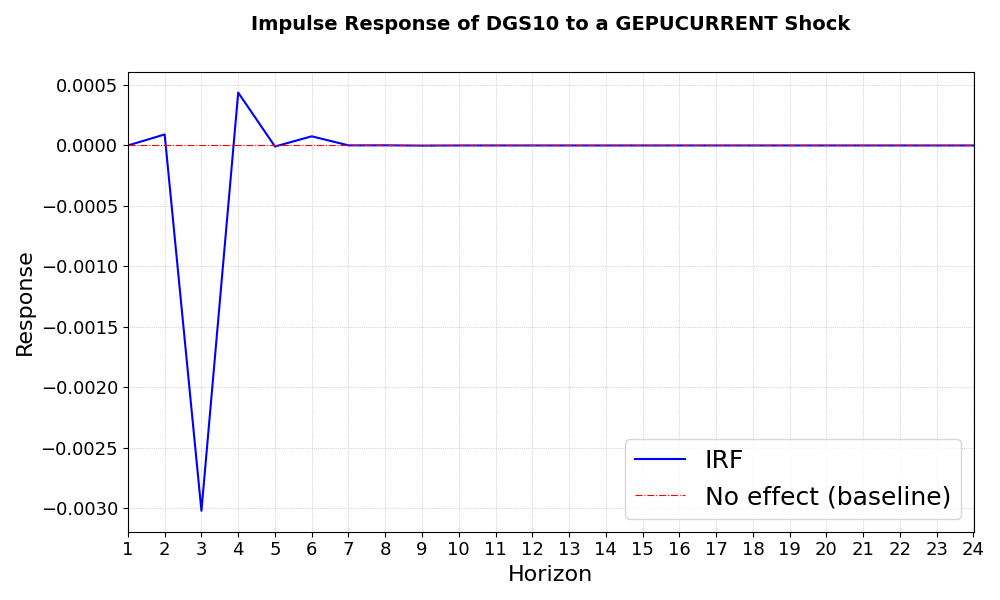}
    \caption{GEPUCURRENT Shock on DGS10}
    \label{figA1_m2}
\end{subfigure}
\hfil
\begin{subfigure}[b]{0.3\textwidth}
    \centering
    \includegraphics[width=0.7\hsize]{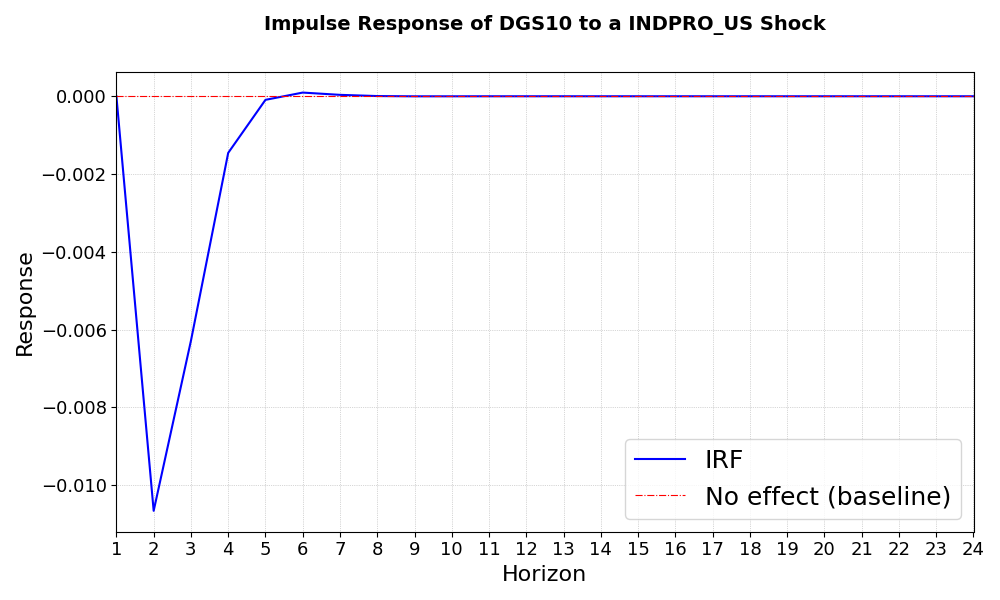}
    \caption{INDPRO$\_$US Shock on DGS10}
    \label{figA1_m3}
\end{subfigure}
\hfil
\begin{subfigure}[b]{0.3\textwidth}
    \centering
    \includegraphics[width=0.7\hsize]{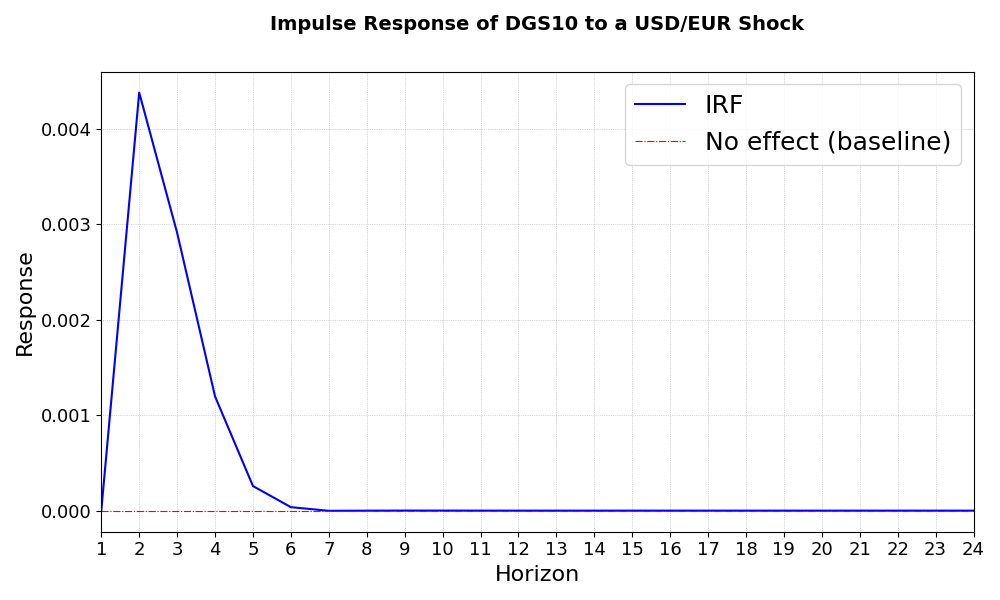}
    \caption{USD/EUR Shock on DGS10}
    \label{figA1_m4}
\end{subfigure}
\hfil
\begin{subfigure}[b]{0.3\textwidth}
    \centering
    \includegraphics[width=0.7\hsize]{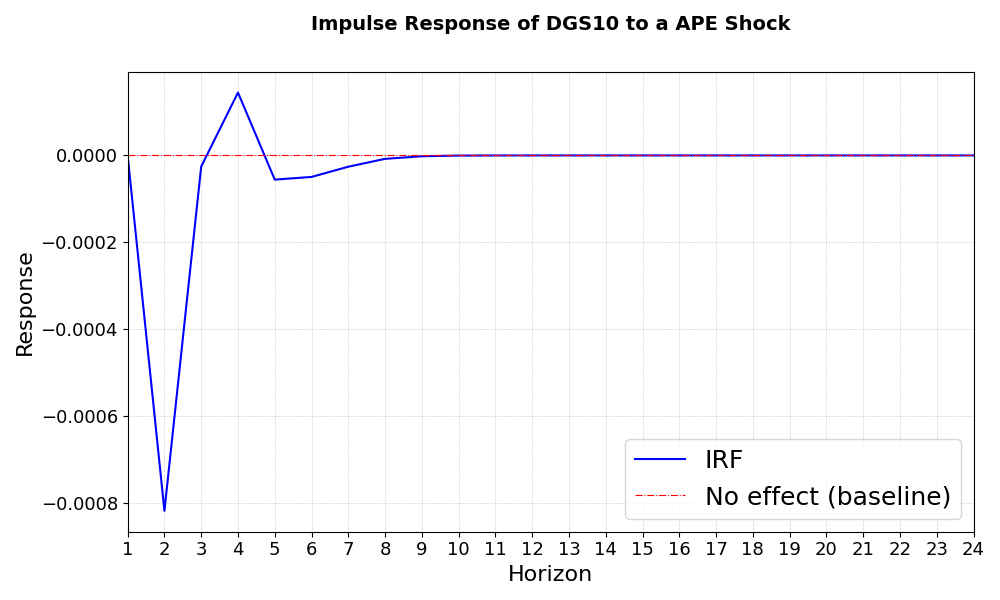}
    \caption{APE Shock on DGS10}
    \label{figA1_m5}
\end{subfigure}
\hfil
\begin{subfigure}[b]{0.3\textwidth}
    \centering
    \includegraphics[width=0.7\hsize]{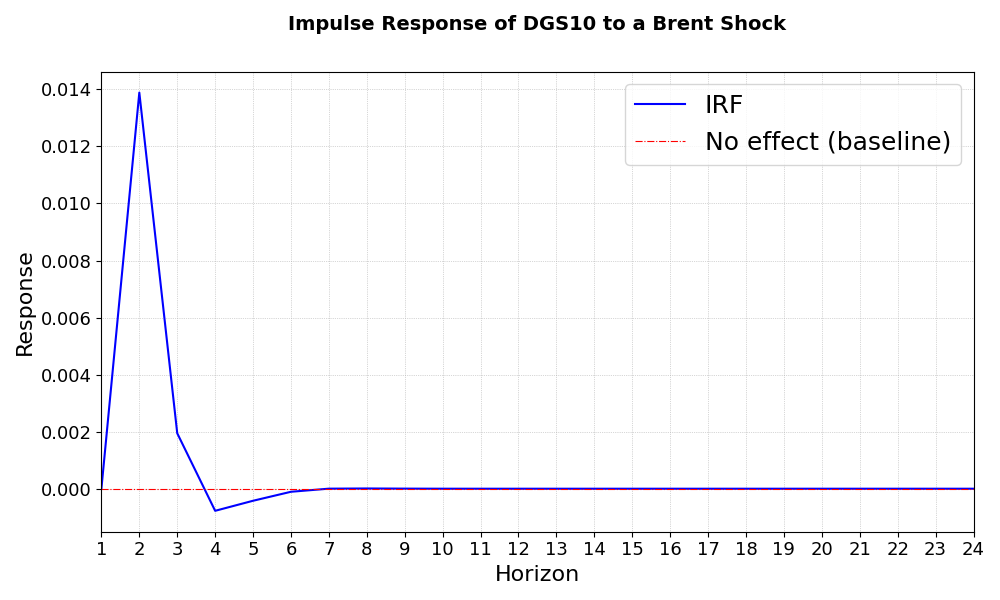}
    \caption{BRENT Shock on DGS10}
    \label{figA1_m6}
\end{subfigure}
%-----------------------------------%
\vspace{1cm}
%-----------------------------------%
\begin{subfigure}[b]{0.3\textwidth}
    \centering
    \includegraphics[width=0.7\hsize]{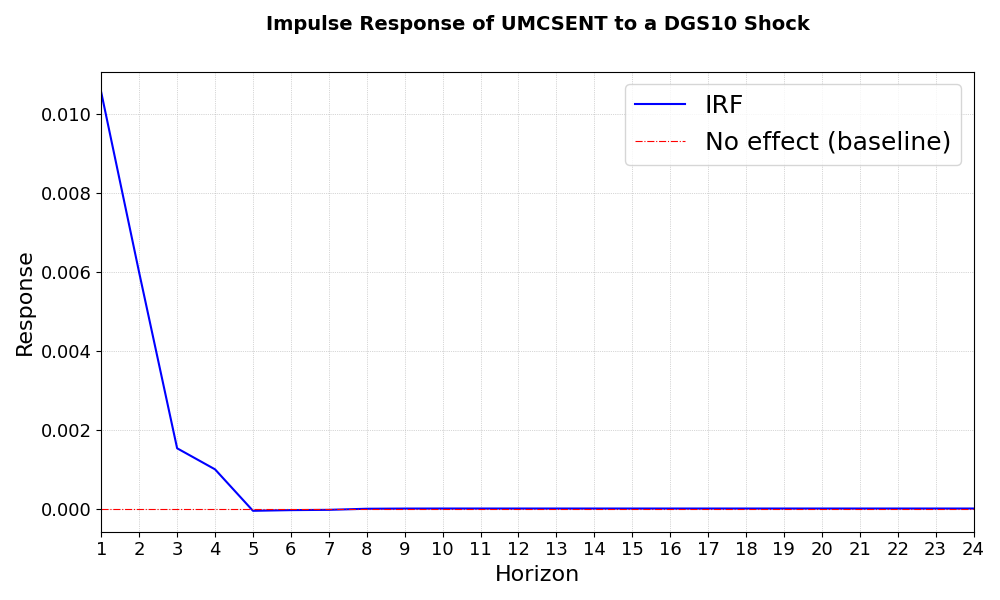}
    \caption{DGS10 Shock on UMCSENT}
    \label{figA1_m7}
\end{subfigure}
\hfil
\begin{subfigure}[b]{0.3\textwidth}
    \centering
    \includegraphics[width=0.7\hsize]{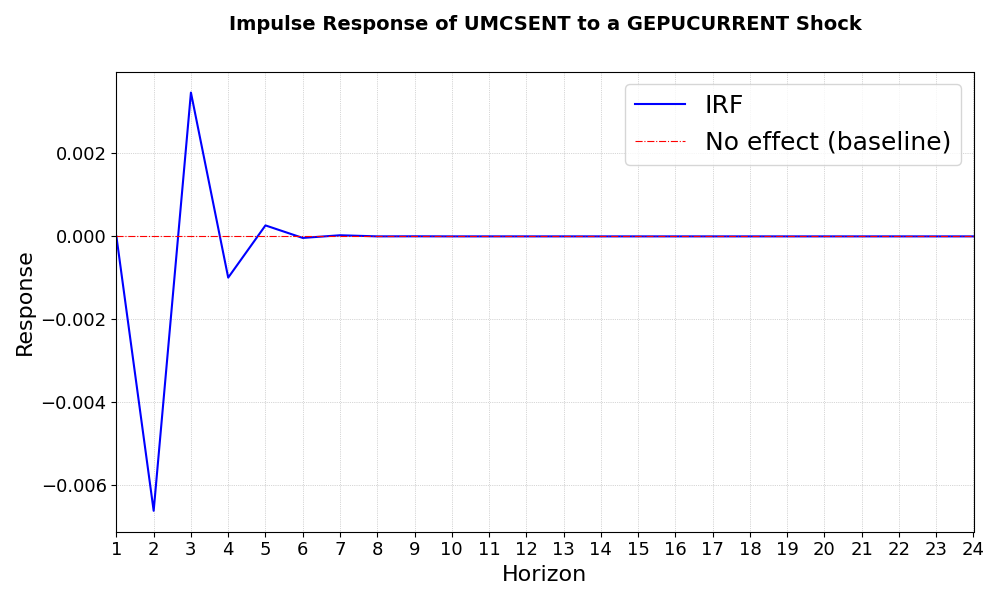}
    \caption{GEPUCURRENT Shock on UMCSENT}
    \label{figA1_m8}
\end{subfigure}
\hfil
\begin{subfigure}[b]{0.3\textwidth}
    \centering
    \includegraphics[width=0.7\hsize]{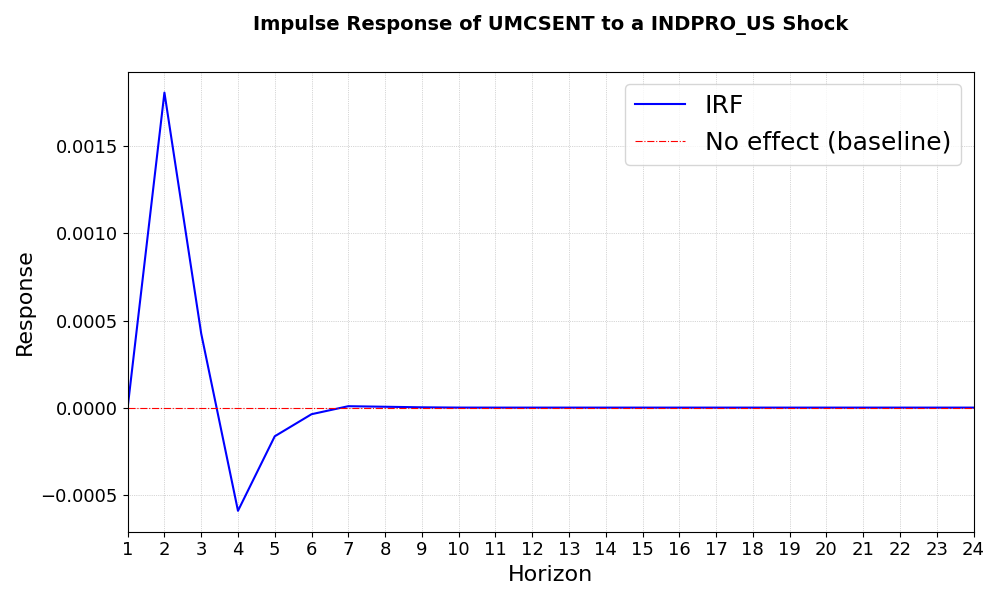}
    \caption{INDPRO$\_$US Shock on UMCSENT}
    \label{figA1_m9}
\end{subfigure}
\hfil
\begin{subfigure}[b]{0.3\textwidth}
    \centering
    \includegraphics[width=0.7\hsize]{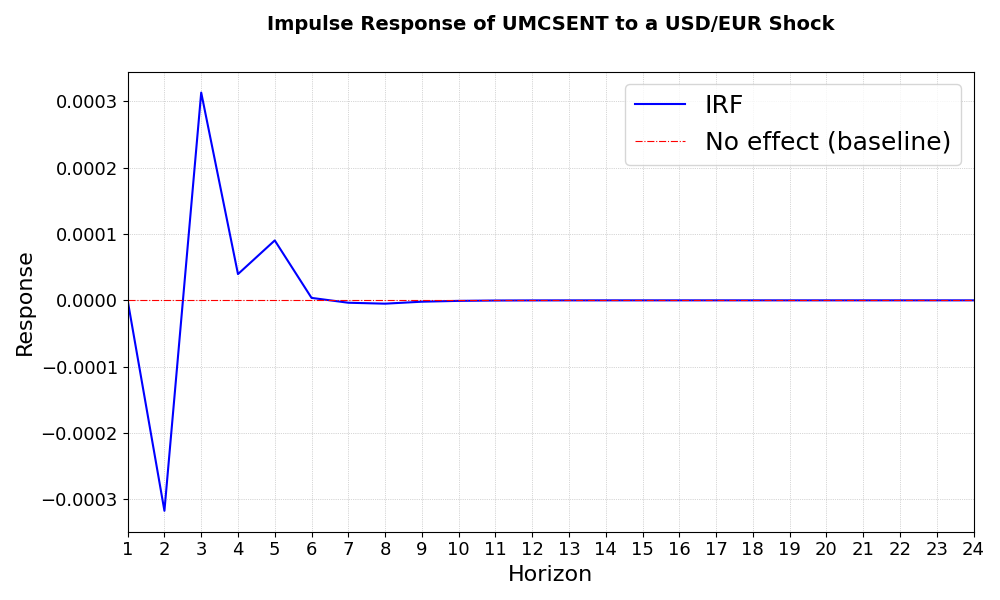}
    \caption{USD/EUR Shock on UMCSENT}
    \label{figA1_m10}
\end{subfigure}
\hfil
\begin{subfigure}[b]{0.3\textwidth}
    \centering
    \includegraphics[width=0.7\hsize]{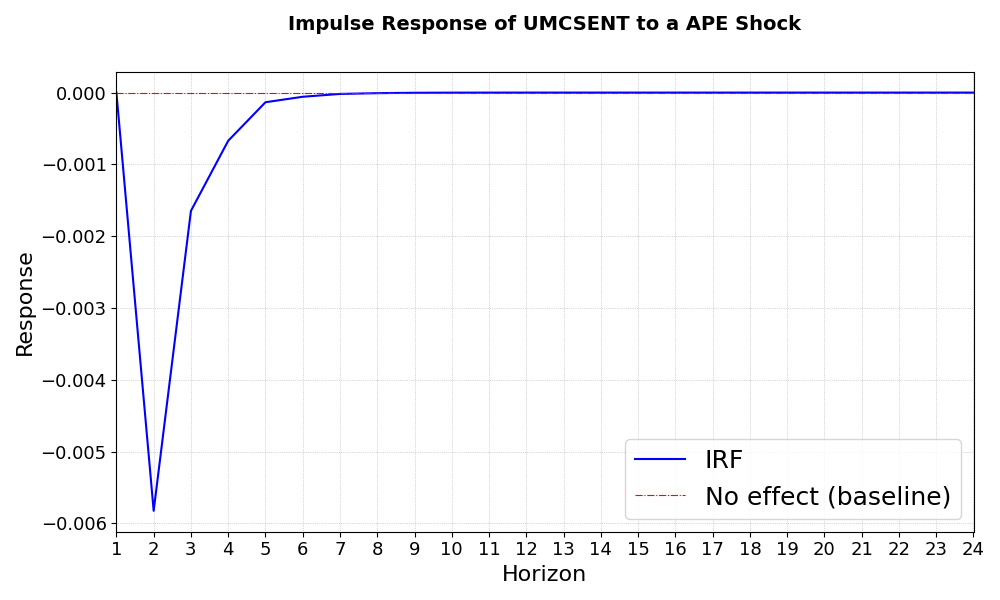}
    \caption{APE Shock on UMCSENT}
    \label{figA1_m11}
\end{subfigure}
\hfil
\begin{subfigure}[b]{0.3\textwidth}
    \centering
    \includegraphics[width=0.7\hsize]{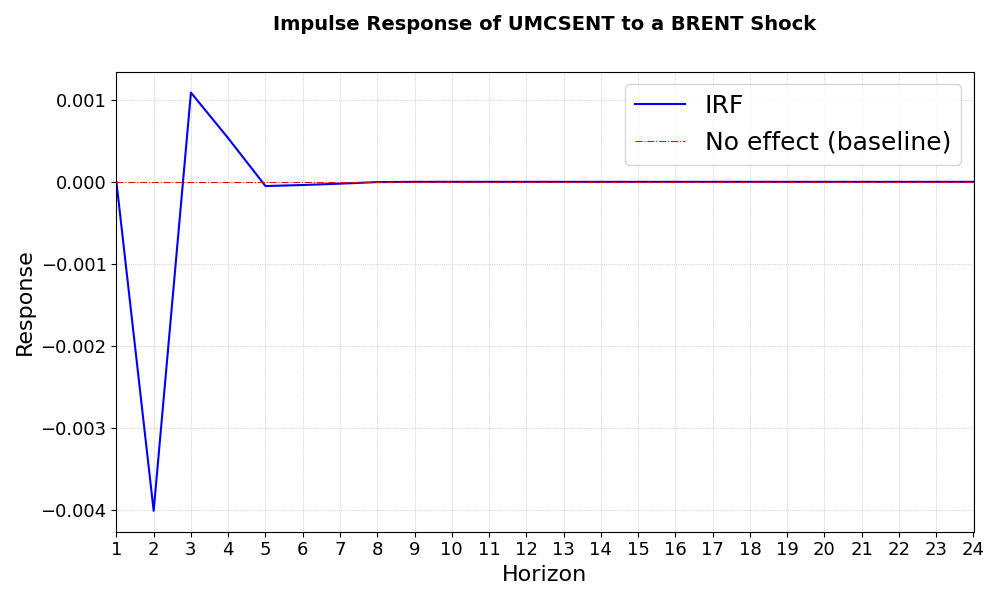}
    \caption{BRENT Shock on UMCSENT}
    \label{figA1_m12}
\end{subfigure}
%-----------------------------------%
\caption{Impulse Response for DGS10 and UMCSENT}
\label{figA1}
\end{sidewaysfigure}

%

%--------------------%
%     Figure A2      %
%--------------------%
\begin{sidewaysfigure}
%\begin{figure}
\centering
\begin{subfigure}[b]{0.3\textwidth}
    \centering
    \includegraphics[width=0.7\hsize]{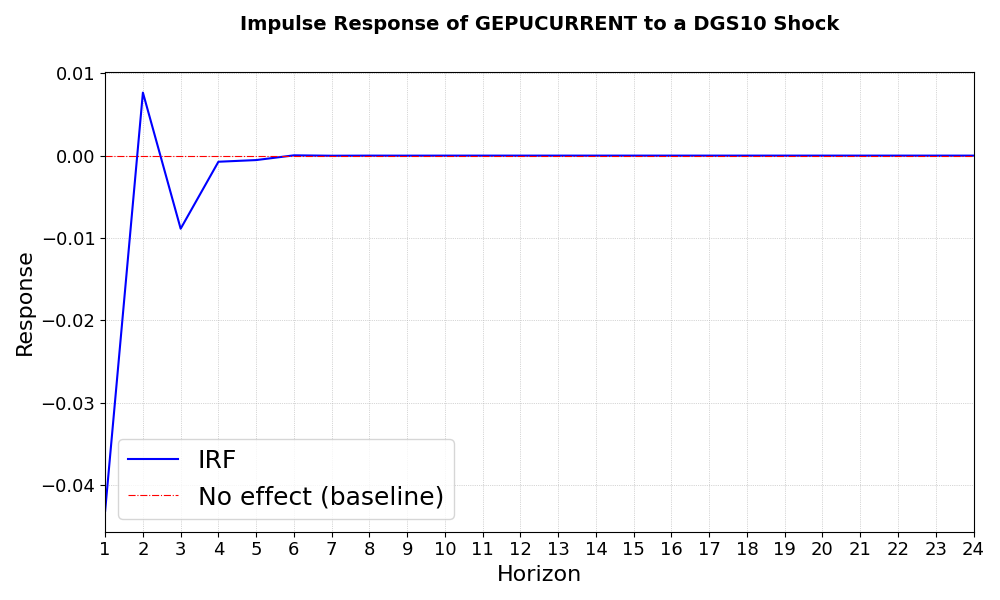}
    \caption{DGS10 Shock on GEPUCURRENT}
    \label{figA2_m1}
\end{subfigure}
\hfil
\begin{subfigure}[b]{0.3\textwidth}
    \centering
    \includegraphics[width=0.7\hsize]{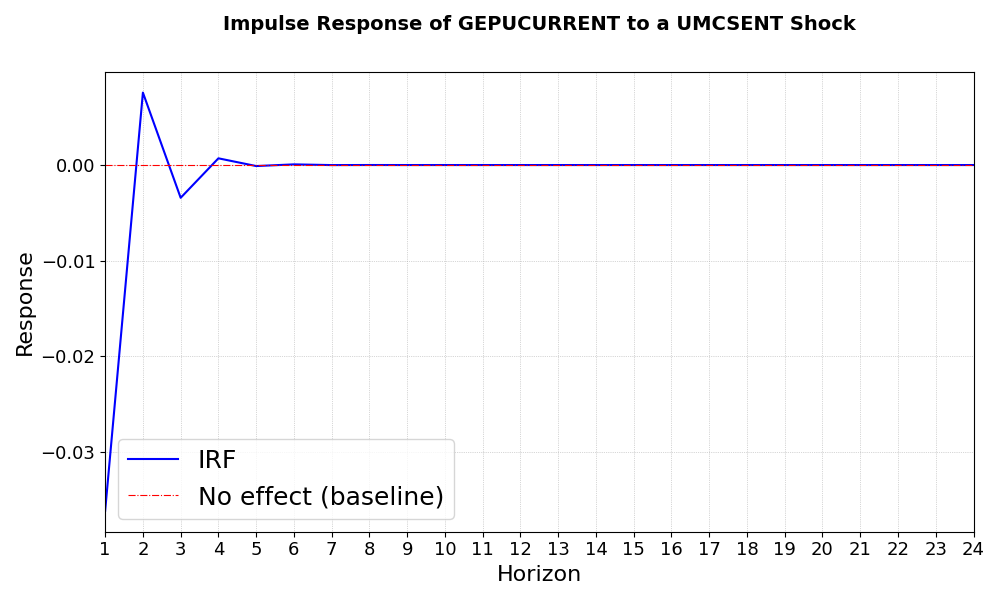}
    \caption{UMCSENT Shock on GEPUCURRENT}
    \label{figA2_m2}
\end{subfigure}
\hfil
\begin{subfigure}[b]{0.3\textwidth}
    \centering
    \includegraphics[width=0.7\hsize]{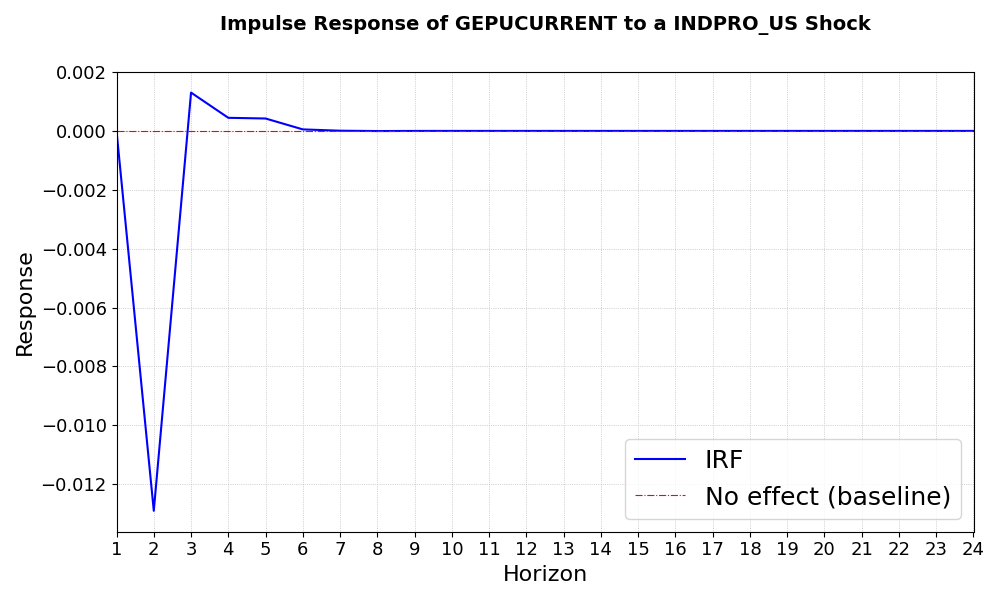}
    \caption{INDPRO$\_$US Shock on GEPUCURRENT}
    \label{figA2_m3}
\end{subfigure}
\hfil
\begin{subfigure}[b]{0.3\textwidth}
    \centering
    \includegraphics[width=0.7\hsize]{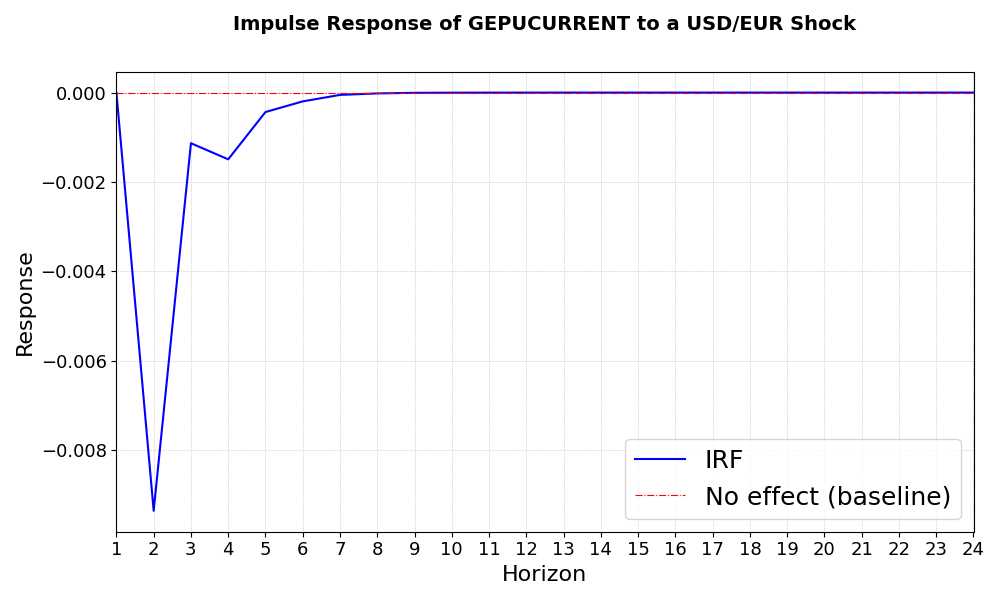}
    \caption{USD/EUR Shock on GEPUCURRENT}
    \label{figA2_m4}
\end{subfigure}
\hfil
\begin{subfigure}[b]{0.3\textwidth}
    \centering
    \includegraphics[width=0.7\hsize]{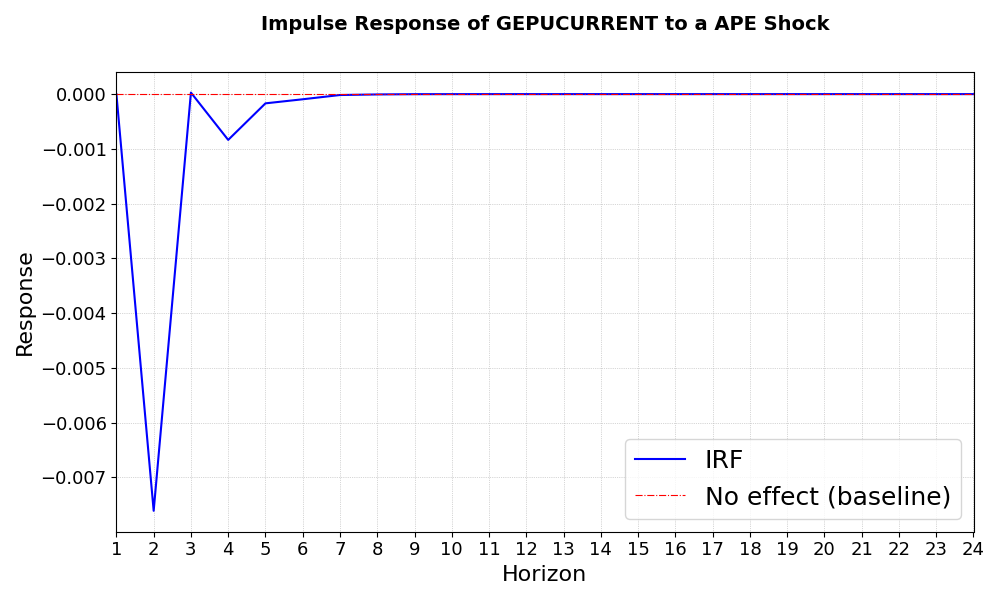}
    \caption{APE Shock on GEPUCURRENT}
    \label{figA2_m5}
\end{subfigure}
\hfil
\begin{subfigure}[b]{0.3\textwidth}
    \centering
    \includegraphics[width=0.7\hsize]{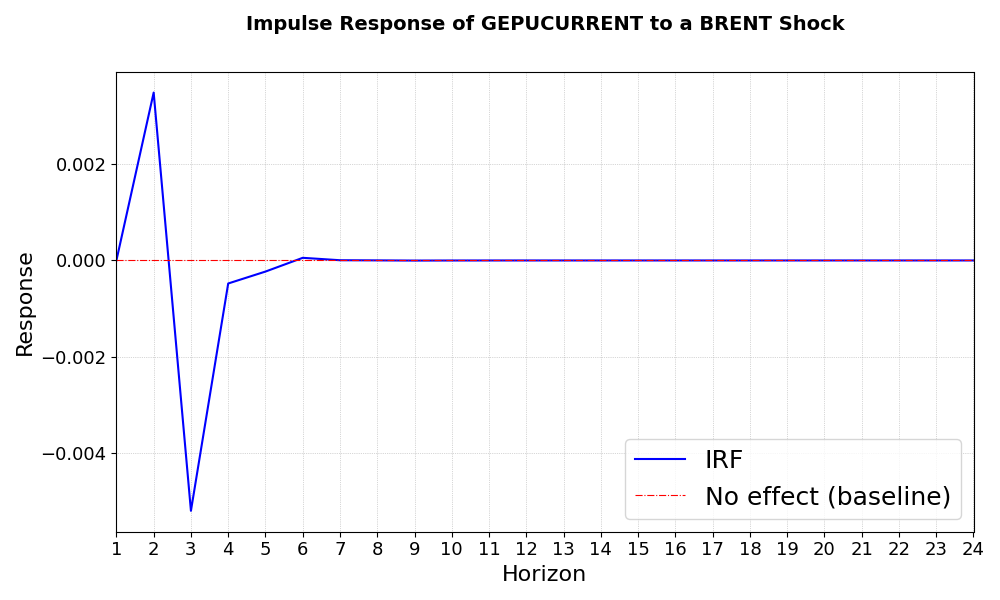}
    \caption{BRENT Shock on GEPUCURRENT}
    \label{figA2_m6}
\end{subfigure}
%-----------------------------------%
\vspace{1cm}
%-----------------------------------%
\begin{subfigure}[b]{0.3\textwidth}
    \centering
    \includegraphics[width=0.7\hsize]{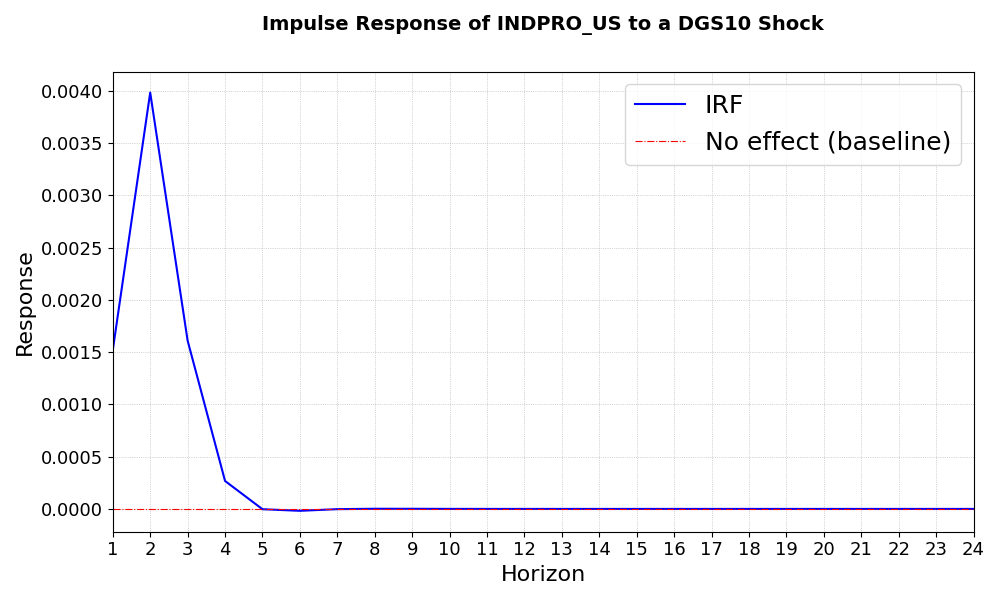}
    \caption{DGS10 Shock on INDPRO$\_$US}
    \label{figA2_m7}
\end{subfigure}
\hfil
\begin{subfigure}[b]{0.3\textwidth}
    \centering
    \includegraphics[width=0.7\hsize]{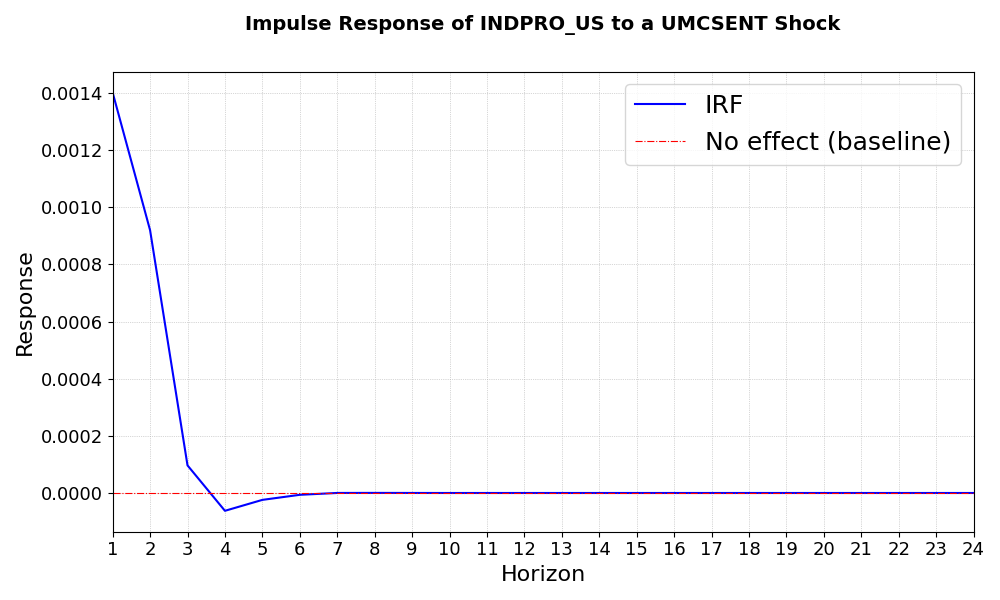}
    \caption{UMCSENT Shock on INDPRO$\_$US}
    \label{figA2_m8}
\end{subfigure}
\hfil
\begin{subfigure}[b]{0.3\textwidth}
    \centering
    \includegraphics[width=0.7\hsize]{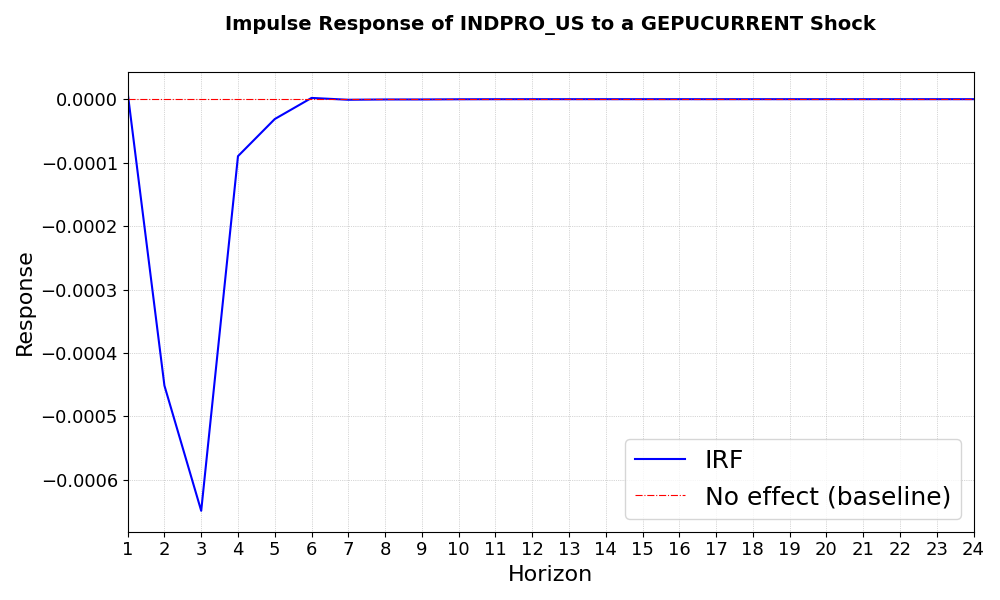}
    \caption{GEPUCURRENT Shock on INDPRO$\_$US}
    \label{figA2_m9}
\end{subfigure}
\hfil
\begin{subfigure}[b]{0.3\textwidth}
    \centering
    \includegraphics[width=0.7\hsize]{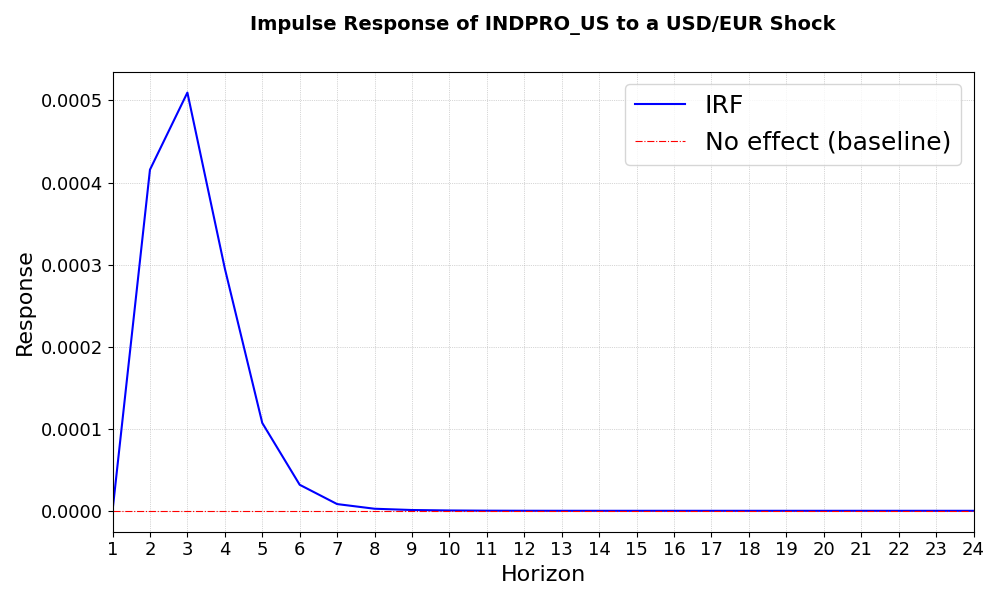}
    \caption{USD/EUR Shock on INDPRO$\_$US}
    \label{figA2_m10}
\end{subfigure}
\hfil
\begin{subfigure}[b]{0.3\textwidth}
    \centering
    \includegraphics[width=0.7\hsize]{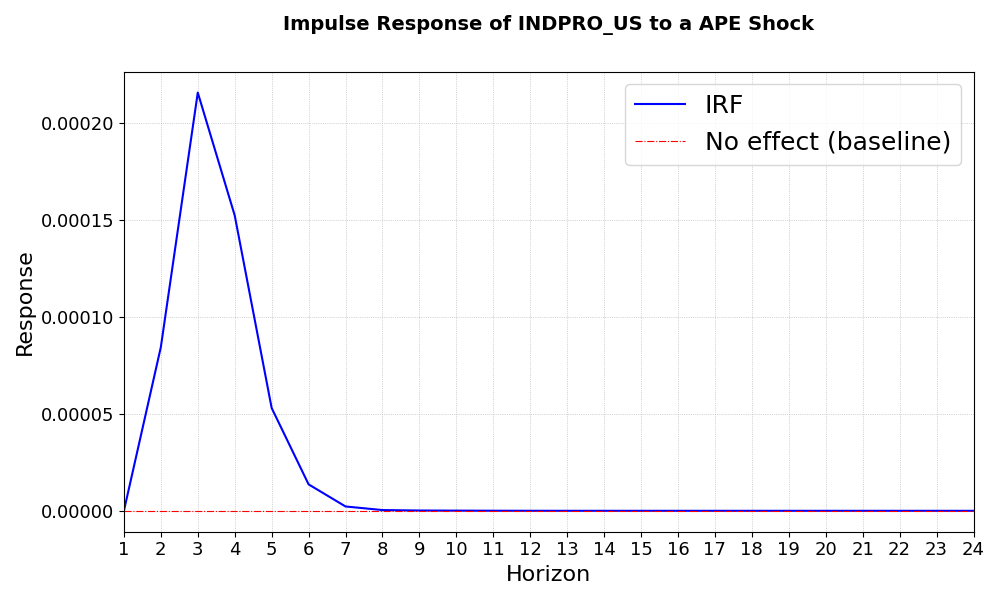}
    \caption{APE Shock on INDPRO$\_$US}
    \label{figA2_m11}
\end{subfigure}
\hfil
\begin{subfigure}[b]{0.3\textwidth}
    \centering
    \includegraphics[width=0.7\hsize]{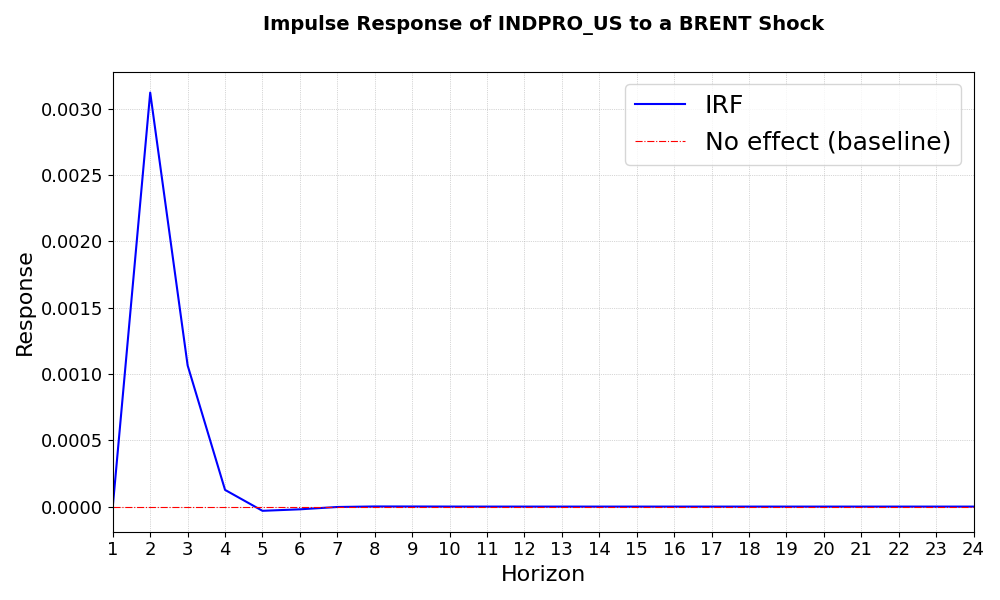}
    \caption{BRENT Shock on INDPRO$\_$US}
    \label{figA2_m12}
\end{subfigure}
%-----------------------------------%
\caption{Impulse Response for GEPUCURRENT and INDPRO$\_$US}
\label{figA2}
\end{sidewaysfigure}

%--------------------%
%     Figure A3      %
%--------------------%
\begin{sidewaysfigure}
%\begin{figure}
\centering
\begin{subfigure}[b]{0.3\textwidth}
    \centering
    \includegraphics[width=0.7\hsize]{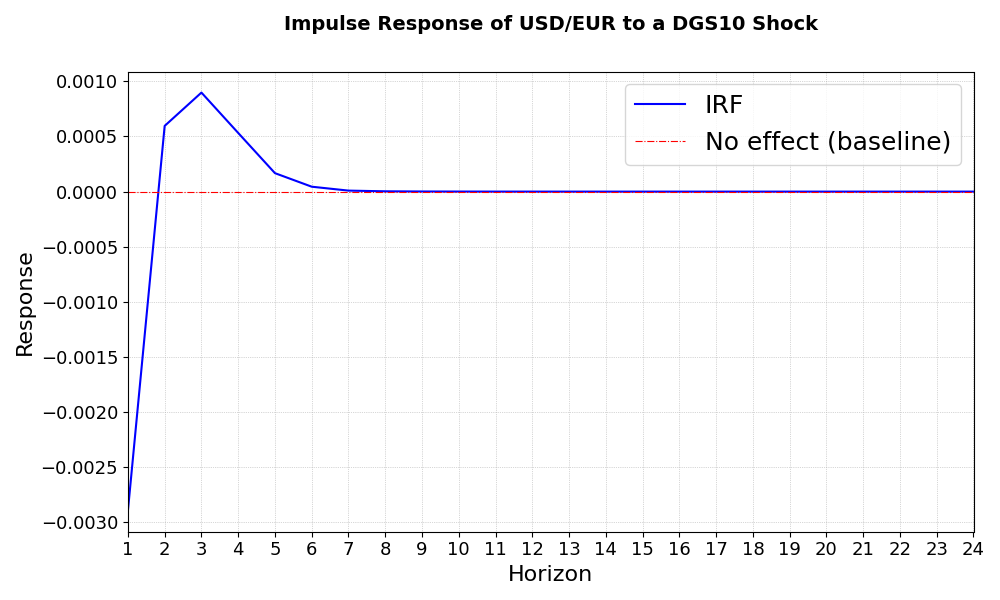}
    \caption{DGS10 Shock on USD/EUR}
    \label{figA3_m1}
\end{subfigure}
\hfil
\begin{subfigure}[b]{0.3\textwidth}
    \centering
    \includegraphics[width=0.7\hsize]{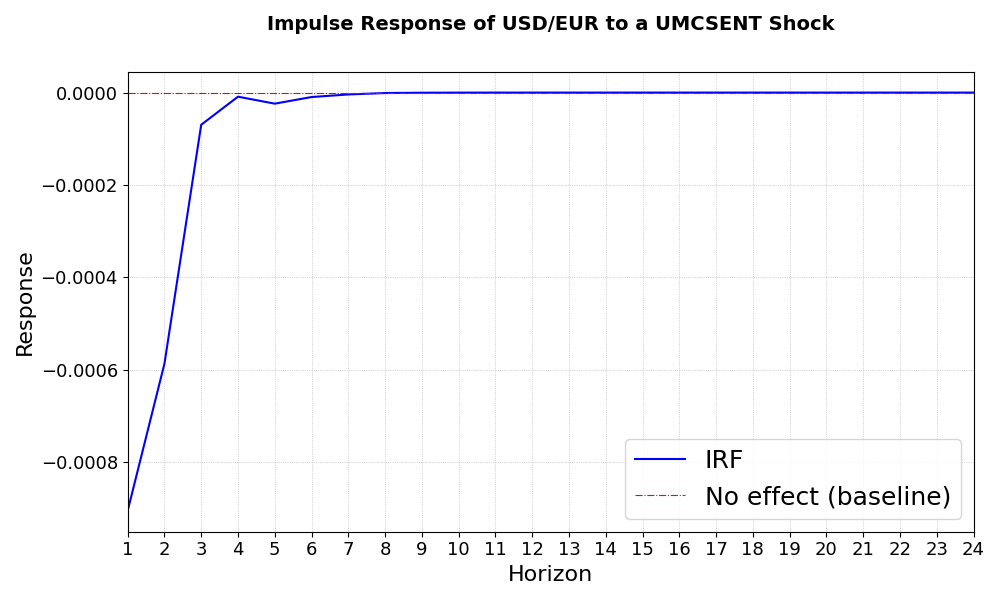}
    \caption{UMCSENT Shock on USD/EUR}
    \label{figA3_m2}
\end{subfigure}
\hfil
\begin{subfigure}[b]{0.3\textwidth}
    \centering
    \includegraphics[width=0.7\hsize]{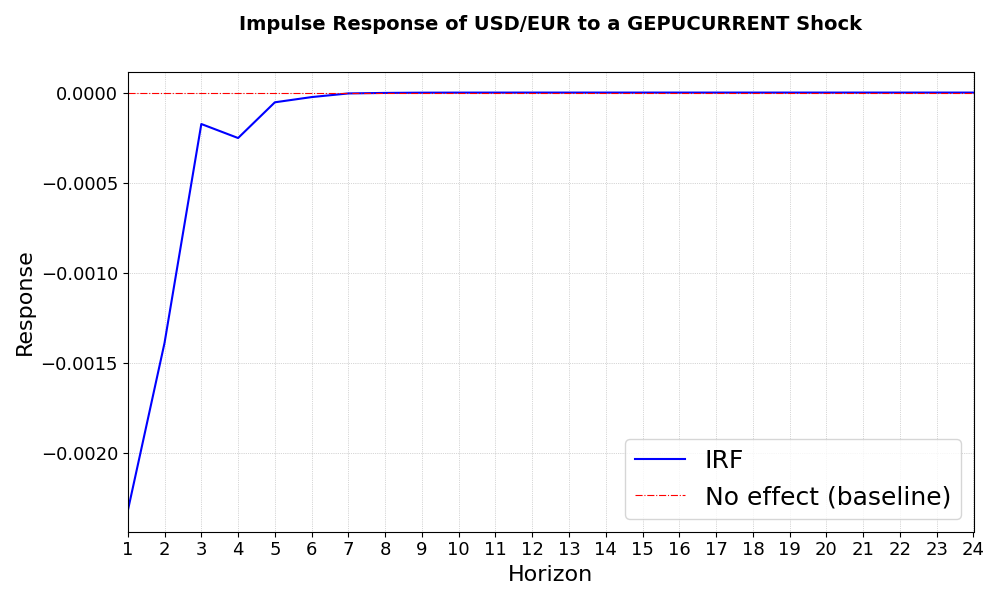}
    \caption{GEPUCURRENT Shock on USD/EUR}
    \label{figA3_m3}
\end{subfigure}
\hfil
\begin{subfigure}[b]{0.3\textwidth}
    \centering
    \includegraphics[width=0.7\hsize]{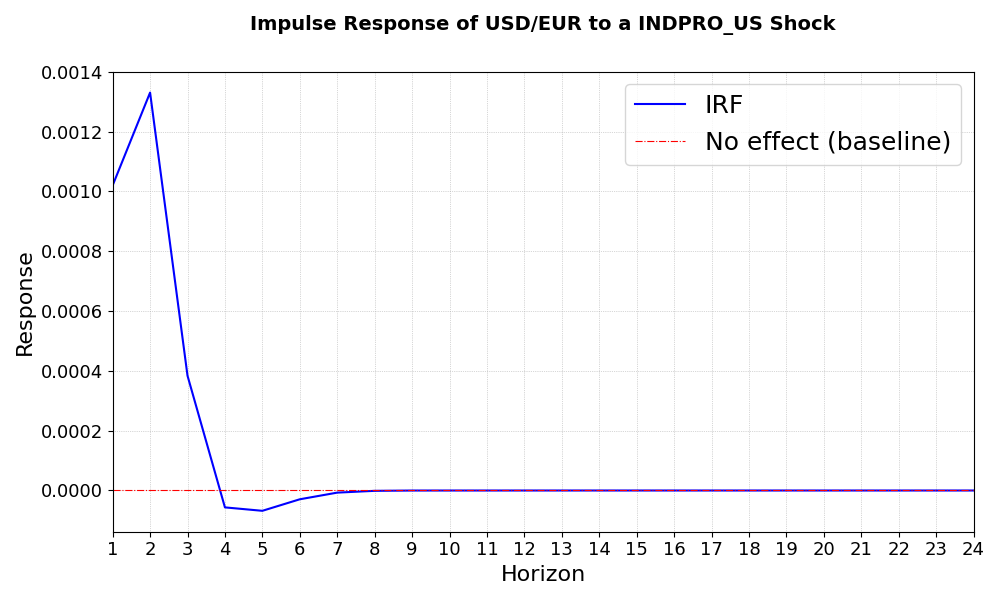}
    \caption{INDPRO$\_$US Shock on USD/EUR}
    \label{figA3_m4}
\end{subfigure}
\hfil
\begin{subfigure}[b]{0.3\textwidth}
    \centering
    \includegraphics[width=0.7\hsize]{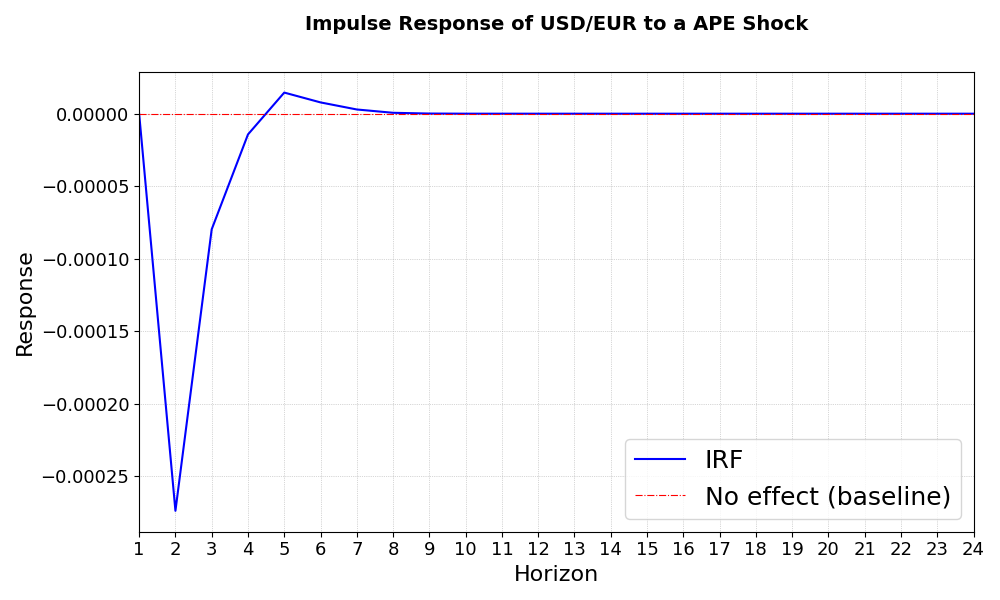}
    \caption{APE Shock on USD/EUR}
    \label{figA3_m5}
\end{subfigure}
\hfil
\begin{subfigure}[b]{0.3\textwidth}
    \centering
    \includegraphics[width=0.7\hsize]{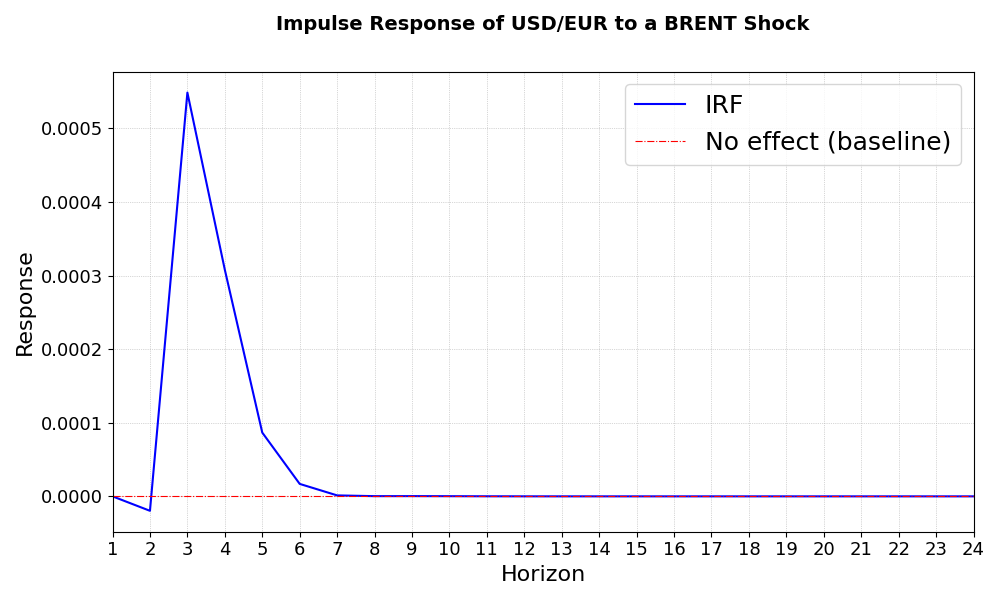}
    \caption{BRENT Shock on USD/EUR}
    \label{figA3_m6}
\end{subfigure}
%-----------------------------------%
\vspace{1cm}
%-----------------------------------%
\begin{subfigure}[b]{0.3\textwidth}
    \centering
    \includegraphics[width=0.7\hsize]{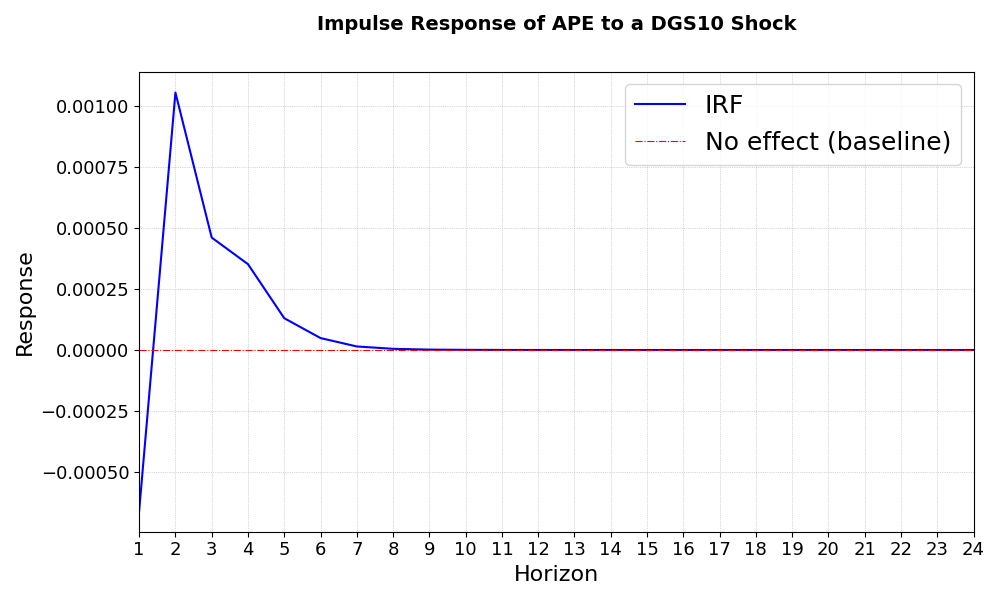}
    \caption{DGS10 Shock on APE}
    \label{figA3_m7}
\end{subfigure}
\hfil
\begin{subfigure}[b]{0.3\textwidth}
    \centering
    \includegraphics[width=0.7\hsize]{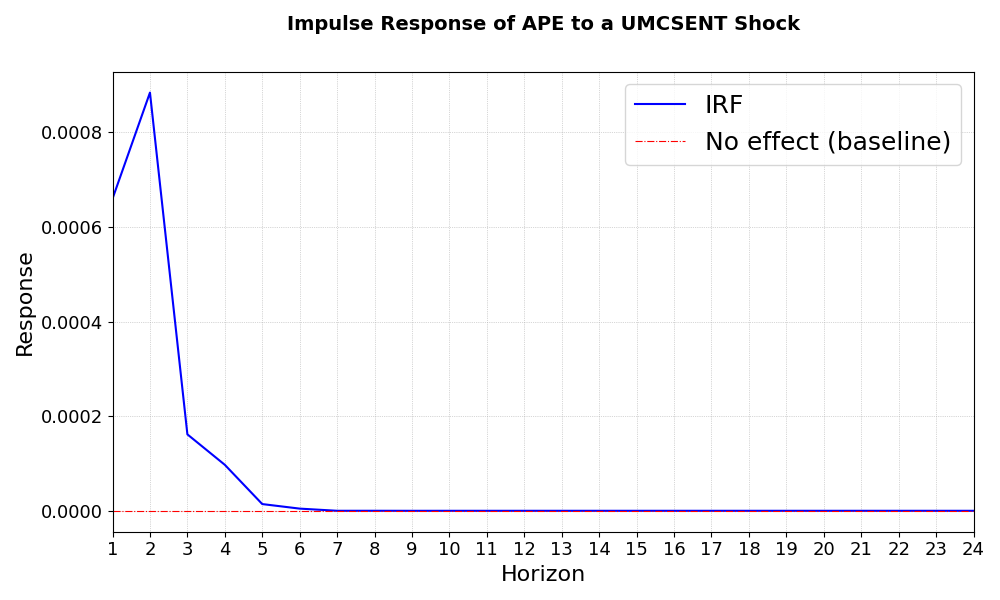}
    \caption{UMCSENT Shock on APE}
    \label{figA3_m8}
\end{subfigure}
\hfil
\begin{subfigure}[b]{0.3\textwidth}
    \centering
    \includegraphics[width=0.7\hsize]{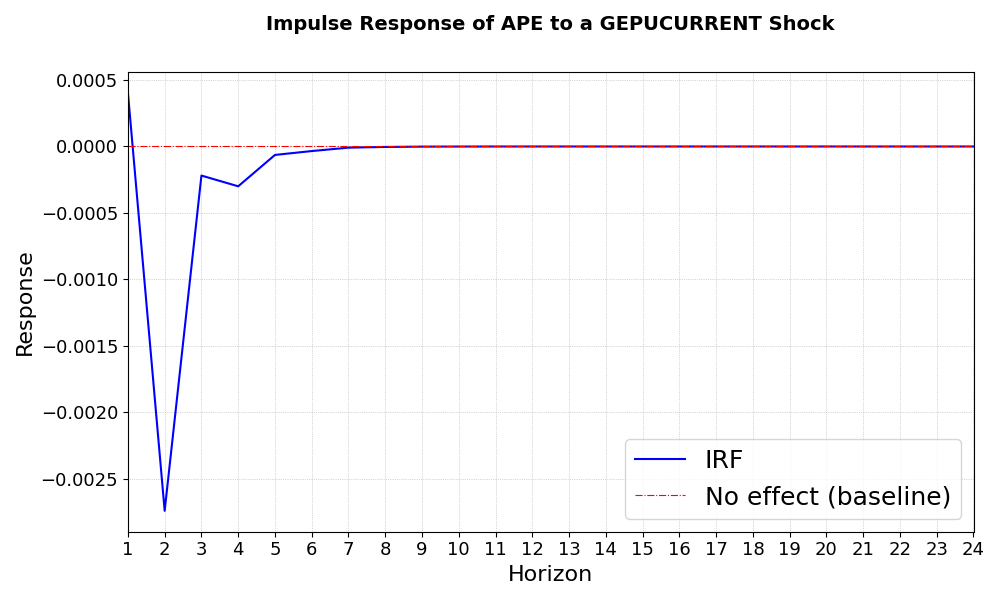}
    \caption{GEPUCURRENT Shock on APE}
    \label{figA3_m9}
\end{subfigure}
\hfil
\begin{subfigure}[b]{0.3\textwidth}
    \centering
    \includegraphics[width=0.7\hsize]{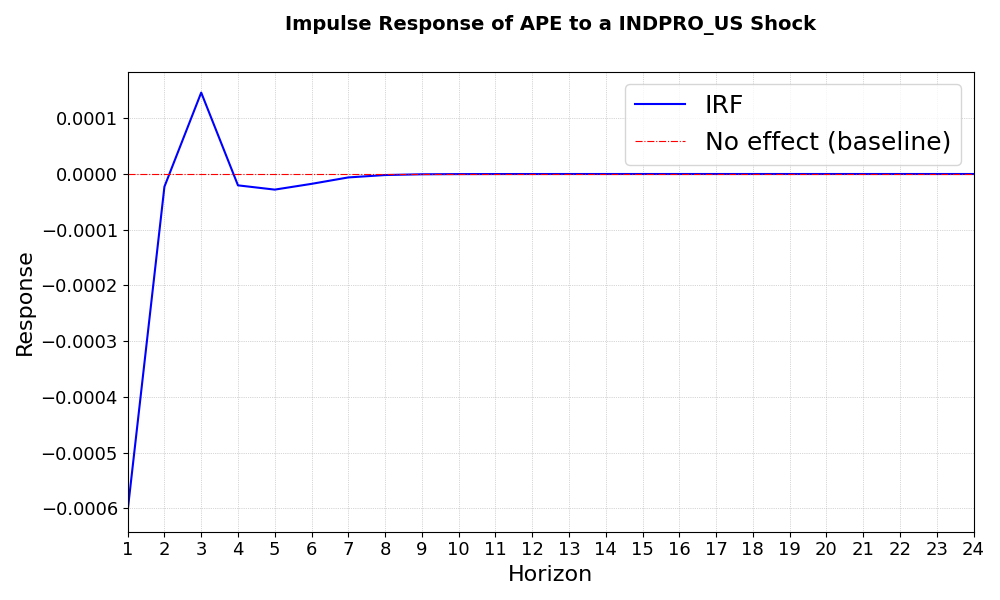}
    \caption{INDPRO$\_$US Shock on APE}
    \label{figA3_m10}
\end{subfigure}
\hfil
\begin{subfigure}[b]{0.3\textwidth}
    \centering
    \includegraphics[width=0.7\hsize]{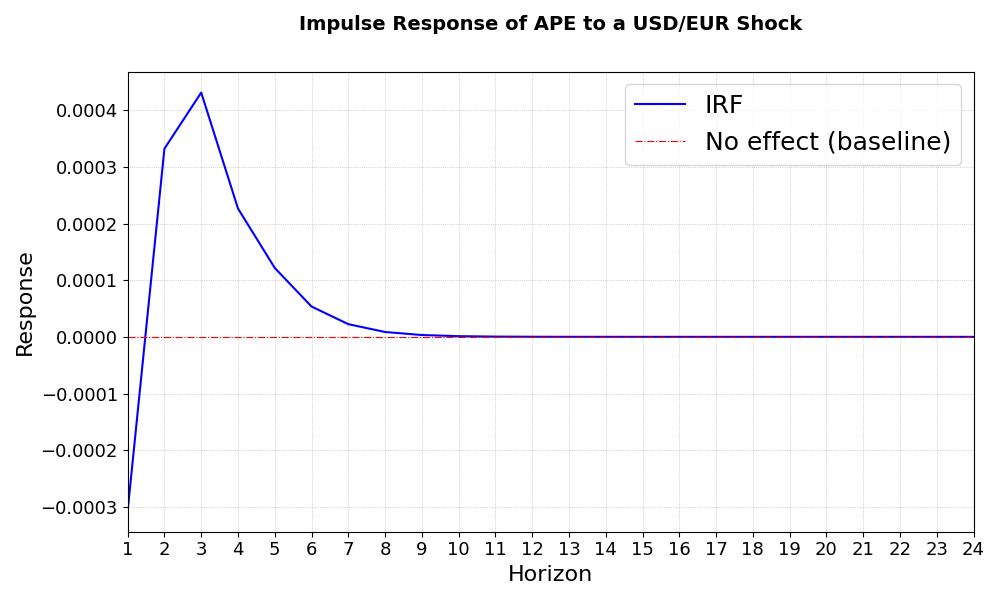}
    \caption{USD/EUR Shock on APE}
    \label{figA3_m11}
\end{subfigure}
\hfil
\begin{subfigure}[b]{0.3\textwidth}
    \centering
    \includegraphics[width=0.7\hsize]{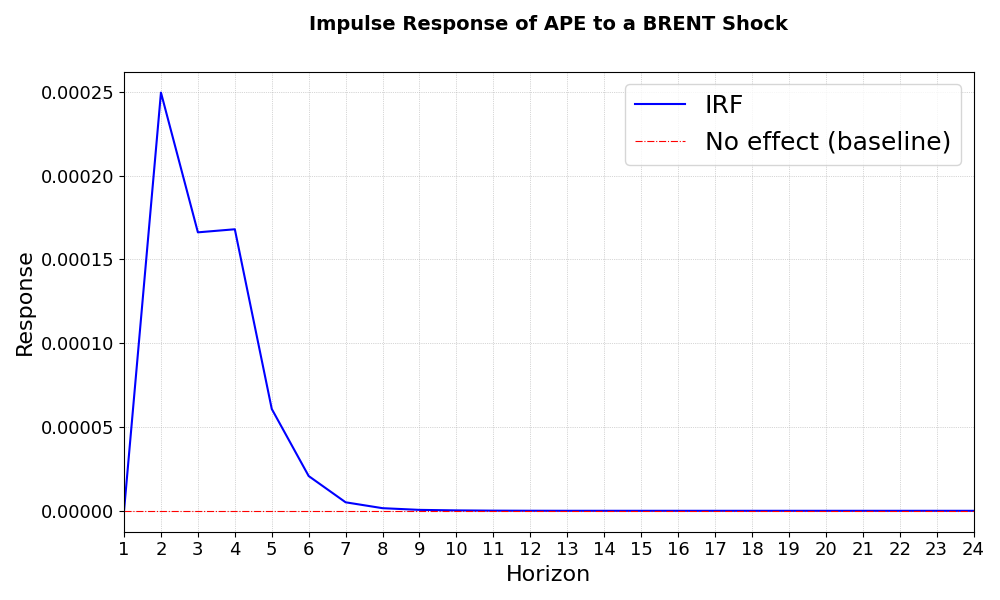}
    \caption{BRENT Shock on APE}
    \label{figA3_m12}
\end{subfigure}
%-----------------------------------%
\caption{Impulse Response for USD/EUR and APE}
\label{figA3}
\end{sidewaysfigure}

%--------------------%
%     Figure A4      %
%--------------------%
\begin{sidewaysfigure}
%\begin{figure}
\centering
\begin{subfigure}[b]{0.3\textwidth}
    \centering
    \includegraphics[width=0.7\hsize]{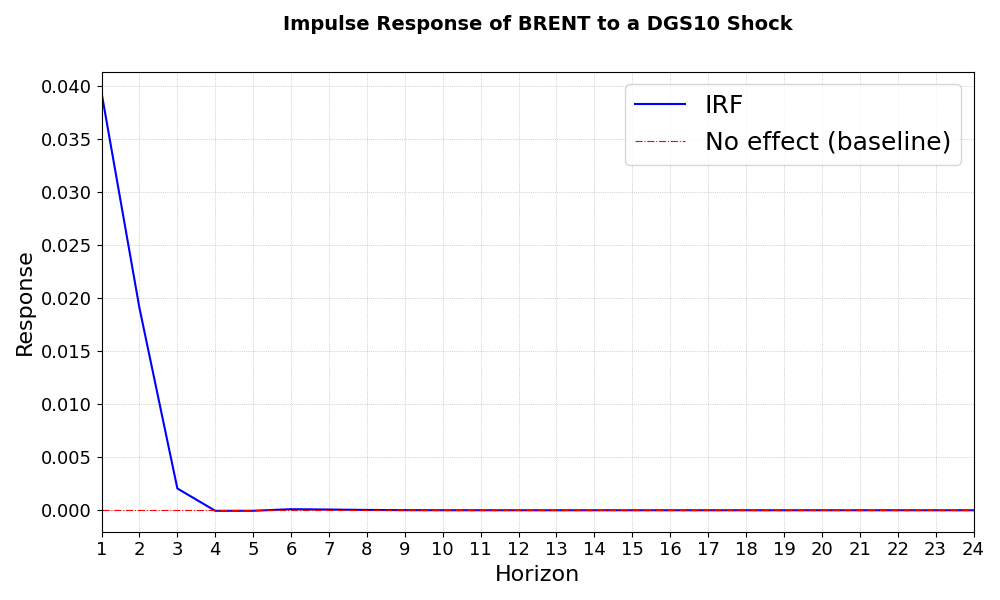}
    \caption{DGS10 Shock on BRENT}
    \label{figA4_m1}
\end{subfigure}
\hfil
\begin{subfigure}[b]{0.3\textwidth}
    \centering
    \includegraphics[width=0.7\hsize]{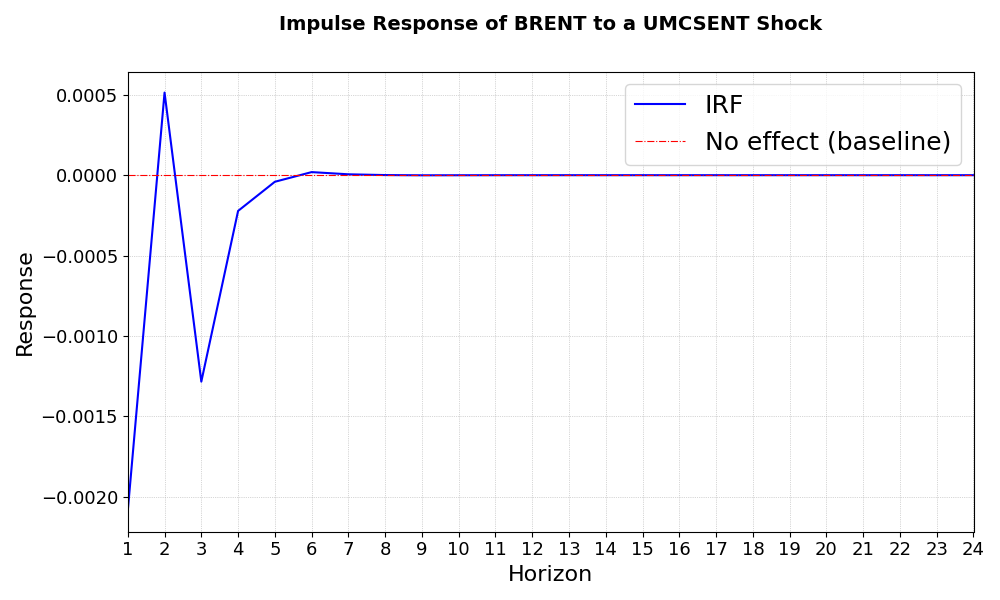}
    \caption{UMCSENT Shock on BRENT}
    \label{figA4_m2}
\end{subfigure}
\hfil
\begin{subfigure}[b]{0.3\textwidth}
    \centering
    \includegraphics[width=0.7\hsize]{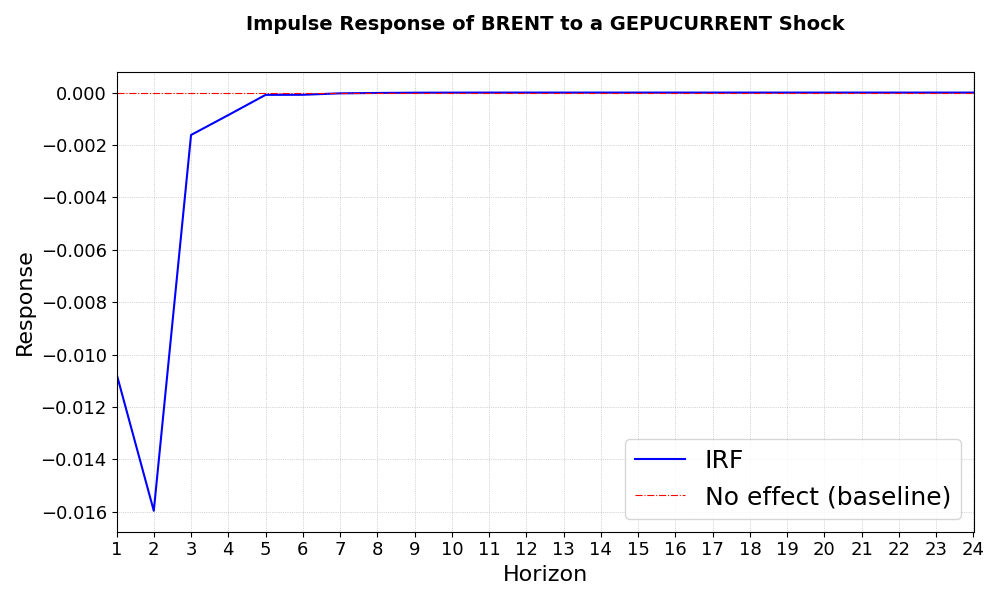}
    \caption{GEPUCURRENT Shock on BRENT}
    \label{figA4_m3}
\end{subfigure}
\hfil
\begin{subfigure}[b]{0.3\textwidth}
    \centering
    \includegraphics[width=0.7\hsize]{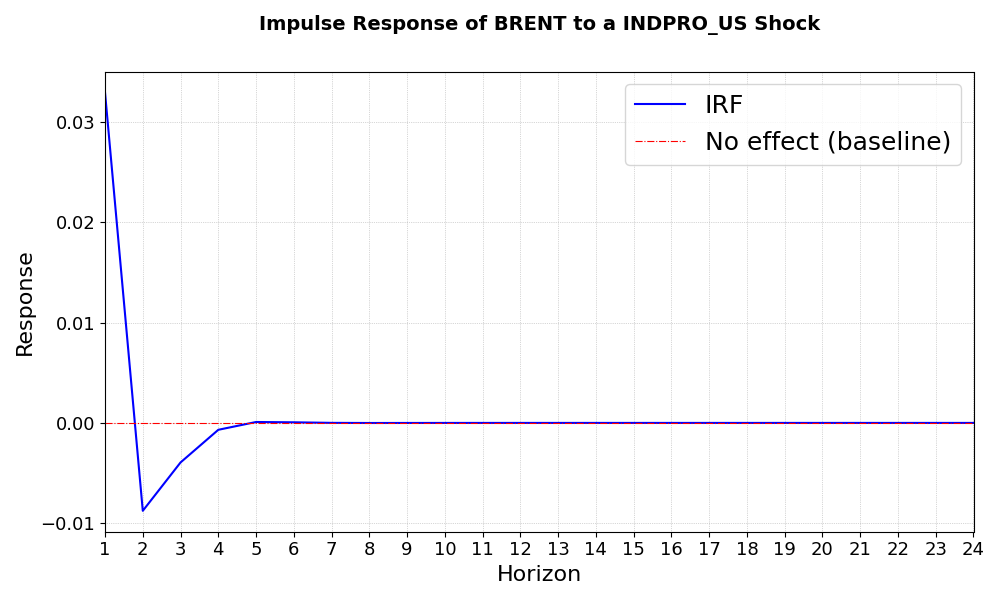}
    \caption{INDPRO$\_$US Shock on BRENT}
    \label{figA4_m5}
\end{subfigure}
\hfil
\begin{subfigure}[b]{0.3\textwidth}
    \centering
    \includegraphics[width=0.7\hsize]{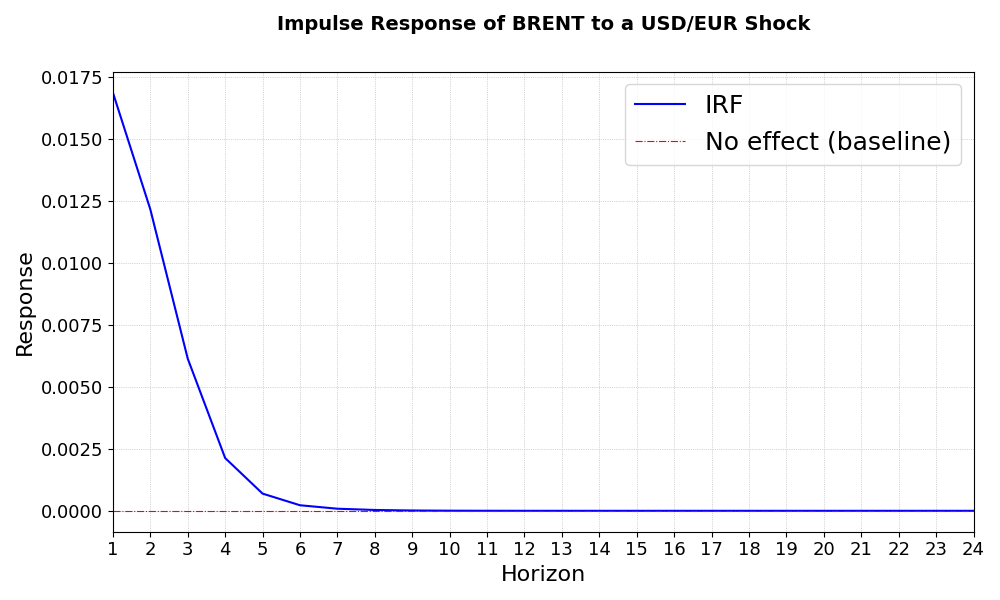}
    \caption{APE Shock on BRENT}
    \label{figA4_m6}
\end{subfigure}
\hfil
\begin{subfigure}[b]{0.3\textwidth}
    \centering
    \includegraphics[width=0.7\hsize]{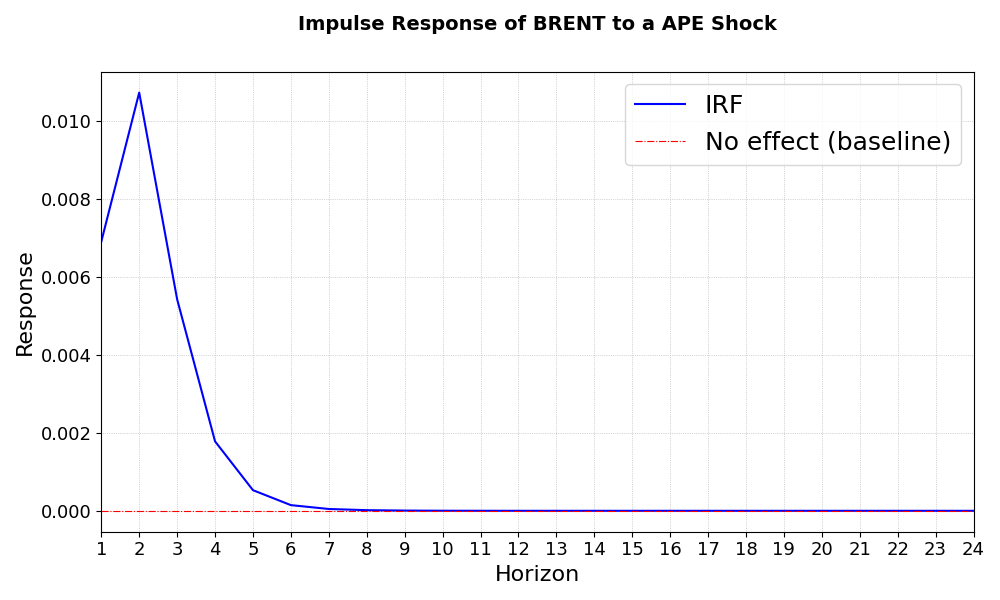}
    \caption{APE Shock on BRENT}
    \label{figA4_m7}
\end{subfigure}
%-----------------------------------%
\caption{Impulse Response for BRENT}
\label{figA4}
\end{sidewaysfigure}

%--------------------%
%     Figure A5      %
%--------------------%
\begin{figure}
%\begin{figure}
\centering
\begin{subfigure}[b]{0.4\textwidth}
    \centering
    \includegraphics[width=1.15\hsize]{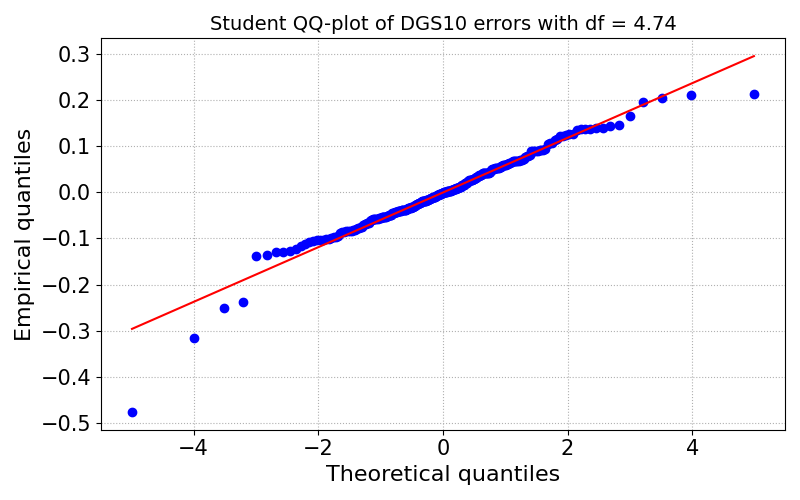}
    \caption{DGS10}
    \label{figA5_m1}
\end{subfigure}
\hfil
\begin{subfigure}[b]{0.4\textwidth}
    \centering
    \includegraphics[width=1.15\hsize]{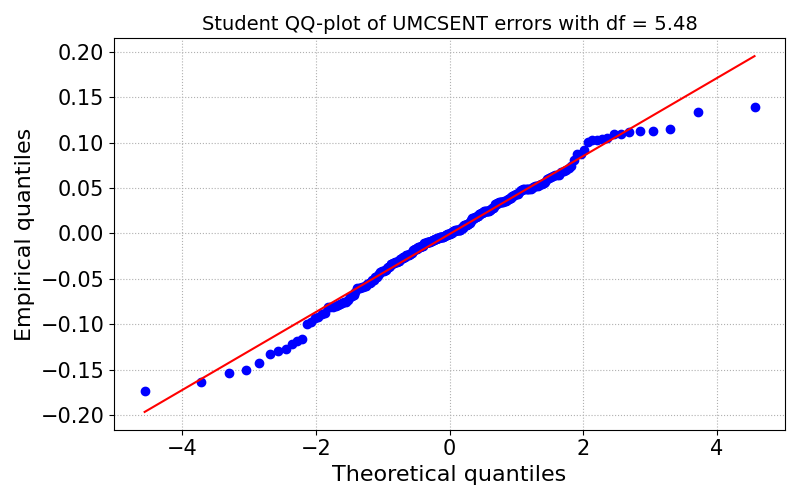}
    \caption{UMSCENT}
    \label{figA5_m2}
\end{subfigure}
\hfil
\begin{subfigure}[b]{0.4\textwidth}
    \centering
    \includegraphics[width=1.15\hsize]{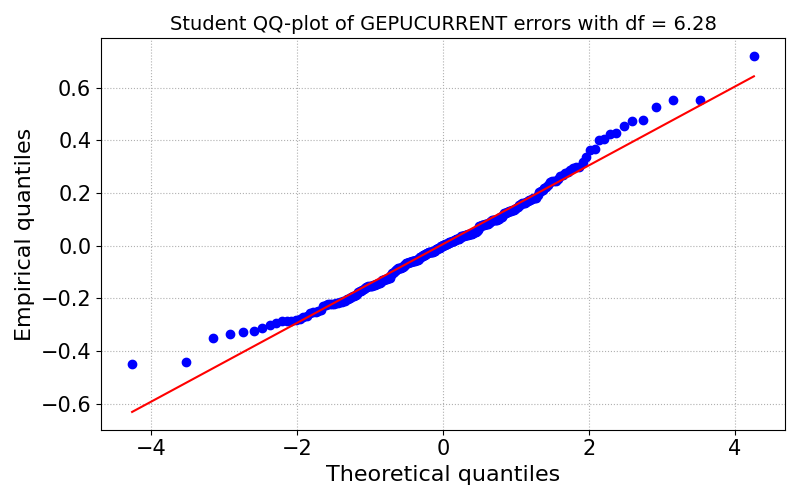}
    \caption{GEPUCURRENT}
    \label{figA5_m3}
\end{subfigure}
\hfil
\begin{subfigure}[b]{0.4\textwidth}
    \centering
    \includegraphics[width=1.15\hsize]{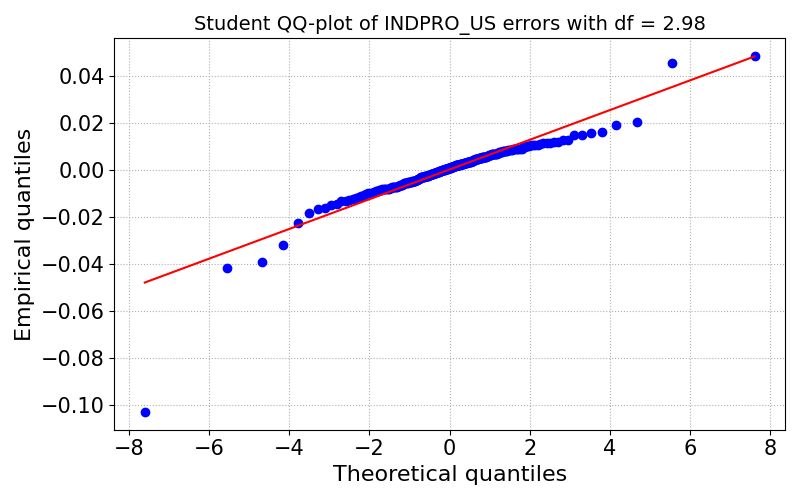}
    \caption{INDPRO$\_$US}
    \label{figA5_m4}
\end{subfigure}
\hfil
\begin{subfigure}[b]{0.4\textwidth}
    \centering
    \includegraphics[width=1.15\hsize]{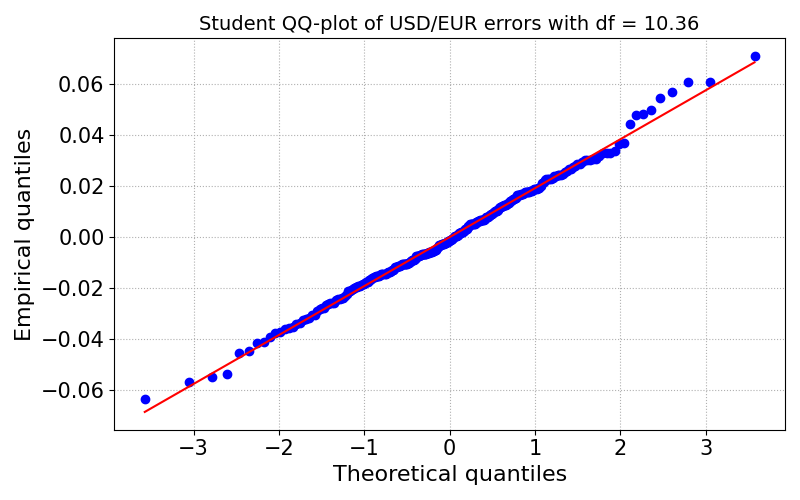}
    \caption{USD/EUR}
    \label{figA5_m5}
\end{subfigure}
\hfil
\begin{subfigure}[b]{0.4\textwidth}
    \centering
    \includegraphics[width=1.15\hsize]{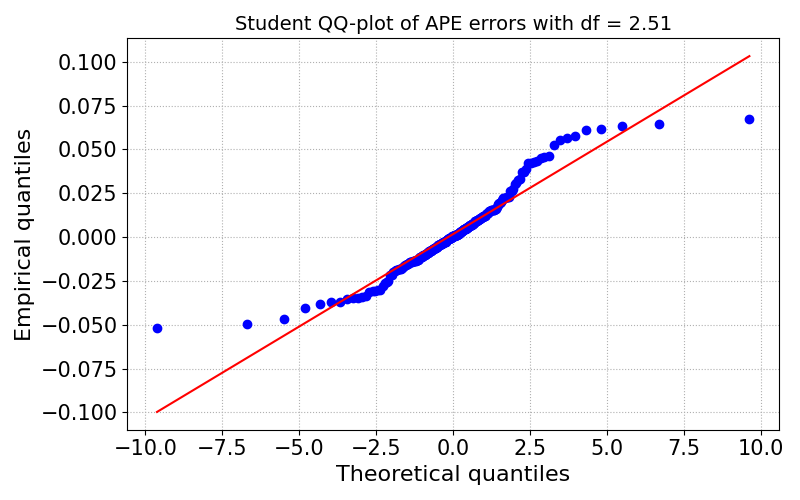}
    \caption{APE}
    \label{figA5_m6}
\end{subfigure}
\hfil
\begin{subfigure}[b]{0.4\textwidth}
    \centering
    \includegraphics[width=1.15\hsize]{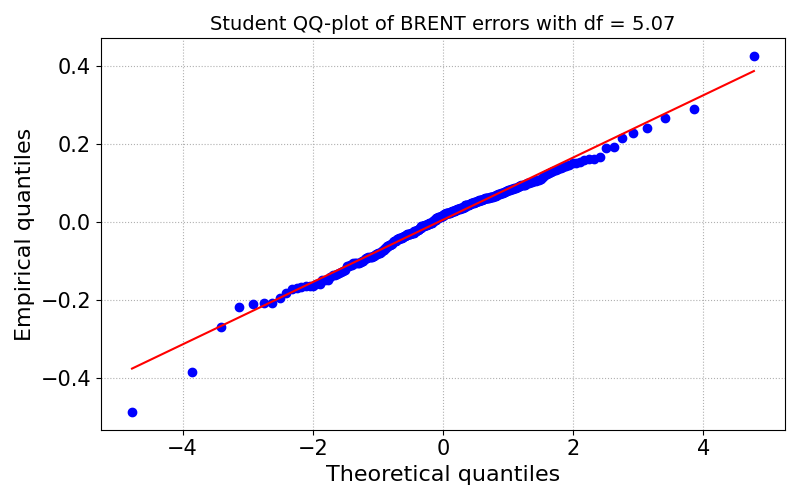}
    \caption{BRENT}
    \label{figA5_m7}
\end{subfigure}
%------------
\caption{Quantile-quantile plots of VAR model residuals}
\label{fig::qqplotVAR}
%------------
\end{figure}

%--------------------%
%     Figure A6      %
%--------------------%
\begin{figure}
%\begin{figure}
\centering
\begin{subfigure}[b]{0.4\textwidth}
    \centering
    \includegraphics[width=1.15\hsize]{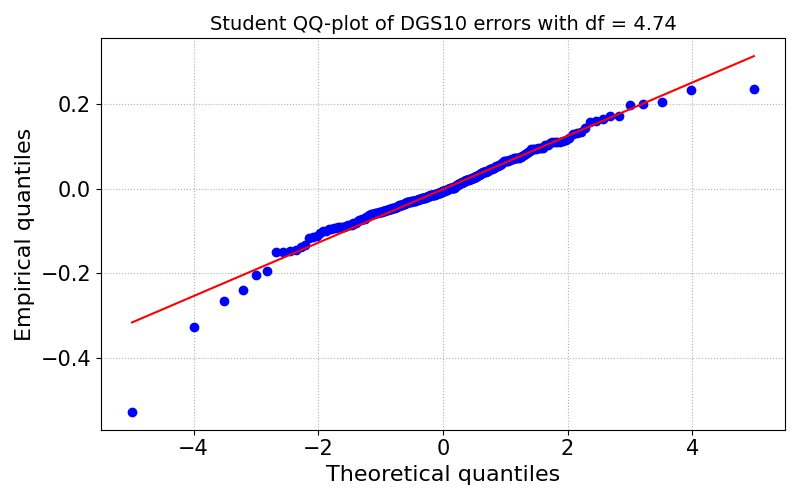}
    \caption{DGS10}
    \label{figA6_m1}
\end{subfigure}
\hfil
\begin{subfigure}[b]{0.4\textwidth}
    \centering
    \includegraphics[width=1.15\hsize]{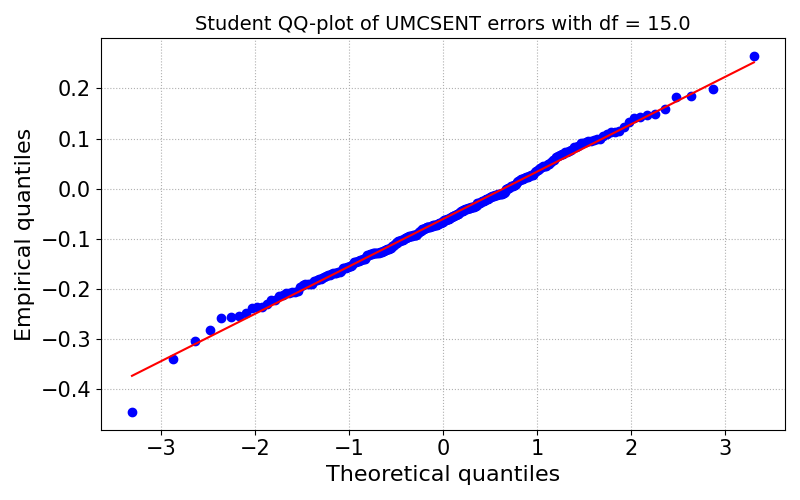}
    \caption{UMSCENT}
    \label{figA6_m2}
\end{subfigure}
\hfil
\begin{subfigure}[b]{0.4\textwidth}
    \centering
    \includegraphics[width=1.15\hsize]{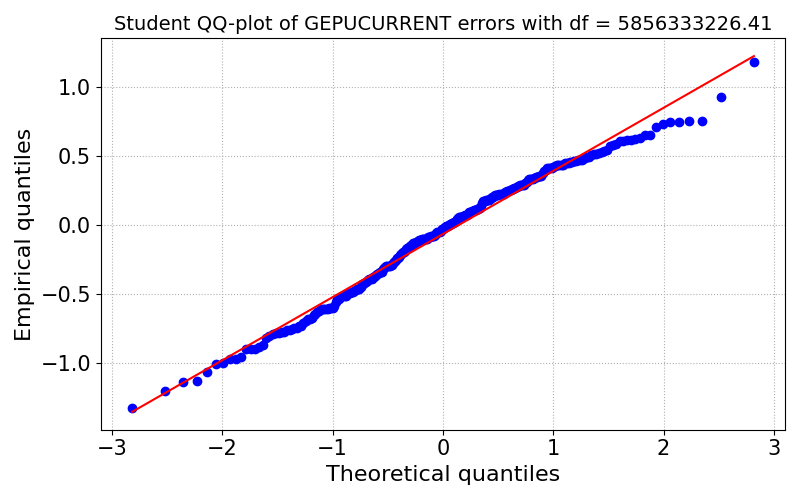}
    \caption{GEPUCURRENT}
    \label{figA6_m3}
\end{subfigure}
\hfil
\begin{subfigure}[b]{0.4\textwidth}
    \centering
    \includegraphics[width=1.15\hsize]{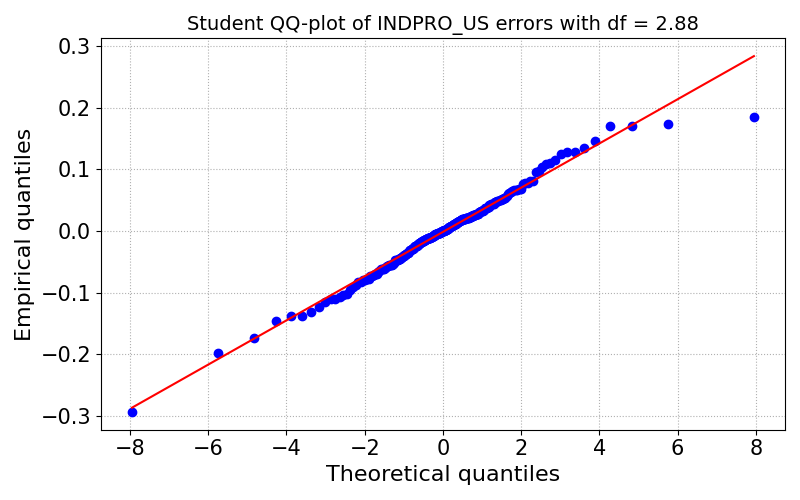}
    \caption{INDPRO$\_$US}
    \label{figA6_m4}
\end{subfigure}
\hfil
\begin{subfigure}[b]{0.4\textwidth}
    \centering
    \includegraphics[width=1.15\hsize]{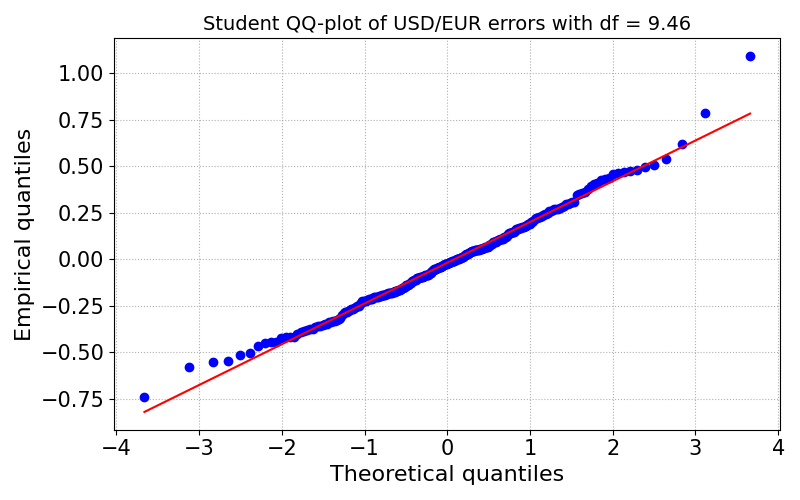}
    \caption{USD/EUR}
    \label{figA6_m5}
\end{subfigure}
\hfil
\begin{subfigure}[b]{0.4\textwidth}
    \centering
    \includegraphics[width=1.15\hsize]{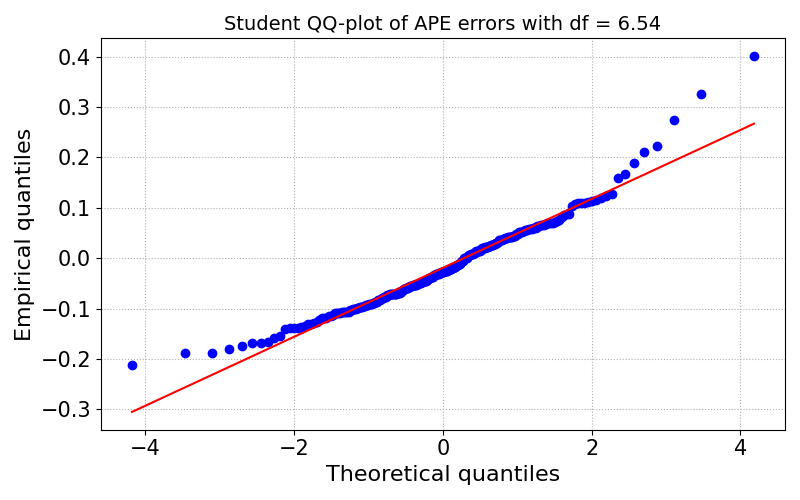}
    \caption{APE}
    \label{figA6_m6}
\end{subfigure}
\hfil
\begin{subfigure}[b]{0.4\textwidth}
    \centering
    \includegraphics[width=1.15\hsize]{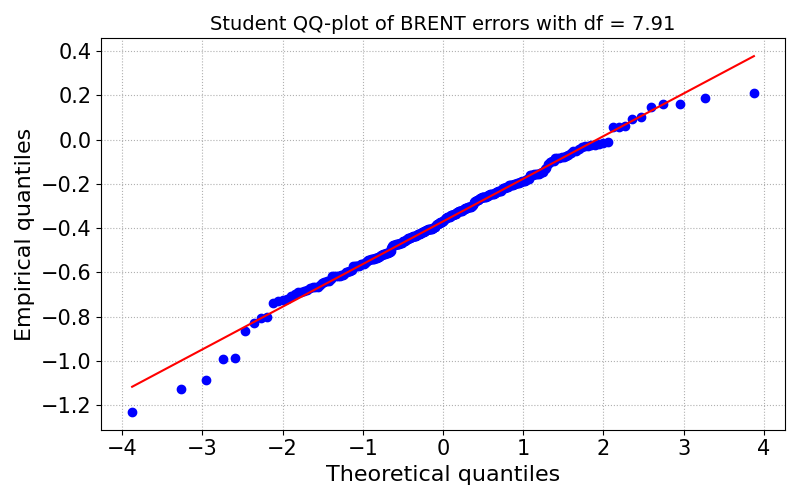}
    \caption{BRENT}
    \label{figA6_m7}
\end{subfigure}
%------------
\caption{Quantile-quantile plots of TVP-SVAR model residuals}
\label{fig::qqplottvpSVAR}
%------------
\end{figure}

\end{appendix}

\end{document}